\RequirePackage{etoolbox}
\csdef{input@path}{%
	{img/}
}%

\documentclass[12pt]{article}
\usepackage{amsmath}
\usepackage{graphicx,psfrag,epsf}
\usepackage{enumerate}
\usepackage{natbib}
\usepackage{url} 
\usepackage[skip=5pt]{caption}
\usepackage[normalem]{ulem}
\usepackage{amsthm}
\usepackage{natbib}
\usepackage{xr-hyper}
\usepackage[colorlinks,citecolor=blue,urlcolor=blue,filecolor=blue,backref=page]{hyperref}
\usepackage{graphicx}
\usepackage{algorithm2e}
\usepackage{amssymb}
\usepackage{amsbsy}
\usepackage{float}
\usepackage{comment}
\usepackage{mathtools}
\usepackage{color,soul}
\usepackage{blkarray}
\usepackage[dvipsnames]{xcolor}
\usepackage{tikz}
\usepackage{color, colortbl}
\usepackage{mathbbol}
\usepackage[T1]{fontenc}
\usepackage{mathrsfs}
\usepackage{hhline}
\usepackage{caption,subcaption}
\usepackage{afterpage}
\usepackage{multirow}
\usepackage{adjustbox}
\usepackage{algorithm2e}
\usepackage{rotating}
\usepackage{tabularx}

\definecolor{Gray}{gray}{0.9}

\allowdisplaybreaks

\usepackage{arydshln}
\makeatletter
\renewcommand*\env@matrix[1][*\c@MaxMatrixCols c]{%
	\hskip -\arraycolsep
	\let\@ifnextchar\new@ifnextchar
	\array{#1}}
\makeatother

\RequirePackage{doi}

\newcommand{\iidsim}{\overset{iid}{\sim}} 

\newcommand{\gpname}{MGP}

\renewcommand{\tilde}[1]{\widetilde{#1}}

\newcommand{\sigmasq}{\sigma^2}

\usepackage{bbm} 

\newcommand{\bolds}[1]{\boldsymbol{#1}}

\newcommand{\calA}{{\cal A}}
\newcommand{\calD}{{\cal D}}
\newcommand{\calG}{{\cal G}}
\newcommand{\calI}{{\cal I}}

\newcommand{\calL}{{\cal L}}
\newcommand{\calM}{{\cal M}}
\newcommand{\calP}{{\cal P}}
\newcommand{\calQ}{{\cal Q}}
\newcommand{\calS}{{\cal S}}
\newcommand{\calT}{{\cal T}}
\newcommand{\calU}{{\cal U}}

\newcommand{\Cov}{\bolds{C}}

\newcommand{\ba}{\bolds{a}}
\newcommand{\bA}{\bolds{A}}
\newcommand{\bb}{\bolds{b}}
\newcommand{\bB}{\bolds{B}}

\newcommand{\bD}{\bolds{D}}

\newcommand{\bE}{\bolds{E}}

\newcommand{\bh}{\bolds{h}}
\newcommand{\bH}{\bolds{H}}

\newcommand{\bI}{\bolds{I}}

\newcommand{\bl}{\bolds{\ell}}

\newcommand{\bO}{\bolds{O}}

\newcommand{\bR}{\bolds{R}}
\newcommand{\bs}{\bolds{s}}

\newcommand{\bt}{\bolds{t}}
\newcommand{\bT}{\bolds{T}}
\newcommand{\bu}{\bolds{u}}

\newcommand{\bv}{\bolds{v}}
\newcommand{\bV}{\bolds{V}}
\newcommand{\bw}{\bolds{w}}

\newcommand{\bX}{\bolds{X}}
\newcommand{\by}{\bolds{y}}

\newcommand{\bz}{\bolds{z}}
\newcommand{\bZ}{\bolds{Z}}

\newcommand{\bzero}{\mathbf{0}}

\newcommand{\bbeta}{\bolds{\beta}}
\newcommand{\beps}{\bolds{\varepsilon}}
\newcommand{\btheta}{\bolds{\theta}}
\newcommand{\bSigma}{\bolds{\Sigma}}
\newcommand{\bGamma}{\bolds{\Gamma}}
\newcommand{\bmu}{\bolds{\mu}}

\newcommand{\bwpa}[1]{\bw_{[{#1}]}}
\newcommand{\bwpaone}[1]{\bw_{[{#1}]_1}}
\newcommand{\bwpatwo}[1]{\bw_{[{#1}]_2}}
\newcommand{\Spa}[1]{\calS_{[{#1}]}}
\newcommand{\pa}[1]{\text{Pa}[{#1}]}
\newcommand{\pazero}[1]{\text{Pa}_0[{#1}]}
\newcommand{\paone}[1]{\text{Pa}_1[{#1}]}
\newcommand{\patwo}[1]{\text{Pa}_2[{#1}]}
\newcommand{\ch}[1]{\text{Ch}[{#1}]}
\newcommand{\bdiag}{\text{blockdiag}}
\newcommand{\tp}{\widetilde{p}}

\newcommand{\matrow}[2]{{#1}\texttt{[{#2},:]}}

\newcommand{\given}{\,|\,}

\newcommand{\meshgpurl}{\if0\blind
	{
		\url{github.com/mkln/meshgp}
	} \fi
	\if1\blind
	{ 
		\url{github.com/}\texttt{[url redacted in blinded version]}
	} \fi}

\usepackage{scalerel,stackengine}
\stackMath
\newcommand\reallywidehat[1]{%
	\savestack{\tmpbox}{\stretchto{%
			\scaleto{%
				\scalerel*[\widthof{\ensuremath{#1}}]{\kern-.6pt\bigwedge\kern-.6pt}%
				{\rule[-\textheight/2]{1ex}{\textheight}}
			}{\textheight}%
		}{0.5ex}}%
	\stackon[1pt]{#1}{\tmpbox}%
}

\usepackage{amsthm}

\theoremstyle{definition}

\newcommand{\blind}{0}

\usepackage[hmargin=1.0in,vmargin=1.0in]{geometry}

\def\mathcolor#1#{\@mathcolor{#1}}
\def\@mathcolor#1#2#3{%
	\protect\leavevmode
	\begingroup
	\color#1{#2}#3%
	\endgroup
}

\newcommand{\footremember}[2]{%
    \footnote{#2}
    \newcounter{#1}
    \setcounter{#1}{\value{footnote}}%
}
\newcommand{\footrecall}[1]{%
    \footnotemark[\value{#1}]%
} 

\date{ }

\begin{document}
	
	\def\spacingset#1{\renewcommand{\baselinestretch}%
		{#1}\small\normalsize} \spacingset{1}

	
	\if0\blind
	{
		\title{\bf Highly Scalable Bayesian Geostatistical Modeling via Meshed Gaussian Processes on Partitioned Domains}
		\author{%
  Michele Peruzzi\footremember{alley}{Department of Forestry, Michigan State University}\footnote{Department of Statistical Science, Duke University}%
  \and Sudipto Banerjee\footnote{Department of Biostatistics, UCLA Fielding School of Public Health}%
  \and Andrew O. Finley\footrecall{alley}
  }
		\maketitle
	} \fi
	
	\if1\blind
	{
		\bigskip
		\bigskip
		\bigskip
		\begin{center}
			{\Large \bf Highly Scalable Bayesian Geostatistical Modeling via Meshed Gaussian Processes on Partitioned Domains}
		\end{center}
		\medskip
	} \fi
	
	\bigskip
	\begin{abstract}
We introduce a class of scalable Bayesian hierarchical models for the analysis of massive geostatistical datasets. The underlying idea combines ideas on high-dimensional geostatistics by partitioning the spatial domain and modeling the regions in the partition using a sparsity-inducing directed acyclic graph (DAG). We extend the model over the DAG to a well-defined spatial process, which we call the Meshed Gaussian Process (MGP). A major contribution is the development of a MGPs on tessellated domains, accompanied by a Gibbs sampler for the efficient recovery of spatial random effects. In particular, the cubic MGP (Q-MGP) can harness high-performance computing resources by executing all large-scale operations in parallel within the Gibbs sampler, improving mixing and computing time compared to sequential updating schemes. Unlike some existing models for large spatial data, a Q-MGP facilitates massive caching of expensive matrix operations, making it particularly apt in dealing with spatiotemporal remote-sensing data. We compare Q-MGPs with large synthetic and real world data against state-of-the-art methods. We also illustrate using Normalized Difference Vegetation Index (NDVI) data from the Serengeti park region to recover latent multivariate spatiotemporal random effects at millions of locations. The source code is available at \meshgpurl.

	\end{abstract}
	
	\noindent%
	
	\spacingset{1.45}
	
	\section{Introduction} \label{spatial:intro}
\begin{figure}
		\centering
		\includegraphics[width=.85\textwidth]{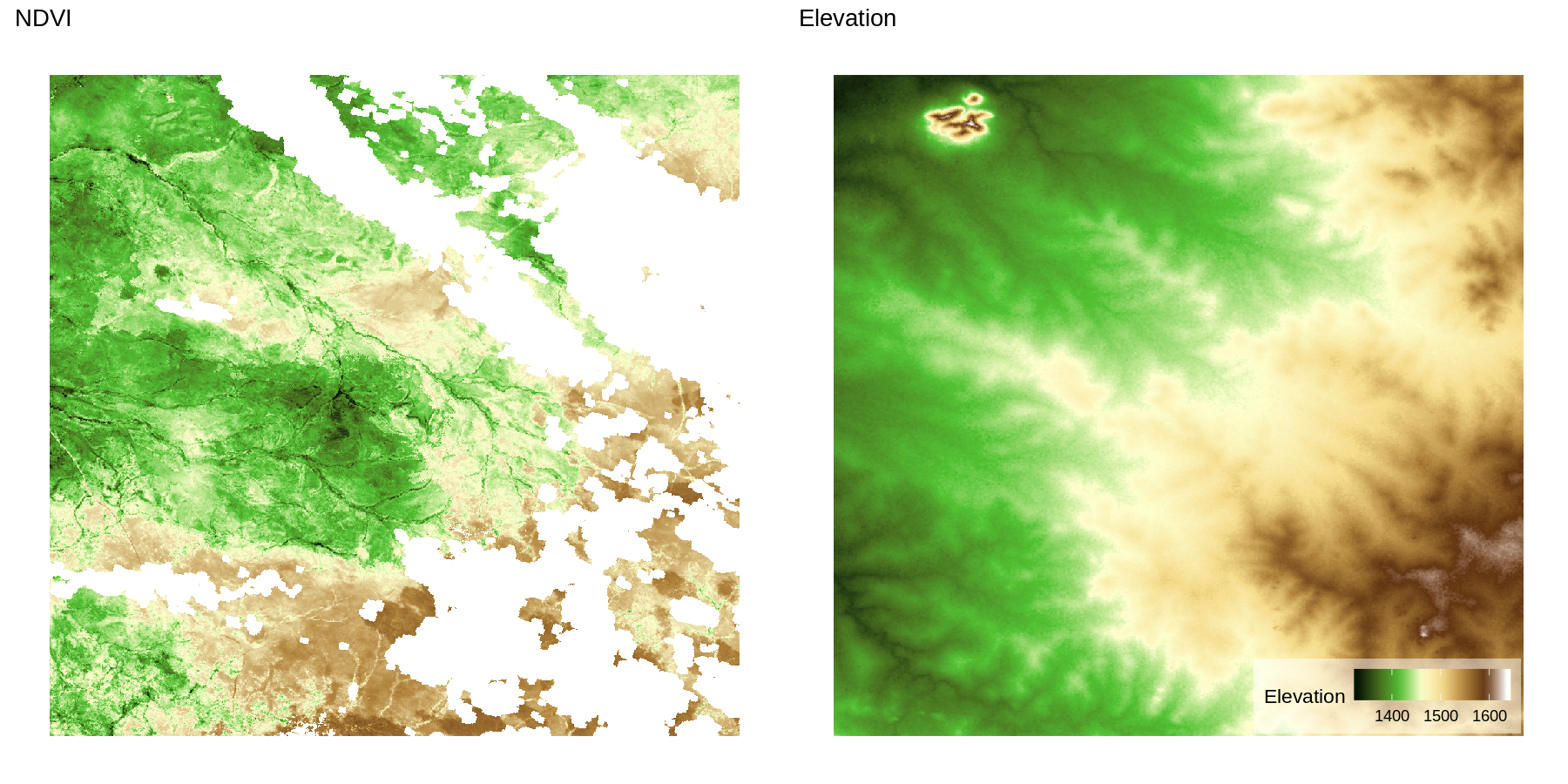}
		\caption{Left: NDVI in the Serengeti region on 2016-12-17. White areas correspond to missing data due to cloud cover. Right: elevation data for the same region.}
		\label{fig:Figure1} 
	\end{figure}

	Collecting large quantities of spatial and spatiotemporal data is now commonplace in many fields. In ecology and forestry, massive datasets are collected using satellite imaging and other remote sensing instruments such as LiDAR that periodically record high-resolution images. Unfortunately, clouds frequently obstruct the view resulting in large regions with missing information. Figure~\ref{fig:Figure1} shows this phenomenon in Normalized Difference Vegetation Index (NDVI) data from the Serengeti region. Filling such gaps in the data is an important goal as is quantifying uncertainty in predictions. This goal is achieved through stochastic modeling of the underlying phenomenon, which involves the specification of a spatial or spatiotemporal process characterizing dependence from a finite realization. Gaussian processes (GP) are a customary choice to characterize spatial dependence, but their implementation is notoriously burdened by their $O(n^3)$ computational complexity. Consequently, intense research has been devoted in recent years to developing scalable models for large spatial datasets -- see detailed reviews by \cite{sunligenton} and \cite{sudipto_ba17}.
	
	Computational complexity can be reduced by considering low-rank models; among these, knot-based methods motivated by ``kriging'' ideas enjoy some optimality properties but oversmooth the estimates of spatial random effects unless the number of knots is large, and require corrections to avoid overestimation of the nugget \citep{gp_predictive_process, frk, gp_pp_biasadj, pp_adaptive_knots, pp_spacetime}. 
	Other methods reduce the computational burden by introducing sparsity in the covariance matrix; strategies include tapering \citep{taper1, taper2} or partitioning of the spatial domain into regions with a typical assumption of independence across regions \citep{fsa, stein2014}. These can be improved by considering a recursive partitioning scheme, resulting in a multi-resolution approximation (MRA; \citealt{katzfuss_jasa17}). Other assumptions on conditional independence assumptions also have a good track record in terms of scalability to large spatial datasets: Gaussian random Markov random fields \citep[GMRF;][]{grmfields}, composite likelihood methods \citep{block_composite_likelihood}, and neighbor-based likelihood approximations \citep{vecchia88} belong to this family. 
	
	The recent literature has witnessed substantial activity surrounding the so called Vecchia approximation \citep{vecchia88}. This approximation can be regarded as a special case of the GMRF approximations with a simplified neighborhood structure motivated from a directed acyclic graphical (DAG) representation of a Gaussian process likelihood. Extensions leading to well-defined spatial processes to accommodate inference at arbitrary locations by extending the DAG representation to the entire domain include Nearest neighbor Gaussian processes (NNGP; \citealt{nngp, nngp_aoas}) and further generalizations by constructing DAGs over the augmented space of outcomes and spatial effects \citep{katzfuss_vecchia}. These approaches render computational scalability by introducing sparsity in the precision matrix. The DAG relies upon a specific topological ordering of the locations, which also determine the construction of neighborhood sets, and certain orderings tend to deliver improved performance of such models \citep{katzfuss_vecchia, guinness_techno}. 
	
	When inference on the latent process is sought, Bayesian inference has the benefits of providing direct probability statements based upon the posterior distribution of the process. Inference based on asymptotic approximations are avoided, but there remain challenges in computing the posterior distribution given that inference is sought on a very high-dimensional parameter space (including the realizations of the latent process). One possibility, available for Gaussian first-stage likelihoods, is to work with a collapsed or marginalized likelihood by integrating out the spatial random effects. However, Gibbs samplers and other MCMC algorithms for the collapsed models can be inexorably slow and are impractical when data are in the millions. A sequential Gibbs sampler that updates the latent spatial effects \citep{nngp} is faster in updating the parameters but suffers from high autocorrelation and slow mixing. Another possibility emerges when interest lies in prediction or imputation of the outcome variable only and not the latent process. Here, a so called ``response'' model that models the outcome itself using an NNGP can be constructed. This model is much faster and enjoys superior convergence properties, but we lose inference on the latent process and its predictive performance tends to be inferior to the latent process model. Furthermore, these options are unavailable in non-Gaussian first-stage hierarchical models or when the focus is not uniquely on prediction. A detailed comparison of different approaches for computing Bayesian NNGP models is presented in \cite{nngp_algos}.

	Our current contribution introduces a class of \emph{Meshed Gaussian Process} (MGP) models for Bayesian hierarchical modeling of large spatial datasets. This class builds upon the aforementioned works that build upon \cite{vecchia88} and other DAG based models. The inferential focus remains within the context of massive spatial datasets over very large domains. We exploit the demonstrated benefits of the DAG based models, but we now adapt them to partitioned domains. We describe dependence across regions of a partitioned domain using a small, patterned DAG which we refer to as a \textit{mesh}. Within each region, some locations are selected as \textit{reference} and collectively mapped to a single node in the DAG. Relationships among nodes are governed by kriging ideas. In the resulting MGP, regions in the spatial domain depend on each other through the reference locations. Realizations at all other locations are assumed independent, conditional upon the reference locations. This construction leads to a valid standalone spatial process. 
	
	As a particular subclass of MGPs, we propose a novel partitioning and graph design based on domain tessellations. Unlike methods that build sparse DAGs by limiting dependence to $m$ nearest neighbors, our approach shapes the underlying DAG with a known, repeating pattern corresponding to the chosen tessellation geometry. The underlying sparse DAG enables scaling computations to large data settings and its known pattern guarantees the availability of block-parallel sampling schemes; furthermore, large computational savings can be achieved at no additional approximation cost if data are collected on patterned lattices. Finally, extensions to spatiotemporal and/or multivariate data are straightforward once a suitable covariance function has been defined. We use axis-parallel domain partitioning and the corresponding cubic DAG -- resulting in cubic MGPs or Q-MGPs -- to show substantial improvements in computational time and inferential performance relative to other models with data sizes ranging from the thousands to the several millions, for both spatial and spatiotemporal data and using multivariate spatial processes.
	
	The present work may appear to share similarities with the block-NNGP model of \cite{prates}, who advocate building sparse DAGs on grouped locations based on their ordering and subsequent identification of $m$ ``past'' neighbors. Unlike block-NNGPs, our tessellated GPs consider the domain tessellation as generating the DAG; the number of parents of any node is thus fixed and depends on the geometry of the chosen tessellation rather than on a user-defined parameter. Inclusion of more distant locations in the parent set of any location will, therefore, not proceed by increasing the number of neighbors $m$, but rather by increasing the regions' size and/or modifying their shape. Central to tessellated GPs is the idea of forcing a DAG with known \textit{coloring} on the data, resulting in guaranteed efficiencies when recovering the latent spatial effects. This strategy is analogous in spirit to multi-resolution approximations \citep{katzfuss_jasa17, treed}, which also force a DAG on the data, resulting in conditional independence patterns that are known in advance and that can be used to improve computations. However, while multi-resolution approximations are defined by branching graphs associated to recursive domain partitioning, tessellated GPs use a single domain partition, with each region connected in the DAG only to its immediate neighbors. Compared to treed graphs, tessellated GPs are associated to DAGs with fewer conditionally-independent groups and whose repeated patterns facilitate the identification of redundant matrix operations arising when one or more coordinate margins are gridded. 
	We also note that while the idea of partitioning domains to create approximations is not new, construction of the DAG-based approximation over partitioned domains has received considerably less attention. Finally, our focus here is in developing tessellated GPs as a methodology that enables the efficient recovery of the latent spatial random effects and the Bayesian estimation of covariance parameters via MCMC; we are thus not focusing on alternative computational algorithms (see e.g. \citealt{nngp_algos}), which have been developed for NNGPs but can nonetheless all be adapted to general MGP models.
	
	The balance of this paper proceeds as follows. Section~\ref{section:general} introduces our general framework for hierarchical Bayesian modeling of spatial processes using networks of grouped spatial locations. The MGP is outlined in Section~\ref{section:general:gaussian}, where we provide a general, scalable computing algorithm in Section~\ref{section:general:estimation}. Tessellation-based schemes and the specific case of Q-MGPs are outlined in Section~\ref{section:qmeshgp}, which highlights their properties and computational advantages. We illustrate the performance of our proposed approach in Section~\ref{section:data} using simulation experiments and an application on a massive dataset with millions of spatiotemporal locations. We conclude the paper with a discussion and pointers to further research. Supplementary material accompanying this manuscript as an Appendix is available online and contains further comparisons of Q-MGPs with several state-of-the-art methods for spatial data.
	
	\section{Spatial processes on partitioned domains} \label{section:general}
	A $q \times 1$ spatial process assigns a probability law on $\{\bw(\bl) : \bl\in\calD\}$, where $\bw(\bl)$ is a $q\times 1$ random vector with elements $w_i(\bl)$ for $i=1,2,\ldots,q$. In the following general discussion we will not distinguish between spatial ($\calD \subset \Re^d$) and spatiotemporal domains ($\calD \subset \Re^{d+1}$), and denote spatial or spatiotemporal locations as $\bl, \bs$, or $\bu$. 
	
	For any finite set of  spatial locations $\{\bl_1,\bl_2, \ldots, \bl_{n_{\calL}}\} = \calL \subset \calD$ of size $n_{\calL}$, let $P(\cdot)$ denote the probability law of the $n_{\calL}q\times 1$ random vector $\bw_{\calL} = (\bw(\bl_1)^{\top},\bw(\bl_2)^{\top},\ldots,\bw(\bl_{n_{\calL}})^{\top})^{\top}$ with probability density $p(\cdot)$. The joint density of $\bw_{\calL}$ can be expressed as a DAG (or a Bayesian network model) with respect to the ordered set of locations $\calL$ as
	\begin{align} \label{section:general:eq:joint_density_indiv}
	p(\bw_{\calL}) &= \prod_{i=1}^{n_{\calL}} p( \bw(\bl_{i}) \given \bw(\bl_1), \ldots, \bw(\bl_{i-1})),
	\end{align}
	where the conditional set for each $\bw(\bl_i)$ can be interpreted as the set of its parents in a large, dense Bayesian network. Defining a simplified valid joint density on $\calL$ by reducing the size of the conditioning sets is a popular strategy for fast likelihood approximations in the context of large spatial datasets. One typically limits dependence to ``past'' neighboring locations with respect to the ordering in (\ref{section:general:eq:joint_density_indiv}) \citep{vecchia88, gramacy_apley14,steinetal2004,nngp,katzfuss_vecchia}. The neighbors are defined and fixed and model performance may benefit from the addition of some distant locations \citep{steinetal2004}. The ordering in $\calL$ is also fixed and inferential performance may benefit from the use of some fixed permutations \citep{guinness_techno}. The result of shrinking the conditional sets to a smaller set of neighbors from the past yields a sparse DAG or Bayesian network, which yields potentially massive computational gains. 
	
	We proceed in a similar manner, but instead of defining a sparse DAG at the level of each individual location, we map entire groups of locations to nodes in a much smaller graph; the same graph will be used to model the dependence between any location in the spatial domain and, therefore, to define a spatial process. Let $\calP = \{\calD_1, \dots, \calD_M\}$ be a partition of $\calD$ into $M$ mutually exclusive subsets so that $\calD = \cup_{i=1}^M \calD_i$ and $\calD_i \cap \calD_j = \emptyset$ whenever $i\neq j$. Similar to the nomenclature in the NNGP, we fix a \emph{reference set} $\calS = \{ \bs_1, \dots, \bs_{n_{\calS}} \} \subset \calD$, which itself is partitioned using $\calP$ by letting $\calS_j = \calD_j \cap \calS$. The set of non-reference locations is similarly partitioned with $\calU_j = \calD_j \setminus \calS_j$ so that $\calD_j = \calS_j\cup \calU_j$ for each $j=1,2,\ldots,M$. We now construct a DAG to model dependence within and between $\calS$ and $\calU$. Let $\calG = \{ \bV, \bE \}$ be a graph with nodes $\bV = \bA \cup \bB$, where we refer to $\bA = \{ \ba_1, \dots, \ba_M \}$ as the \textit{reference} nodes and to $\bB = \{ \bolds{b}_1, \dots, \bolds{b}_M\} $ as the \textit{non-reference}, or simply ``other'', nodes. Let $\bA \cap \bB = \emptyset$. We introduce a map  $\eta: \calD \rightarrow \bV$ such that 
	\begin{align}\label{eq: map}
	\eta(\bl) &= \left\{\begin{array}{l}
	\ba_j \in \bA\; \mbox{ if }\; \bl \in \calS_j\\
	\bolds{b}_j \in \bB \; \mbox{ if }\; \bl \in \calU_j
	\end{array} \right. \;.
	\end{align}
	This surjective many-to-one map links each location in $\calS_j$ and $\calU_j$ to a node in $\calG$. The edges connecting nodes in $\calG$ are $\bE = \{ \pa{\bv_1}, \dots, \pa{\bv_{2M}} \}$ where $\pa{\bv} \subset \bV$ denotes the set of parents of any $\bv \in \bV$ and, hence, identifies the directed edges pointing to $\bv$. We let $\mathcal{G}$ be acyclic, i.e., there is no chain $\{\bv_{i_1}\to\bv_{i_2}\to \cdots \to\bv_{i_t}\}$ of elements of $\bV$ such that $\bv_{i_j} \in \pa{\bv_{i_{j+1}}}$ and $\bv_{i_{j+1}} \in \pa{\bv_{i_1}}$. Crucially, we assume that $\pa{\bv} \subset \bA$ for all $\bv \in \bV$, i.e., that only reference nodes have children, to distinguish the reference nodes $\bA$ from the other nodes $\bB$. Apart from the assumption that $\ba_j \in \pa{\bolds{b}_j}$, we refrain from defining the parents of a node, thereby retaining flexibility. In general, however, all locations in $\calU_j$ will share the same parent set. 
	In Section~\ref{section:qmeshgp} we will consider meshes associated to domain tessellations.
	
	Consider the enumeration $\calS_i = \{\bs_{i_1}, \dots, \bs_{i_{n_i}} \}$, where $\{i_1,i_2,\ldots,i_{n_i}\}\subset \{1,2,\ldots,n_{\calS}\}$, and let $\bw_i = (\bw(\bs_{i_1})^{\top}, \bw(\bs_{i_2})^{\top}, \ldots, \bw(\bs_{i_{n_i}})^{\top})^{\top}$ be the $n_iq\times 1$ random vector listing elements of $\bw(\bs)$ for each $\bs\in \calS_i$. We now rewrite (\ref{section:general:eq:joint_density_indiv}) as a product of $M$ conditional densities
	\begin{align} \label{section:general:eq:joint_density_group}
	p(\bw_{\calS}) &= p(\bw_1, \bw_2, \ldots, \bw_M) = \prod_{i=1}^{M} p(\bw_i \given \bw_1, \ldots, \bw_{i-1}).
	\end{align}
	The conditioning sets are then reduced based on the graph $\calG$:
	\begin{align} \label{section:general:eq:pS}
	\tilde{p}(\bw_{\calS}) &= \prod_{i=1}^{M} p( \bw_i \given \bwpa{i})\;,
	\end{align}
	where we denote $\bwpa{i} = \{\bw_j : \ba_j \in \pa{\ba_i}\}$, and $\pa{\ba_i} \subset \{ \ba_1, \dots, \ba_{i-1} \} \subset \bA$. This is a proper multivariate joint density since the graph is acyclic \citep{lauritzen}. 
	It is instructive to note how the above approximation behaves when the size of the parent set shrinks, for a given domain partitioning scheme. To this end, we adapt a result in \cite{sudipto_ss20} and show that sparser DAGs correspond to a larger Kullback-Leibler (KL) divergence from the base density $p$. This result has been proved earlier for Gaussian likelihoods by \cite{guinness_techno}, but the argument given below is free of distributional assumptions and is linked to the submodularity of entropy and the ``information never hurts'' principle \citep[see e.g.][]{coverthomas91}. 
	
	Consider random vector $\bw$ and some partition of the domain $\calP$ corresponding to nodes $\bV = \{ \bv_1, \dots, \bv_M \}$ via map $\eta$. Let the base process correspond to graph $\calG_0 = \{ \bV, \bE_0 \}$ where $\bE_0 = \{ \pazero{\bv_1}, \dots, \pazero{\bv_M}\}$. Then, let $\calG_1 = \{ \bV, \bE_1 \}$ where $\bE_1 = \{ \paone{\bv_1}, \dots, \paone{\bv_M} \}$ and $\paone{\bv_{i}} \subseteq \pazero{\bv_{i}}$ for all $i \in \{1, \dots, M\}$. Finally construct $\calG_2 = \{ \bV, \bE_2 \}$ by letting $\patwo{\bv_{i^*}} = \paone{\bv_{i^*}} \setminus \{ \bv^* \} $ for some $\bv^* \in \paone{\bv_{i^*}}$. In other words, graph $\calG_2$ is obtained by removing the directed edge $\bv^* \to \bv_{i^*}$ from $\calG_1$. We approximate $p$ using densities $p_1$ and $p_2$ based on $\calG_1$ and $\calG_2$, respectively, obtaining
	\begin{equation}\label{eq:ratio_densities} \frac{p_1(\bw)}{p_2(\bw)} = \prod_{i=1}^{M} \frac{ p( \bw_i \given \bwpaone{i}) }{ p( \bw_i \given \bwpatwo{i}) } = \frac{ p(\bw_{i^*} \given \bwpaone{i^*}) }{ p(\bw_{i^*} \given \bwpatwo{i^*}) }.
	\end{equation}
	Considering the Kullback-Leibler divergence of each density from $p$, and denoting $\bV^* = \bV \setminus \{\{ i^*\} \cup \paone{i^*}\}$, we find
	
	\begin{equation}\label{KL_approx}
	\begin{aligned}
	KL(p_2 \| p) - KL(p_1 &\| p) = \int \left\{ \log\left(\frac{p(\bw)}{p_2(\bw)}\right) - \log\left(\frac{p(\bw}{p_1(\bw} \right) \right\} p(\bw) d\bw \\
	&= \int \log\left(\frac{p_1(\bw)}{p_2(\bw)}\right) p(\bw) d\bw = \int \log \left(\frac{ p(\bw_{i^*} \given \bwpaone{i^*}) }{ p(\bw_{i^*} \given \bwpatwo{i^*}) }\right) p(\bw) d\bw\\
	&= \int \log \left(\frac{ p(\bw_{i^*} \given \bwpaone{i^*}) }{ p(\bw_{i^*} \given \bwpatwo{i^*}) }\right) p(\bw_{i^*}, \bwpaone{i^*}) d\bw_{i^*} d\bwpaone{i^*} \\
	&= \int \left\{ \int \log \left(\frac{ p(\bw_{i^*} \given \bwpaone{i^*}) }{ p(\bw_{i^*} \given \bwpatwo{i^*}) }\right)  p(\bw_{i^*} \given \bwpaone{i^*})  d\bw_{i^*} \right\} p(\bwpaone{i^*}) d\bwpaone{i^*} \geq 0, 
	\end{aligned}
	\end{equation} 
	where we use (\ref{eq:ratio_densities}), the fact that $\bV^*$ and $ \{i^* \} \cup \paone{i^*}$ are disjoint, and Jensen's inequality. This result implies that larger parent sets are preferrable as they correspond to better approximations to the full model; the choice of sparser graphs will be driven by computational considerations -- see Section \ref{section:general:estimation:complexity}.
	
	We construct the spatial process over arbitrary locations by enumerating other locations as $\calU = \{\bu_1, \dots, \bu_{n_{\calU}}\} \subset \calD \setminus \calS$ and extending (\ref{section:general:eq:pS}) to the non-reference locations. Given the partition of $\calU$ defined earlier with components $\calU_j$ for $j=1,2,\ldots,M$, for each $\bu \in \calU_j$ we set $\eta(\bu) = \bolds{b}_j$ and recall that $\pa{\bolds{b}_i} \subset \bA$ by construction. For each $i=1, \dots, n_{\calU}$, we denote $\bwpa{\bu_i} = \{\bw_j : \bolds{a}_j \in \pa{\eta(\bu_i)}\} \subset \bw_{\calS}$ 
	and define the conditional density of $\bw_{\calU}$ given $\bw_{\calS}$ as
	\begin{align} \label{section:general:process:eq:pUgivenS}
	\tilde{p}(\bw_{\calU} \mid \bw_{\calS}) &= \prod_{\bu_i \in \calU} p(\bw({\bu_i}) \mid \bwpa{\bu_i}) = \prod_{j=1}^{M} p( \bw_{\calU_j} \mid \bwpa{\bolds{b}_j}).
	\end{align}
	Therefore, for any finite subset of spatial locations $\mathcal{L} \subset \mathcal{D}$ we can let $\calU = \mathcal{L} \setminus \calS$ and obtain
	\begin{align*}
	\tilde{p}(\bw_\mathcal{L}) &= \int \tilde{p}(\bw_{\calU} \mid \bw_{\calS}) \tilde{p}(\bw_{\calS}) \prod_{ \bs_i \in \calS \setminus \mathcal{L}} d(\bw(\bs_i))\; .
	\end{align*}
	We show 
	(see Appendix A, available online) that this is a well-defined process by verifying the Kolmogorov consistency conditions. This new process can be built starting from a base process, a fixed reference set, domain partition $\calP$ and a graph $\calG$. Next, we elucidate with Gaussian processes.
	
	\section{Meshed Gaussian Processes} \label{section:general:gaussian}
	Let $\{\bw(\bl) : \bl \in \calD\}$ be a $q$-variate multivariate Gaussian process, denoted as $\bw(\bl) \sim GP(\bzero, \Cov(\cdot, \cdot \mid \btheta))$. The \emph{cross-covariance} $\Cov(\cdot, \cdot \mid \btheta)$ indexed by parameters $\btheta$ is a function $\Cov: \calD\times\calD \rightarrow {\cal M}_{q\times q}$, where ${\cal M}_{q\times q}$ is a subset of $\Re^{q\times q}$ (the space of all $q\times q$ real matrices) such that the $(i,j)$-th entry of $\Cov(\bl,\bl'\mid \btheta)$ evaluates the covariance between the $i$-th and $j$-th elements of $\bw(\bl)$ at $\bl$ and $\bl'$, respectively, i.e., $\mbox{cov}(w_i(\bl),w_j(\bl'))$. We omit dependence on $\btheta$ to simplify notation. The cross-covariance function itself needs to be neither symmetric nor positive-definite, but must satisfy the following two properties: (i) $\Cov(\bl,\bl') = \Cov(\bl',\bl)^{\top}$; and (ii) $\sum_{i=1}^n\sum_{j=1}^n \bz_i^{\top}\Cov(\bl_i,\bl_j)\bz_j > 0$ for any integer $n$ and any finite collection of points $\{\bl_1,\bl_2,\ldots,\bl_n\}$ and for all $\bz_i \in \Re^{q}\setminus \{\bolds{0}\}$. See \cite{genton_ccov} for a review of cross-covariance functions for multivariate processes. The (partial) realization of the multivariate process over any finite set $\calL$ has a multivariate normal distribution $\bw_{\calL} \sim N(0, \Cov_{\calL})$ where $\bw_{\calL}$ is the $q n_{\calL} \times 1$ column vector and $\Cov_{\calL}$ is the $qn_{\calL} \times qn_{\calL}$ block matrix with the $q\times q$ matrix $\Cov(\bl_i, \bl_j)$ as its $(i,j)$ block for $ i,j=1, \dots, n_{\calL}$. 
	
	We construct the MGP from a base, or \emph{parent}, multivariate GP for $\bw(\bl)$ and then, using the graph $\calG$ defined in Section~\ref{section:general}, represent the joint density at the reference set $\calS$ as
	\begin{align} \label{section:general:eq:pS:gaussian}
	\tilde{p}(\bw_{\calS}) &= \prod_{j=1}^{M} N( \bw_j \mid \bH_{j} \bwpa{j}, \bR_{j}),
	\end{align}
	where $\bH_1 = \bO_{n_1 \times 1}$, $\bR_1 = \Cov_{\calS_j}$ and for $j>1$, $\bH_{j} = \Cov_{\calS_j, \Spa{j}} \Cov^{-1}_{\Spa{j}} $ and $\bR_{j} = \Cov_{\calS_j} - \Cov_{\calS_j, \Spa{j}}\Cov^{-1}_{\Spa{j}} \Cov_{\Spa{j}, \calS_j}$.
	The resulting joint density $\tilde{p}(\bw_{\calS})$ is multivariate normal with covariance $\tilde{\Cov}_{\calS}$ and a precision matrix $\tilde{\Cov}^{-1}_{\calS}$. 
	The precision matrix for Gaussian graphical models is easily derived using customary linear model representations for each conditional regression. Consider the DAG in (\ref{section:general:eq:pS}). Each $\bw_i$ is $n_iq\times 1$ and let $J_i = |\pa{\ba_i}|$ be the number of parents for $\ba_i$ in the graph ${\cal G}$. Furthermore, let $\Cov_{i,j}$ be the $n_iq\times n_jq$ covariance matrix between $\bw_i$ and $\bw_j$, $\Cov_{i,[i]}$ be the $n_iq\times J_iq$ covariance matrix between $\bw_i$ and $\bw_{[i]}$, and $\Cov_{[i],[i]}$ be the $J_iq\times J_iq$ covariance matrix between $\bwpa{i}$ and itself. Representing each conditional density in (\ref{section:general:eq:pS}) as a linear regression on $\bw_i$, we get
	\begin{equation}\label{section:general:eq:cond_reg}
	\bw_1 = \bolds{\omega}_1 \sim N(\bolds{0},\bR_1)\;; \quad \bw_i = \sum_{\{j : \ba_j\in \pa{\ba_i}\}} \bH_{ij}\bw_j + \bolds{\omega}_i\;,\;\; i=2,3,\ldots,M\;,
	\end{equation}
	where each $\bH_{ij}$ is an $n_iq \times n_jq$ is a coefficient matrix representing the multivariate regression of $\bw_j$ given $\bw_{[i]}$, $\bolds{\omega}_i \stackrel{ind}{\sim} N(\bolds{0}, \bR_i)$ for $i=1,2,\ldots,M$, and each $\bR_i$ is an $n_iq\times n_iq$ residual covariance matrix. We set $\bH_{ii}=\bO$ and $H_{ij}=\bO$, where $\bO$ is the matrix of zeros, whenever $j \in \{j : \ba_j \notin \pa{\ba_i}\}$. For $j\in \{j : \ba_j \in \pa{\ba_i}\}$, let $\{j_1, j_2, \ldots, j_{J_i}\}$ be the indices in $\pa{\ba_i}$ and let $\bH_{i,[i]} = \left[\bH_{i,j_1},\bH_{i,j_2},\ldots,\bH_{i,j_{J_i}}\right]$ be the $n_iq\times (\sum_{k=1}^{J_i}n_{j_k})q$ block matrix formed by stacking $\bH_{i,j_k}$ side by side for each $\ba_{j_k}\in \pa{\ba_i}$. Since $\mbox{E}[\bw_i\given\bw_{[i]}] = \bH_{i,[i]}\bw_{[i]} = \Cov_{i,[i]}\Cov_{[i][i]}^{-1}\bw_{[i]}$, we obtain $\bH_{i,[i]} = \Cov_{i,[i]}\Cov_{[i][i]}^{-1}$  and each $\bH_{ij_K}$ can be obtained from the respective submatrix of $\bH_{i[i]}$. We also obtain $\bR_i = \mbox{var}\{\bw_i\given \bw_{[i]}\} = \Cov_{i,i} - \Cov_{i,[i]}\Cov_{[i][i]}^{-1}\Cov_{[i],i}$. Therefore, all the $\bH_{ij}$'s and $\bR_i$'s can be computed from the base cross-covariance function. 
	
	The distribution of $\bw = [\bw_1^{\top}, \bw_2^{\top},\ldots,\bw_M^{\top}]^{\top}$ can be obtained by noting that $\bw = \bH\bw + \bolds{\omega}$, where $\bH = \{\bH_{ij}\}$ is the $(\sum_{i=1}^Mn_iq)\times(\sum_{i=1}^Mn_iq)$ block matrix with $\{\bH_{ij}\}$ as $(i,j)$-th block. Therefore, $\tilde{\Cov}_{\calS} = \mbox{var}(\bw) = (\bI - \bH)^{-1}\bR(\bI - \bH)^{-\top}$, where $\bR$ is block-diagonal with $\bR_i$ as the $(i,i)$-th block. Note that $\bI-\bH$ is block lower-triangular with $1$'s on the diagonal, hence non-singular. Also, the precision matrix $\tilde{\Cov}^{-1}_{\calS} = (\bI - \bH)^{\top}\bR^{-1}(\bI-\bH)$ is sparse because of $\bH_{ij}=\bO$ whenever $\ba_j\notin \pa{\ba_i}$. Block-sparsity of $\tilde{\Cov}^{-1}_{\calS}$ can be induced by building $\mathcal{G}$ with few, carefully placed directed edges among nodes in $\bA$; Appendix B, available online, 
	contains a more in-depth treatment. 
	We extend (\ref{section:general:eq:pS:gaussian}) to the collection of non-reference locations $\calU \subset \calD\setminus \calS$:
	\begin{align} \label{section:general:eq:pUgivenS:gaussian}
	\tp( \bw_{\calU} \mid \bw_{\calS}) &= \prod_{j=1}^{M} N( \bw_{\calU_j} \mid \bH_{\calU_j}\bwpa{\bolds{b}_j}, \bR_{\calU_j} ) = N(\bw_{\calU} \mid \bH_{\calU} \bw_{\calS}, \bR_{\calU}),
	\end{align}
	where $\bH_{\calU_j} = \Cov_{\calU_j, \Spa{\bolds{b}_j}} \Cov^{-1}_{\Spa{\bolds{b}_j}} $ and $\bR_{\calU_j} = \Cov_{\calU_j} - \Cov_{\calU_j, \Spa{\bolds{b}_j}} \Cov^{-1}_{\Spa{\bolds{b}_j}} \Cov_{\Spa{\bolds{b}_j}, \calU_j}$, analogously to (\ref{section:general:eq:pS:gaussian}), while $ \bH_{\calU}$ and $\bR_{\calU}$ are analogous to $ \bH_{\calS}$ and $\bR_{\calS}$. Clearly, given that all the $\tp$ densities are Gaussian, all finite dimensional distributions will also be Gaussian. We have constructed a Gaussian process with the following cross-covariance function for any two locations $\bl_1, \bl_2 \in \calD$
	\begin{align*}
	Cov_{\tp}(\bw(\bl_1), \bw(\bl_2)) &=
	\begin{cases}
	\tilde{\Cov}_{\bs_i, \bs_j} & \text{if\ } \bl_1 = \bs_i, \bl_2 = \bs_j \text{ and } \bs_i, \bs_j \in \calS\\
	\bH_{\bl_1} \tilde{\Cov}_{\Spa{\bl_1}, \bs_j }  & \text{if\ } \bl_1 \in \calD\setminus \calS, \bl_2 = \bs_j \text{ and } \bs_j \in \calS\\
	\delta_{(\bl_1 = \bl_2)}\bR_{\bl_1} + \bH_{\bl_1} \tilde{\Cov}_{\Spa{\bl_1}, \Spa{\bl_2}} \bH_{\bl_2}^\top & \text{otherwise.}
	\end{cases}
	\end{align*}
	For a given base Gaussian covariance function $\Cov$, domain partitioning $\calP$, mesh $\calG$, and reference set $\calS$, we denote the corresponding meshed Gaussian process as $\text{MGP}(\calG, \calP, \calS, \Cov)$.
	
	\subsection{Bayesian hierarchical model and Gibbs sampler} \label{section:general:estimation}
	Meshed GPs produce block-sparse precision matrices that are constructed cheaply from their block-sparse Cholesky factors by solving small linear systems. General purpose sparse-Cholesky algorithms \citep{davis, cholmod} can then be used to obtain collapsed samplers as in \cite{nngp_algos}. Unfortunately, these algorithms can only be used on Gaussian first stage models and are computationally impracticable for data in the millions. Hence, we develop a more general scalable Gibbs sampler for the recovery of spatial random effects in hierarchical MGP models that entirely circumvents large matrix computations.
	
	Consider a multivariate spatiotemporally-varying regression model at $\bl \in \calD \subset \Re^{d+1}$,
	\begin{equation}\label{eq:linear_svc}
	\by( \bl ) = \bX(\bl)^\top \bbeta + \bZ(\bl)^\top \bw(\bl) + \beps(\bl),
	\end{equation}
	where $\by(\bl) \in \Re^l$ is the multivariate point-referenced outcome, $\bX(\bl)^\top = \bdiag\{ \bolds{x}_i(\bl)^\top \}_{i=1}^l$ is a $l \times p = l \times \sum p_i$ matrix of spatially referenced predictors linked to constant coefficients $\bbeta$, $\bw(\bl)$ is the spatial process, $\bZ(\bl)$ is a $l \times q$ design matrix, $\beps(\bl)$ is measurement error such that $\beps(\bl) \iidsim N(0, \bD)$ and $\bD = \text{diag}(\tau^2_1, \dots, \tau^2_l)$. A simple univariate regression model with a spatially-varying intercept can be obtained with $l=1$, $\bZ(\bl) = 1$. For observed locations $\calT = \{ \bl_1, \dots, \bl_n \}$, we write the above model compactly $\by = \bX \bbeta + \bZ \bw + \beps$,
	where $\by = ( \by(\bl_1)^\top, \dots, \by(\bl_n)^\top)^\top$, $\bw$ and $\beps$ are similarly defined, $\bX = [\bX(\bl_1) : \dots : \bX(\bl_n) ]^\top$, $\bZ = \bdiag( \{ \bZ(\bl_i)^\top \}_{i=1}^n )$, and  $\bD_n = \bdiag(\{ \bD \}_{i=1}^n)$. 
	
	For subsets $\{\bl_1,\dots,\bl_{n_{\mathcal{A}}} \} = \mathcal{A} \subset \calT $, let
	$\by(\mathcal{A}) = (\by( \bl_1 )^\top, \ldots, \by(\bl_{n_{\mathcal{A}}})^\top)^\top$, with analogous definitions for $\bw(\mathcal{A})$ and $\beps(\mathcal{A})$, $\bX(\mathcal{A}) = [ \bX(\bl_1) : \ldots : \bX(\bl_{n_\mathcal{A}}) ]^\top$, $\bZ_{\calA} = \bdiag( \{ \bZ(\bl_i)^\top \}_{i=1}^{n_{\mathcal{A}}})$ and $\bD_{\calA} = \bdiag(\{ \bD \}_{i=1}^{n_{\calA}})$. After fixing a reference set $\calS$, we obtain $\calS^* = \calT \cap \calS$ and $\calU = \calT\setminus \calS$. 
	We partition the domain as above to obtain $\calS_j, \calS_j^*, \calU_j$ for $j=1, \dots, M$ and model $\bw(\bl)$ using the \gpname\, which yields $\bw \sim N(\bolds{0}, \tilde{\Cov_{\calS}}^{-1})$. We complete the model specification by assigning $\bbeta \sim N(\bbeta \mid \bmu_{\beta}, \bSigma_{\beta})$, $\tau^2_j \sim Inv.Gamma(\tau^2_j \mid a_{\tau_j}, b_{\tau_j})$, $\btheta \sim p(\btheta)$. 
	
	The resulting full conditional distribution for $\bbeta$ is $N(\bSigma_{\beta}^* \bmu_{\beta}^*, \bSigma_{\beta}^*)$, where $\bSigma_{\beta}^* = (\bSigma_{\beta}^{-1} + \bX^{\top} \bD^{-1}_n \bX )^{-1}$, $\bmu_\beta^* = \bSigma_{\beta}^{-1}\bmu_\beta + \bX^\top \bD^{-1}_n (\by - \bZ \bw)$. For $\tau^2_r$, $r=1, \dots, q$, the full conditional is Inverse-Gamma with parameters $a_{\tau_r} + n/2$ and $b_{\tau_r} + \frac{1}{2}\bE_r^\top \bE_r$ where $\bE_r = \by_{\cdot r} - \bX_{\cdot r} \bbeta - \bZ_{\cdot r} \bw $ and $\by_{\cdot r}, \bX_{\cdot r}, \bZ_{\cdot r}$ are the subsets of $\by, \bX, \bZ$ corresponding to outcome $r$ (out of $q$). 
	
	The Gibbs update of the $\bw_{\calU}$ components can proceed simultaneously as all blocks in $\calU$ have no children and their parents are in $\calS$. The full conditional for $\bw_{\calU_j}$ for $j=1, \dots, M$ is thus $N(\bSigma_{\calU_j}^*\bmu_{\calU_j}^*, \bSigma_{\calU_j}^*)$ where $\bSigma_{\calU_j}^* = (\bZ(\calU_j) \bD^{-1} \bZ(\calU_j)^\top + \bR_{\calU_j}^{-1})^{-1}$ and $\bmu_{\calU_j}^* = \bZ(\calU_j) \bD^{-1} (\by(\calU_j) - \bX(\calU_j)^\top \bbeta ) + \bR_{\calU_j}^{-1} \bH_{\calU_j} \bwpa{\bolds{b}_j}$, where $\bwpa{\bolds{b}_j}$ is the spatial process at locations corresponding to the parents of $\bolds{b}_j \in \bB \subset \bV$. 
	
	We update $\bw_{\calS_j} = \bw_{j}$ for $j=1, \dots, M$ via its full conditional $N(\bSigma_j^*\bmu_j^*, \bSigma_j^*) $. Let $\bolds{1}_j = ( In(s_1 \in \calS^*_j), \dots, In(s_{n_j} \in \calS^*_j))^\top$ be the vector of indicators that identify locations with non-missing outputs, and let $\ba_j \in \bV$ be the node in $\calG$ corresponding to $\calS_j$. Then,
	\begin{equation}\label{section:general:estimation:eq:fullconds}
	\begin{aligned}
	\bSigma_j^{*-1} &= \bZ_j^\top \tilde{\bD}_{n_j}^{-1} \bZ_j + \bR_j^{-1} + \sum_{i=1}^{|\ch{\ba_j}|} \bH_i^{^{[j]}{\top}} \bR_i^{^{[j]}{-1}} \bH_i^{^{[j]}}\\
	\bmu_j^* &= \bR_j^{-1}\bH_j\bwpa{j} +  \bZ_j^\top\tilde{\bD}_{n_j}^{-1} \tilde{\by}_j + \sum_{i=1}^{|\ch{\ba_j}|} \bH_i^{^{[j]}\top}  \bR_i^{^{[j]}{-1}} w_i^{^{[j]}}\;,
	\end{aligned}
	\end{equation}
	where $\tilde{\bD}^{-1}_{n_j} = \bI_j \odot \bD^{-1}_{n_j}$ with $\bI_j = \bolds{1}_j \bolds{1}_j^\top$, and $\tilde{\by}_j = \bolds{1}_j \odot (\by_j - \bX_j \beta)$ and $\odot$ denotes the Hadamard or Schur (element-by-element) product. Finally, $\btheta$ is updated via a Metropolis step with target density $p(\btheta) N(\bw_{\calS} \mid \bolds{0}, \tilde{\Cov}_{\calS}) N(\bw_{\calU} \mid \bH_{\calU} \bw_{\calS}, \bR_{\calU})$ using (\ref{section:general:eq:pS:gaussian}) and (\ref{section:general:eq:pUgivenS:gaussian}). The Gibbs sampling algorithm will iterate across the above steps and, upon convergence, will produce samples from $p(\bbeta, \{ \tau^2_j \}_{j=1}^q, \bw \mid \by)$. 
	
	We obtain posterior predictive inference at arbitrary $\bl \in \calD$ by evaluating $p(\by(\bl)\given \by)$. If $\bl \in \calS \cup \calU$, then we draw one sample of $\by(\bl) \sim N(\bX(\bl)^\top \bbeta + \bZ(\bl)^\top \bw(\bl), \bD)$ for each draw of the parameters from $p(\bbeta, \{ \tau^2_j \}_{j=1}^q, \bw \mid \by)$. Otherwise, considering that $\bl \in \calD_j$ for some $j$ and thus $\eta(\bl) = \bolds{b}_j$, with parent nodes $\pa{\bolds{b}_j}$ and children $\ch{\bolds{b}_j} = \emptyset$, we sample $\bw(\bl)$ from the full conditional $N(\bSigma_{\bl}^*\bmu_{\bl}^*, \bSigma_{\bl}^*)$, where $\bSigma_{\bl}^* = (\bZ(\bl) \bD^{-1} \bZ(\bl)^\top + \bR_{\bl}^{-1})^{-1}$ and $\bmu_{\bl}^* = \bZ(\bl) \bD^{-1} (\by(\bl) - \bX(\bl)^\top \bbeta ) + \bR_{\bl}^{-1} \bH_{\bl} \bwpa{\bolds{b}_j}$, then draw $\by(\bl) \sim N(\bX(\bl)^\top \bbeta + \bZ(\bl)^\top \bw(\bl), \bD)$.
	
	\subsection{Non-separable multivariate spatiotemporal covariances} \label{section:general:estimation:complexity}
	
	We provide an account of the computational cost of general MGPs as a starting point to motivate the introduction of more efficient tessellated MGPs, and specifically Q-MGPs, in Section \ref{section:qmeshgp}. We consider (\ref{eq:linear_svc}) and take $l=1$ to simplify our exposition. In the resulting model, $\beta$ is the regression coefficient on the $p$ point-referenced regressors with a static effect on the outcome, whereas the $q$-variate spatiotemporal process $\bw(\cdot)$ captures the dynamic effect of the $\bZ$ regressors. Typically in geostatistical modeling $p$ and $q$ are small, hence sampling $\bbeta$ and $\tau^2$ carries a negligible computational cost. The cost of each Gibbs iteration is dominated by updates of $\btheta$ and $\bw$. Let us assume, solely for expository purposes, that each of the $M$ blocks comprise the same number of locations, i.e. $|\calS_j| = |\calU_j| = m$, for all $j=1, \dots, M$. Thus, $m = \frac{n}{2M}$ and the graph nodes have $J$ or fewer parents and $L$ or fewer children.
	
	The evaluation of $N(\bw_{\calS} \mid \bolds{0}, \tilde{\Cov}_{\calS}) = \prod_{j=1}^{M} N( \bw_j \given \bH_{j} \bwpa{j}, \bR_{j})$ and $ N(\bw_{\calU} \given \bH_{\calU} \bw_{\calS}, \bR_{\calU}) = \prod_{j=1}^{M} N( \bw_{\calU_j} \given \bH_{\calU_j}\bwpa{\bolds{b}_j}, \bR_{\calU_j} )$ dominates the computation. Each term in the product entails $\bR_j^{-1}$ and $\bR_{\calU_j}^{-1}$, both of size $qm\times qm$, and their determinants. These require $\Cov^{-1}_{[j]}$ of size $Jqm \times Jqm$ or less, resulting in $O(2M ( q^3m^3 + J^3q^3m^3 )) = O(2M q^3m^3 (J^3 + 1)) \approx O(2M q^3m^3 J^3) = O(\frac{n^3 q^3J^3 }{M^2})$ flops via Cholesky decomposition. Reasonably, $J$ and $m$ are fixed so $M$ may grow linearly with sample size and the cost is $O(n q^3J^3)$ considering $M \propto n$. The total computing time is $\sim O(\frac{n q^3 J^3}{K})$ with $K$ processors for computing the $2M$ densities.  
	Sampling $\bw_{\calS}$ and $\bw_{\calU}$ from their full conditional distributions requires $O(2Mq^3 m^3 + M L q^2m^2 + M q^2m^2)$ flops, assuming $\bR_{j}^{-1}$ and $\bR_{\calU_j}^{-1}$ are stored in the previous step. The first term in the complexity order is due to the Cholesky decomposition of covariance matrices, the second is due to sampling the reference nodes, and the third comes from sampling other nodes. Without further assumptions, parallelization reduces complexity to $O(\frac{2Mq^3 m^3}{K} + \frac{Mq^2 m^2}{K} + M L q^2m^2 )$, since the covariances can be computed beforehand and the $M$ components of $\bw_{\calU}$ are independent given $\bw_{\calS}$. With fixed block size $m$, the overall complexity for a Gibbs iteration is 
	$O(\frac{2}{K}M q^3m^3 (J^3 + 1) + \frac{1}{K}2M q^3m^3 + \frac{1}{K}M q^2m^2 + M L q^2m^2 ) 
	\approx O(\frac{1}{K}J^3q^3 n + q^2 n) \approx O(n)$, linear in the sample size and cubic in $J$, highlighting the computational speedup of sparse graphs ($J$ small), the negative impact of large $q$, and the serial sampling of $\bw_{\calS}$. 
	
	In terms of storage, $\bH_j$ and $\bR_j$ correspond to a storage requirement of $O(4Mq^2m^2) = O(q^2 n)$. The matrix $\bZ$ of size $qn \times qn$ can be represented as a list of $2M$ block-diagonal (hence sparse) $\bZ_j$ matrices. Furthermore, computing $\bZ \bw$ (dimension $n \times 1$) can be vectorized as the row-wise sum of $\bZ^* \odot \bw^*$ where $\bZ^*$ and $\bw^*$ are $n \times q$ matrices with $j$th column representing the $j$th space-time varying predictor. The cost of storing $\bZ$ is thus $O(2qn)$.
	
	Complexity is further reduced by considering a graph with small $J$ or a finer partition resulting in large $M$ and small $m$, whereas the overall time can be reduced by distributing computations on $K$ processors. Possible choices for $\calG$ include nearest-neighbor graphs and multiresolution trees. 
	In settings with large $q$, adjusting $J$ and $M$ may be insufficient to reduce the computational burden. Covariance functions that are separable in the variables (but perhaps non-separable in space and time) bring the cost of Choesky factorizations of $Jqm \times Jqm$ matrices from $O(J^3 q^3 m^3)$ to $O(J^3m^3 + q^3)$ because $\Cov^{-1} = (\Cov_{\bh, u} \otimes \Cov_{\bv})^{-1} = \Cov^{-1}_{\bh, u} \otimes \Cov^{-1}_{\bv}$, where $\Cov_{\bh, u}$ is the $Jm\times Jm$ space-time component of the cross-covariance, and $\Cov_{\bv}$ the $q\times q$ variable component. Savings accrue when evaluating the likelihood and in sampling from the full-conditionals at the cost of realism in describing the spatial process. 
	
	The next section develops a novel MGP design based on domain tessellations or tiling -- i.e. partitions of the domain forming repeated patterns -- to which we associate similarly patterned meshes. If observations are also located in patterns, the bulk of the largest linear solvers will be redundant, resulting in a significant reduction in computational time. In either scenario, sampling $\bw_{\calS}$ will also proceed in parallel with improved mixing.
	
	\section{MGPs based on domain tessellation or tiling} \label{section:qmeshgp}
\begin{figure}
		\centering
		\includegraphics[width=.8\textwidth]{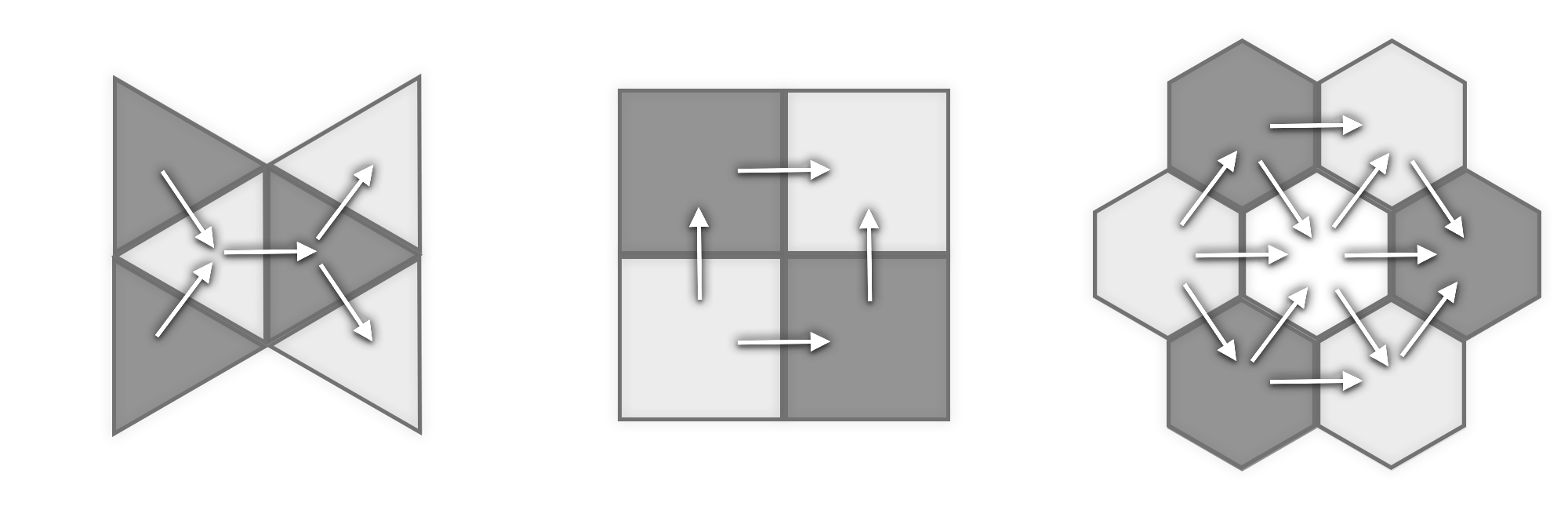}
		\caption{Regular tessellation base units and corresponding MGP graphs for spatial domains.}
		\label{fig:Figure_tiling} 
	\end{figure}

	We construct MGPs based on a tessellation or tiling of the domain. For spatial domains ($d=2$, Figure \ref{fig:Figure_tiling}), regular tiling results in triangular, quadratic, or hexagonal tessellations; mixed designs are also possible. These partition schemes can be linked to a DAG $\calG$ by drawing directed outgoing edges starting from an originating node/tile. The same fixed pattern can be repeated over a surface of any size. In dimensions $d>2$, which may include time, space-filling tessellations or honeycombs can be constructed analogously, along with their corresponding meshes. Constructions of MGPs based on these ideas simply requires partitioning the locations $\calS$ into subsets based on the chosen tessellation. 
	
	This subclass of MGP models corresponds by design to graphs with known \textit{coloring}, with each color linked to a subgraph conditionally independent of all nodes of other colors, regardless of the dimension of the domain. This feature enables large-scale parallel sampling of $\bw_\calS$ and improves mixing without the need to implement heuristic graph-coloring algorithms. Furthermore, regions in a tessellated domain are typically translations and/or rotations of a single geometric shape. Carefully choosing $\calS$, it will be possible to avoid computing the bulk of linear solvers, resulting in substantial computational gains. Subsequently, we focus on axis-parallel partitioning (quadratic or cubic tessellation) and cubic meshes, but analogous constructions and the same properties hold with other tessellation schemes.
	
	A cubic MGP (Q-MGP) is constructed by partitioning each coordinate axis into intervals. In $d+1$ dimensions, splitting each axis into $L$ intervals results in $L^{d+1}$ regions. Consider a spatiotemporal domain $\calD = \bigtimes_{r=1}^{d+1}\calD^{(r)}$, where $\calD^{(d+1)}$ is the time dimension. We partition each coordinate axis into $L_r$ disjoint sets: $\calD^{(r)}=\calI_{r, 1} \cup \cdots \cup \calI_{r, L_r}$, where $\calI_{r, j}\cap \calI_{r, k} = \emptyset$ if $j\neq k$ and $\calI_{r, s}$ denotes the $s$th interval in the $r$th coordinate axis. Solely for exposition, and without loss of generality, assume that $\calD^{(r)} = \calI = [0,1]$ and $L_r = L$ for $r=1, \dots, d+1$. Any location $\bl = (\ell_1, \dots, \ell_{d+1}) \in \calD$ will be such that $\bl \in \calI_{1,i_1} \times \cdots \times \calI_{d+1, i_{d+1}} = \calD_j$ for some $i_1, \dots, i_{d+1}$ and with $j=1, \dots, M$, where $M=L^{d+1}$. We refer to this axis-parallel partition scheme as a cubic tessellation and denote it by $\bT = \{\calI_{r,s}\}_{r=1,\dots,d+1}^{s=1,\dots,L}$. We use $\bT$ to partition the reference set $\calS$ as $\calS_j = \calD_j \cap \calS$ for $j = 1, \dots, L^{d+1}$. 
	
	Next, we define $\eta(\bl) = (\eta_1(\bl), \ldots, \eta_L(\bl)) \in \{1, \dots, L \}^{d+1}$, where $\eta_j = \eta_j(\bl) = r$ if $\ell_j \in \calI_{j,r}$. Then, let $\calQ = (\bV, \bE)$ be a directed acyclic graph with $\bV = \bA \cup \bB$ and reference nodes $\bA = \{ \ba_1, \ldots, \ba_{L^{d+1}}\}$. Therefore, for any $j = 1, \dots, L^{d+1}$ if $\bs \in \calS_j$ then $ \eta(\bs) = \ba_j \in \bA \subset \bV$. We write each node $\bv \in \bV$ as $\bv = (v_{\eta_1}, \ldots, v_{\eta_L}) \in \{1, \dots, L \}^{d+1}$. The directed edges are constructed using a ``line-of-sight'' strategy. Suppose $\pa{\bv} = \{ \bolds{x}^{(1)}, \ldots, \bolds{x}^{(d+1)} \}$. The $h$th parent of $\bv$ is defined as $\bolds{x}^{(h)} = (a_{\eta_1}, \dots, a_{\eta_h} - k, \ldots, a_{\eta_L}) \cap \{1, \ldots, d+1\}^{d+1}$, where $k \geq 1$ is the smallest integer such that $\bolds{x}^{(h)} \in \bA$. Consequently $\bolds{x}^{(h)} = \emptyset$ if $a_h=1$. Thus, the parents of node $\bv = \eta(\bl)$ are the ones that precede it along each of the $d+1$ coordinates. If $\bl \in \calD_j \setminus \calS_j$, then $\eta(\bl) = \bb_j \in \bB$ and $\pa{\bb_j} = \{ \ba_j \} \cup \pa{\ba_j}$ where $\ba_j \in \bA$ is a reference node. To avoid $\pa{\bb_j} = \emptyset$ we set $\pa{\bb_j} = \{ \bolds{x}^{(1)}_1, \bolds{x}^{(1)}_2, \ldots, \bolds{x}^{(d+1)}_1, \bolds{x}^{(d+1)}_2 \}$. The two parents along the $h$th dimension are $\bolds{x}^{(h)}_1 = a_{\eta_h} + k_1$, $\bolds{x}^{(h)}_2 =  a_{\eta_h} - k_2$ where $k_i$ is the smallest positive integer such that $\bolds{x}^{(h)}_i \in \bA$, $i=1,2$. In this setting $J = 2(d+1)$. The construction is finalized by fixing the cross-covariance function $\Cov(\bl, \bl')$; Figure \ref{fig:Figure_networks} shows that the same basic structure can be immediately extended to higher dimensions, including time.

\subsection{Caching redundant expensive matrix operations} \label{section:qmeshgp:caching}
\begin{figure}
	\centering
	\includegraphics[width=.9\textwidth]{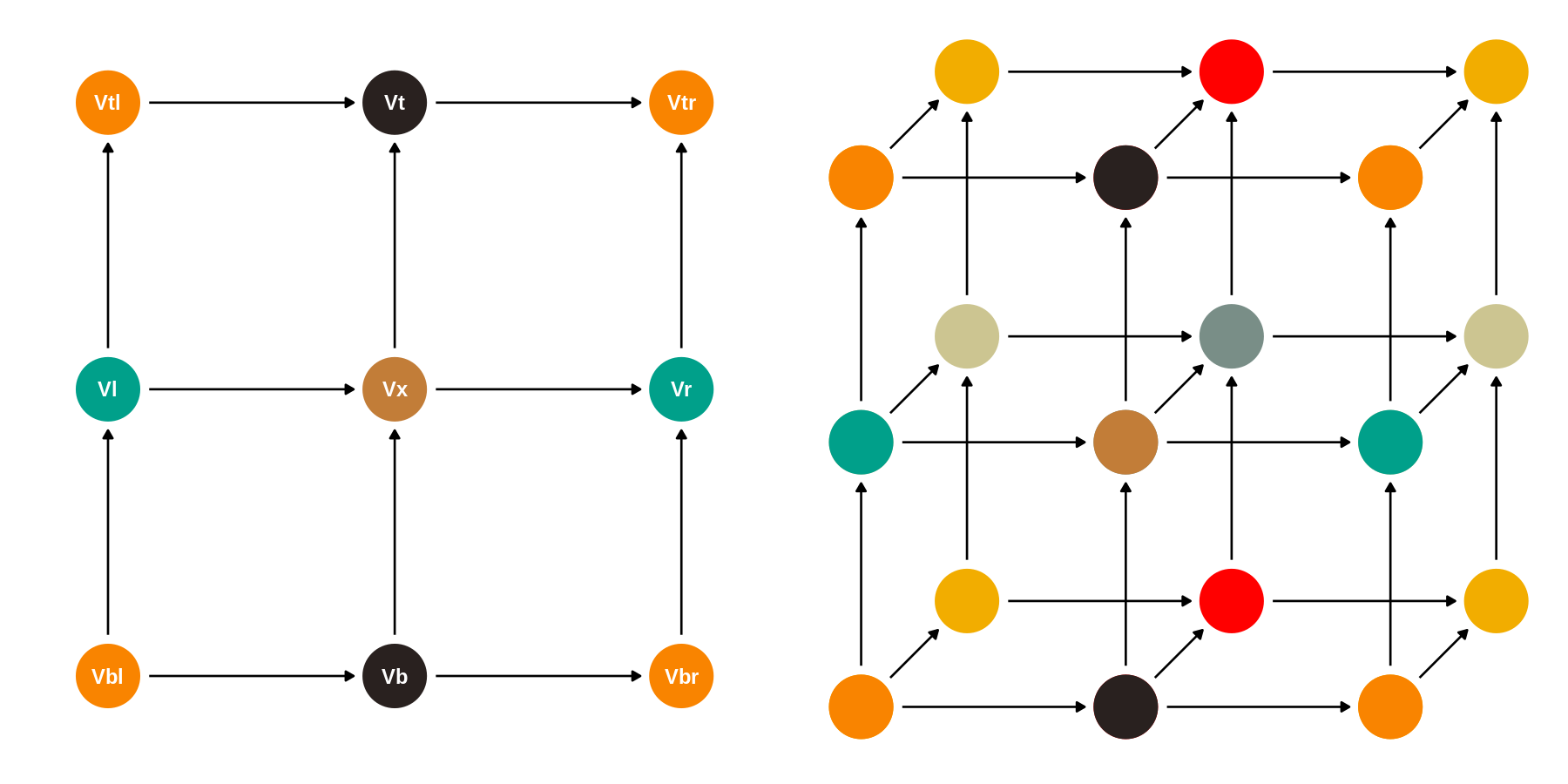} 
	\caption{Q-MGP meshes used for spatial data on $d=2$ (left) can be extended for use on spatiotemporal data $d=3$ (right). Node colors correspond to Gibbs sampler blocks. }\label{fig:Figure_networks} 
    \vspace*{\floatsep}%
	\centering
	\includegraphics[width=.5\textwidth]{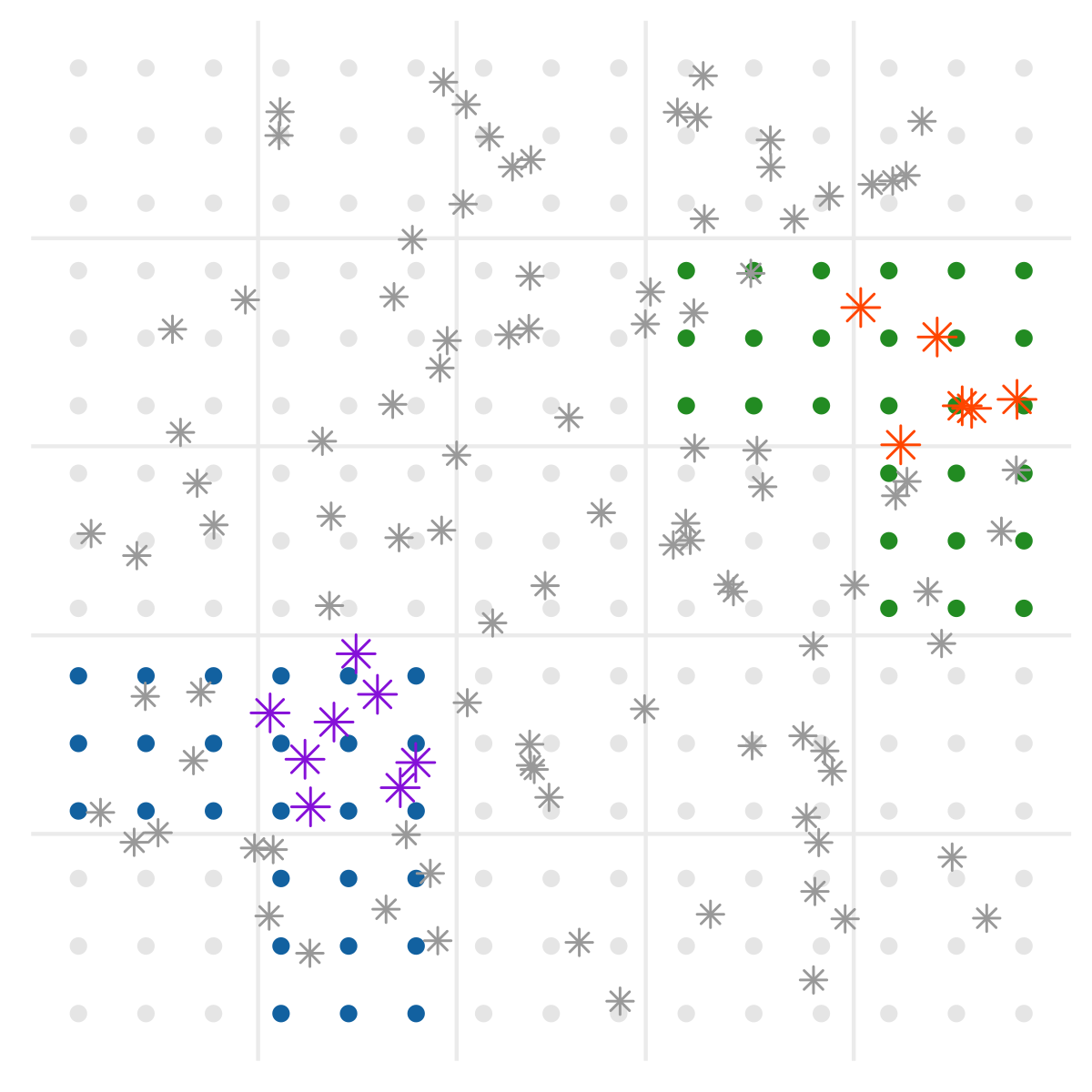}
	\caption{Visualizing redundancies: a spatial domain is partitioned in $M=25$ regions and linked to a quadratic mesh. The reference set $\calS$ is fixed on a regular grid, with $m=9$. Parent locations of the orange (resp. purple) are in green (resp. blue). Using a stationary covariance, $\Cov_{\text{blue}, \text{blue}} = \Cov_{\text{green}, \text{green}}$. Therefore only one inversion is necessary; this can be replicated at no cost across 9 of the 16 regions. }
	\label{fig:Figure_caching} 
	\end{figure}

	The key computational bottleneck for the Gibbs sampler in Section~\ref{section:general:estimation:complexity} is calculating, for $j=1, \dots, 2M$, of (i) $\Cov^{-1}_{[j]}$ ($2MJ^3q^3m^3$ flops) and (ii) $\bR_{j}^{-1}$, $\bSigma_j^{*-1}$ ($4Mq^3 m^3$ flops). The former is costlier than the latter by a factor of $J^ 3/2$. Q-MGPs are designed to greatly reduce this cost. We start with an axis-parallel tessellation of the domain in equally-sized regions $\calD_1, \dots, \calD_M$, storing observed locations in $\calU$ to create $\calU_1, \dots, \calU_M$, which we assume, for simplicity, to be no larger than $m$ in size. Taking a stationary base-covariance function $\Cov$, 
	implies that $\Cov(\calL_1, \calL_2) = \Cov(\calL_1 + \bh, \calL_2 +\bh)$, where $\bh \in \Re^{d+1}$ is used to shift all locations in the sets. Recall that the reference set $\calS$ of MGPs can include unobserved locations. Hence, we can build $\calS$ on a lattice of regularly spaced locations. Since domain partitions have the same size, we have $\calS_j = \calS^* + \bh_j$ for $j=1, \dots, M$, where $\calS^*$ is a single ``prototype set'' using which one can locate all other reference subsets. Also, since $\pa{\ba_j} \subset \pa{\bb_j}$, there will be $4(d+1)$ prototype sets for parents, i.e. $\calS_{\pa{\bv_j}} = \calS^*_{r} + \bh_j$ for some $r \in \{1,\dots, 4(d+1)\}$ and $j=1, \dots, 2M$. 
	Then, we can build maps $\xi_{\calS}: \{1, \dots, M\} \to \{1, \dots, 4(d+1)\}$ and $\xi_{\calU}: \{1, \dots, M\} \to \{1, \dots, 4(d+1)\}$ linking each of $\calS_j$ and $\calU_j$ to a parent prototype. This ensures that $\Cov^{-1}_{[j]} = \Cov^{-1}_{\calS^*_r}$ for each $j=1, \dots, 2M$. One only needs the maps $\xi_{\calS}$ and $\xi_{\calU}$, \textit{cache} the $r$ unique inverses, and reuse them. The same method applies to cache $\bR_{\calS_j}^{-1} = \bR_{\calS^*_r}^{-1}$ on reference sets, but not on other locations since no redundancy arises in $\Cov_{\calU_j}$ for $j=1, \dots, M$. See Figure \ref{fig:Figure_caching} for an illustration. Compared to general MGPs (see Table \ref{table:caching}), the number of large linear system solvers is now constant with sample size and $(d+1) \ll M$ significantly reduces computational cost. 
	
	Furthermore, Q-MGPs automatically adjust to settings where observed locations $\calT$ are on partly regular lattices, i.e., they are located at patterns repeating in space or time which emerge after initial inspections of the data. 
	Appendix G, available online, outlines a simple algorithm to identify such patterns and create maps $\xi_\calS$ and $\xi_\calU$.
	In such cases, we fix $\calS \supseteq \calT$ and $\calU = \emptyset$. In addition to the above mentioned savings, we now do not have to compute $\bR^{-1}_{\calU_j}$ and $\bSigma^{*-1}_{\calU_j}$. If $\calT$ is not a regular lattice over the whole domain, $4(d+1)$ is a lower bound and in general there are $M^* \ll M$ inverses to compute. If $\calT$ is a fully observed regular lattice and if $\bZ(\bl) = I$ (a varying intercept model), then we save in computing the full conditional covariances as well, since all $\bD_j = I$. See Appendix C, available online, 
	for details on choosing $\calS$ and $\calU$.
	
    \begin{table}[]
    \begin{adjustbox}{width=\textwidth}
    \begin{tabular}{|lllllll|}
    \hline
       & $ \Cov^{-1}_{[j], [j]}$ & $\bR_{\calS_j}^{-1}$ & $\bR_{\calU_j}^{-1}$ & $\bSigma_{\calS_j}^{*-1}$ & $\bSigma_{\calU_j}^{*-1}$ & Sampling $\bw_{\calS}, \bw_{\calU}$ \\
       \hline
                             \textit{MGPs} (all cases)               & $2MJ^3q^3m^3$       & $Mq^3m^3$        & $Mq^3m^3$ & $Mq^3m^3$ & $Mq^3m^3$ & $M L q^2m^2  + M q^2m^2$ \\
    \textit{Q-MGPs} & & & &  & & \\
    --- Irregular locations & $4(d+1) J^3q^3 m^3$ & $4(d+1) q^3 m^3$ & $Mq^3m^3$ & $Mq^3m^3$ & $Mq^3m^3$ & $M L q^2m^2 + M q^2m^2$ \\
    --- Pattern lattice w/missing  & $2M^* J^3q^3 m^3$ & $2M^* q^3 m^3$ &           & $ Mq^3m^3$ &           & $M L q^2m^2 $ \\
    --- Lattice w/ missing  & $4(d+1) J^3q^3 m^3$ & $4(d+1) q^3 m^3$ &           & $ Mq^3m^3$ &           & $M L q^2m^2 $ \\
    --- Full lattice and $\bZ(\bl) = I_q$       & $4(d+1) J^3q^3 m^3$ & $4(d+1) q^3 m^3$ &           & $2^{(d+2)}(d+1) q^3m^3$ &           & $M L q^2m^2 $ \\ \hline
    \end{tabular}
    \end{adjustbox}
    \caption{Summary of computational cost of general MGPs and Q-MGPs. Rows are sorted from most expensive (top) to least expensive (bottom).} \label{table:caching}
    \end{table}

	\subsection{Improved mixing via parallel sampling} \label{section:qmeshgp:parallel}
    With caching, a much larger proportion of time is spent on sampling; parallelization may in general be achieved via appropriate node coloring algorithms \citep[see e.g.][]{molloyreed2002,parallel_gibbs, lewis2016}, but this step is unnecessary in Q-MGPs as the colors in $\calQ$ are set in advance independently of the data and result in efficient parallel sampling of the latent effects. Reference nodes $\bA$ of $\calQ$ are colored to achieve independence conditional on realizations of nodes of all other colors. For example, we partition spatial domains ($d=2$) into $M_1 \times M_2$ regions and link each region to a reference node in a quadratic mesh. A ``central'' reference node $\bv_{+}$ will have two parents and two children, i.e. $\pa{\bv_+} = \{ \bv_{l}, \bv_{b}\}$ and $\ch{\bv_+} = \{ \bv_{r}, \bv_{t} \}$, with $l, b, r, t$ respectively denoting \textit{left}, \textit{bottom}, \textit{right}, \textit{top} -- refer to Figure \ref{fig:Figure_networks} (left). We have $\pa{\bv_t} = \{\bv_+, \bv_{tl} \}$ and $\pa{\bv_r} = \{\bv_+, \bv_{br} \}$. The Markov blanket of $\bv_+$, denoted as $\texttt{mb}(\bv_+)$, is the set of neighbors of $\bv_+$ in the undirected ``moral'' graph $\calQ^\calM$, hence $\texttt{mb}(\bv_+) = \pa{\bv_+} \cup \ch{\bv_+} \cup \{ \bv_{tl}, \bv_{br} \}$. The corresponding spatial process is such that $p(\bw_{+} \mid \bw \setminus \bw_{+}) = p(\bw_+ \mid \bw_{\texttt{mb}(\bv_+)})$. Denoting $\bv_{bl} = \pa{v_{l}} \cap \pa{v_{b}}$ and $\bv_{tr} = \ch{v_{r}} \cap \ch{v_t}$, we note that $\{\bv_{bl}, \bv_{tr}\} \cap \texttt{mb}(\bv_+) = \emptyset$. We partition reference nodes $\bA$ into four groups $\{\bA^{(1)}, \bA^{(2)}, \bA^{(3)}, \bA^{(4)}\}$, such that $\{\bv_+\} \subset \bA^{(1)}$, $\{ \bv_b, \bv_{t}\} \subset \bA^{(2)}$, $\{ \bv_l, \bv_{r}\} \subset \bA^{(3)}$, and  $\{\bv_{tl}, \bv_{tr}, \bv_{bl}, \bv_{br}\} \subset \bA^{(4)}$. This $3\times 3$ pattern is repeated over the whole graph. Then, if $\bv \in \bA^{(j)}$, $\texttt{mb}(\bv) \cap \bA^{(j)} = \emptyset$. Denoting by $\mathscr{D}$ the other variables in the Gibbs sampler, we get:  
    \begin{align*}
	p(\bw_j \mid \bw_{-j}, \mathscr{D}) &=p(\bw_j \mid \bw_{\texttt{mb}(\bv_j)}, \mathscr{D}) = \prod_{\bv_i \in \bA^{(j)}} p(\bw_i \mid \bw_{\bA^{(-j)}}, \mathscr{D}).
	\end{align*}
	Since parallelization is possible within each of the groups, only be four serial steps are needed; time savings are due to $M/4$ typically being orders of magnitude larger than the number of available processors.
	Extensions to other tessellation schemes and higher dimensional domains and the associated graphs follow analogously.

	\section{Data analysis} \label{section:data}
	Satellite imaging and remote sensing data are nowadays frequently collected in large quantities and processed to be used in geology, ecology, forestry, and other fields, but clouds and atmospheric conditions obstruct aerial views and corrupt the data creating gaps. Recovery of the underlying signal and quantification of the associated uncertainty are thus the major goals to enable practitioners in the natural sciences to fully exploit these data sources. Several scalable geostatistical models based on Gaussian processes have been implemented on tens or hundreds of thousands of data points, with few exceptions. In considering larger data sizes, one must either have a large time budget -- usually several days -- or reduce model flexibility and richness. Scalability concerns become the single most important issue in multivariate spatiotemporal settings. In fact, repeated collection of aerial images and multiple spatially-referenced predictors modeled to have a variable effect on the outcome have a multiplicative effect on data size. With no separability assumptions, the dimension of the latent spatial random effects that one may wish to recover will be huge even when individual images would be manageable when considered individually. 
	
	The lack of software to implement scalable models for spatiotemporal data makes it difficult to compare our proposed approach with others in these settings. On the other hand, a recent article \citep{Heaton2019} pins many state-of-the-art models against each other in a spatial ($d=2$) prediction contest. On the same data, we show in Appendix E, 
	available online, that Q-MGPs can outperform all competitors in terms of predictive performance and coverage while using a similar computational budget.
	
	\subsection{Non-separable multivariate spatiotemporal base covariance}
	In our analyses, we choose a class of multivariate space-time cross-covariances that models the covariance between variables $i$ and $j$ at the $(\bh, u) \in \Re^{d+1}$ space-time lags as:
	\begin{equation}\label{eq:apanasovich_genton_covariance}
    \begin{aligned}
    \Cov_{ij}(\bh, u) &= \frac{\sigmasq}{ \left( \psi_1 \left( \frac{|u|^2}{\psi_2\left( \delta_{ij}^2 \right)} \right) \right)^{d/2} \left(\psi_2\left( \delta_{ij}^2 \right) \right)^{1/2} } \phi_1 \left( \frac{\|\bh \|^2}{\psi_1 \left( \frac{|u|^2}{\psi_2\left( \delta_{ij}^2\right)} \right)} \right),
    \end{aligned}
	\end{equation}
	where $\delta_{ij} > 0$ (and with $\delta_{ij}=\delta_{ji}$) is the latent dissimilarity between variables $i$ and $j$. In the resulting cross-covariance function $\Cov(\bh, u, \bv)$ in $\Re^{d+1+k}$, each component of the $q$-variate spatial process is represented by a point in a $k$-dimensional latent space, $k\leq q$. Refer to \cite{apanasovich_genton2010} for a more in-depth discussion. We set $\phi_1(x)=\exp(-c x)$ and $\psi_j(x) = (a_j x^{\alpha_j} + 1)^{\beta_j}$, $j=1,2$; see \cite{gneiting2002} for alternatives. We also fix $\alpha_1 = \alpha_2 = \frac{1}{2}$, and seek to estimate $\btheta = (\sigmasq, c, a_1, \beta_1, a_2, \beta_2, \{ \delta_{ij}\}_{i<j, j=1,\dots,q})$ a posteriori. The usual exponential covariance arises in univariate spatial settings. 
	
	\begin{figure}
\centering
\includegraphics[width=.8\textwidth]{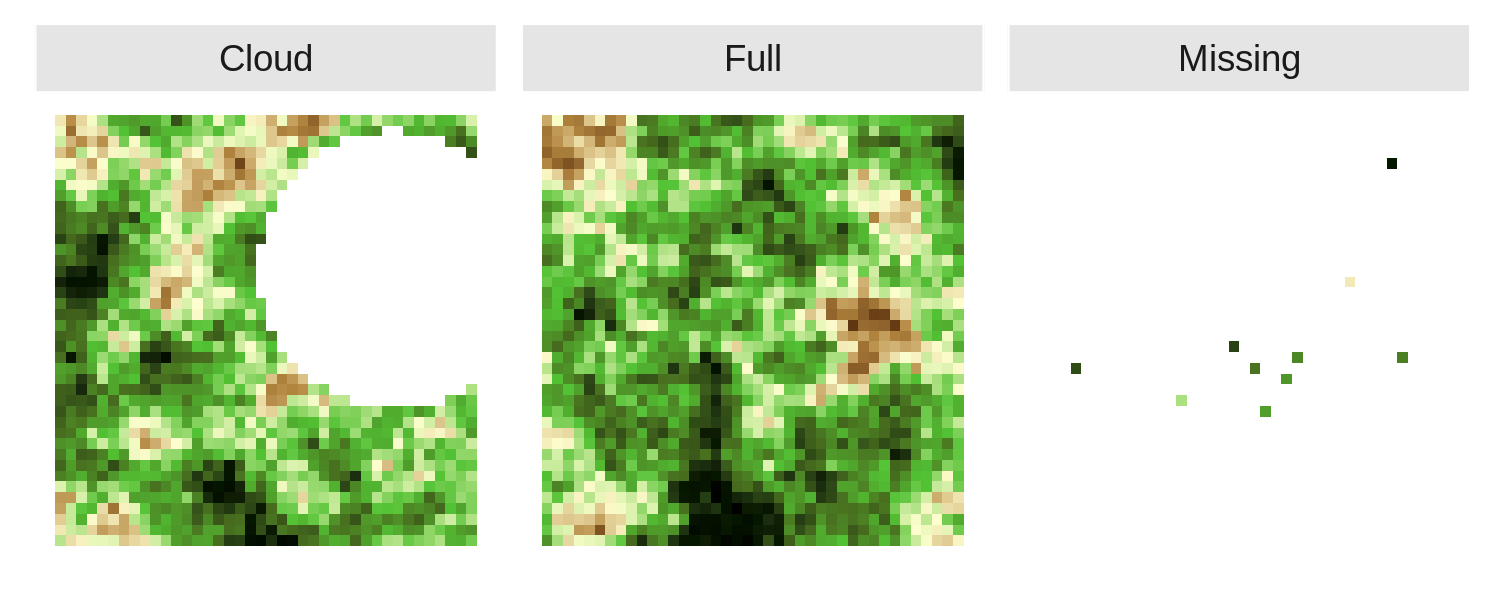}
\caption{Artificial cloud covering in synthetic data.}
\label{section:data:simulation:artificialclouds} 
\end{figure}

	\subsection{Synthetic data} \label{section:data:simulation}
	We mimick real world satellite imaging data analyzed later in Section \ref{section:data:serengeti} at a much smaller scale by generating 81 datasets from the model $\by(\bl) = \bZ(\bl)^\top\bw(\bl) + \beps(\bl)$, where $\beps(\bl) \sim N(0, \tau^2)$ with $\bl \in \calT$ and  $\calT$ is a regular grid of size $40 \times 40 \times 10$, resulting in $n_{\text{all}}=16,000$ total locations. We take $\bw \sim GP(\bzero, \Cov)$ where $\Cov$ is as in (\ref{eq:apanasovich_genton_covariance}), $\psi_2 \equiv 1$ and $\sigma^2=1$. We generate one dataset for each combination of $\tau^2 \in \{1/1000, 1/20, 1/10 \}$, temporal range $\alpha \in \{ 5, 50, 500 \}$, space-time separability $\beta \in \left\{ 1/20, 1/2, 1-\frac{1}{20} \right\}$, and spatial range $c \in \{ 1, 5, 25\}$.

    We compare Q-MGPs with the similarly-targeted Gapfill method of \cite{gapfill} as implemented in the \texttt{R} package \texttt{gapfill}. We create ``synthetic clouds'' of radius $\sqrt{0.1}$ and with center $(c_{1,t}, c_{2,t}) \in [0,1/20]^2$ where $c_{1,t}, c_{2,t} \iidsim U[0,1]$ to cover the outcomes at six randomly selected times for each of the 81 datasets. Outcomes at two of the remaining four time periods were then randomly selected to be completely unobserved at all but 10 locations in order to avoid errors from \texttt{gapfill}. Refer to Figure \ref{section:data:simulation:artificialclouds} for an illustration.
	
\begin{figure}
\centering
\includegraphics[width=.7\textwidth]{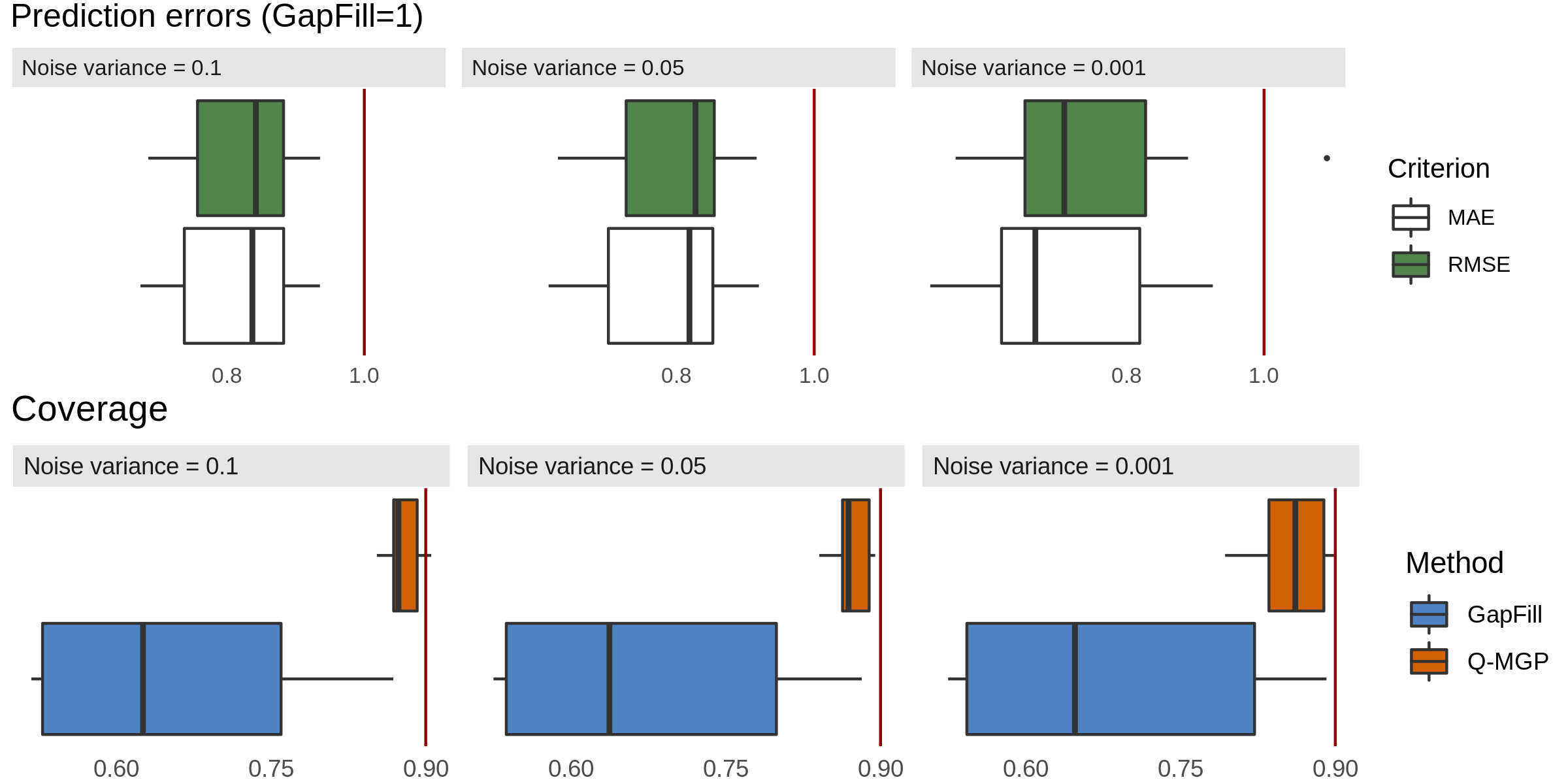}
\caption{Performance of Q-MGP and Gapfill in out-of-sample predictions in 81 spatiotemporal datasets, at the three tested levels of noise variance $\tau^2$. }
\label{section:data:simulation:boxplots} 
\end{figure}

	A Q-MGP model with $M=500$ was fit by partitioning each spatial axis into 10 intervals and the time axis into 5 intervals. The priors were $\tau^2 \sim Inv.G.(2, 1)$, $\sigmasq \sim Inv.G.(2, 1)$, $\beta \sim U(0,1)$, $\alpha \sim U(0, 10^4), c \sim U(0, 10^4)$; 7000 iterations of Gibbs sampling were run, of which 5000 used for burn-in and thinning the remaining 2000 to obtain a posterior sample of size 1000. For each of the 81 datasets we calculate the mean absolute prediction error (MAE) and the root mean squared prediction error (RMSE). Figure~\ref{section:data:simulation:boxplots} compares Gapfill's 90\% intervals with 90\% posterior equal-tailed credible intervals for the Q-MGP predictions obtained from 1000 posterior samples. In terms of MAE, the Q-MGP model outperformed Gapfill in all datasets; in terms of RMSE, it outperformed Gapfill in all but one dataset. The average MAE of Q-MGP was 0.4094 against Gapfill's 0.5366; the average RMSE was 0.5308 against Gapfill's 0.6820. The Q-MGP also yielded improved coverage of the prediction intervals, although some under-coverage was observed possibly due to the large $M$. This comparison may favor Q-MGPs as the data were generated from a Gaussian process. Appendix K, available online, 
	confirms similar findings on non-Gaussian data (a GIF image).
	
	\subsection{NDVI data from the Serengeti ecosystem} \label{section:data:serengeti}
	
	Time series of Normalized Difference Vegetation Index (NDVI) derived from satellite imagery are used to understand spatial patterns in vegetation phonology. For such studies, image pixel-level NDVI values are observed over time to assess seasonal trends in vegetation green-up, growing season length and peak, and senescence. These analyses typically require NDVI values for all pixels over the region and time period of interest. As noted in the beginning of this section, atmospheric conditions, e.g., cloud cover, and sensor malfunction cause missing NDVI pixel values and hence predicting such values, i.e., gap-filling, is of key interest to the remote sensing community. Here, we consider NDVI data derived from the LandSat 8 sensor (which provides a $\sim$30$\times$30 m spatial resolution pixel) taken over Serengeti National Park, Tanzania, Africa. These data were part of a larger study conducted by \cite{Desanker2019} that looked at environmental drivers in vegetation phonology change. The data cover an area of 30$\times$30km and 34 months, and correspond to 64 images of size 1000$\times$1000 collected at 16-day intervals. Data on NDVI are complemented with elevation and soil moisture data, for a total of three spatially referenced predictors. We are thus interested in understanding their varying effect in space and time, in addition to predicting NDVI output at missing locations. We achieve both these goals by implementing model (\ref{eq:linear_svc}),
	where $\bZ(\bl)=\bX(\bl)$ includes the intercept and three predictors; their varying effect will be represented by $\bw(\bl)$, which we recover by implementing Q-MGP models. Storing posterior samples of the multivariate spatially-varying coefficients for the full data with $q=4$ is impossible using our computing resources as each sample would be of size $1000\times1000\times64\times4 = 2.56\textsc{e+8}$. For this reason, we consider two feasible setups. Denote by $n_{\text{all}}$ the number of observed and missing locations. In model (1), we subsample each image to obtain 64 frames of size 500 $\times$ 500, and fit a regression model with $\bZ(\bl) = 1$ resulting in a spatially-varying intercept model on $n_{\text{obs}} = 12,582,484$ observed locations, a total of $n_{\text{all}} = 16,000,000$ locations for prediction, and a latent spatial random effect $\bw$ of the same size. The Q-MGP was fit using $M=328,125$ space-time regions of size $\sim$48. 
	
	The base covariance of (\ref{eq:apanasovich_genton_covariance}) becomes a univariate non-separable spatiotemporal covariance as in \cite{gneiting2002}. In model (2), we aim to estimate the varying effect of elevation on NDVI. We subsample each image to obtain 64 frames of size 278 $\times$ 278, each covering an area of 25$\times$25km, and take $\bZ(\bl) = (1\ \bX_{\text{elev}}(\bl))$ resulting in $q=2$ and targeting the recovery of latent effects of size $9,892,352$. Considering the additional computational burden of the multivariate latent effects, in this case we used $M=156,800$, corresponding to smaller space-time regions of average size $\sim$31. In this model there is a single unknown $\delta_{ij}$ in (\ref{eq:apanasovich_genton_covariance}) which corresponds to the latent dissimilarity between the intercept and elevation. We thus consider $\psi_2 = (a_2 \delta_{ij} + 1)^{\beta_2}$ as the unknown parameter. 
	We assign priors $\beta_r \sim N(0, 100)$ for $r=1, \dots, q$, $\sigmasq \sim Inv.G.(2,1)$, $\tau^2 \sim Inv.G(2,1)$, and uniform priors to other covariance parameters (their support is reported in Table \ref{tab:serengeti:postsummary}).
	
	\begin{figure}
		\centering
		\includegraphics[width=.75\textwidth]{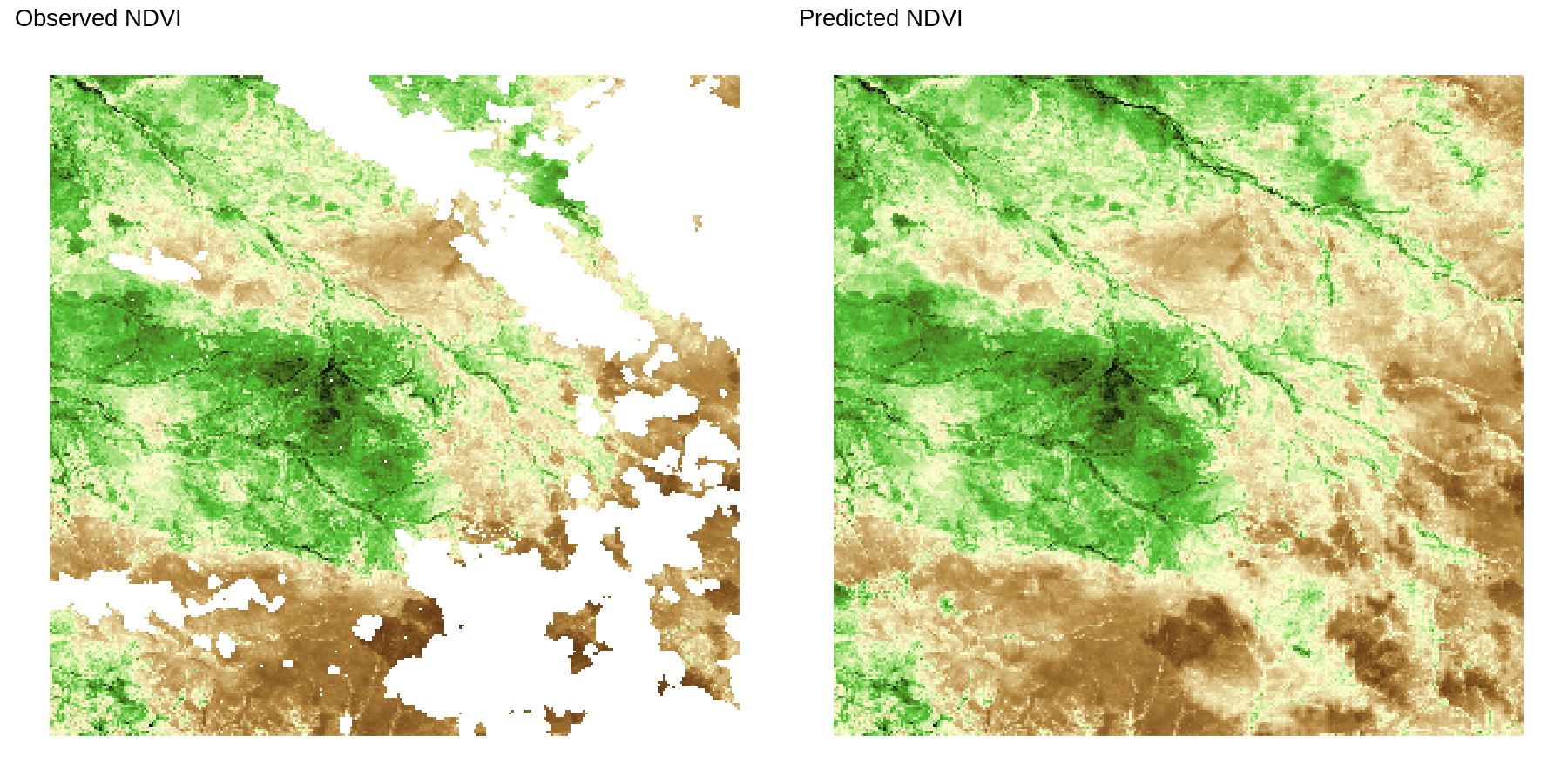}
		\caption{NDVI predictions from Q-MGP model (2) at time 60 (2016-12-17).}
		\label{fig:ndvi_predict} 
	\end{figure}
	
	\begin{figure}
		\centering
		\includegraphics[width=.9\textwidth]{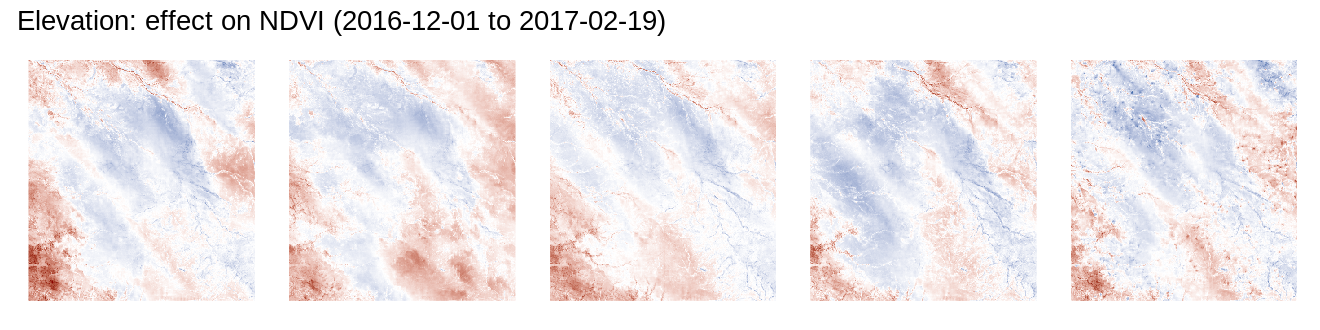}
		\includegraphics[width=.9\textwidth]{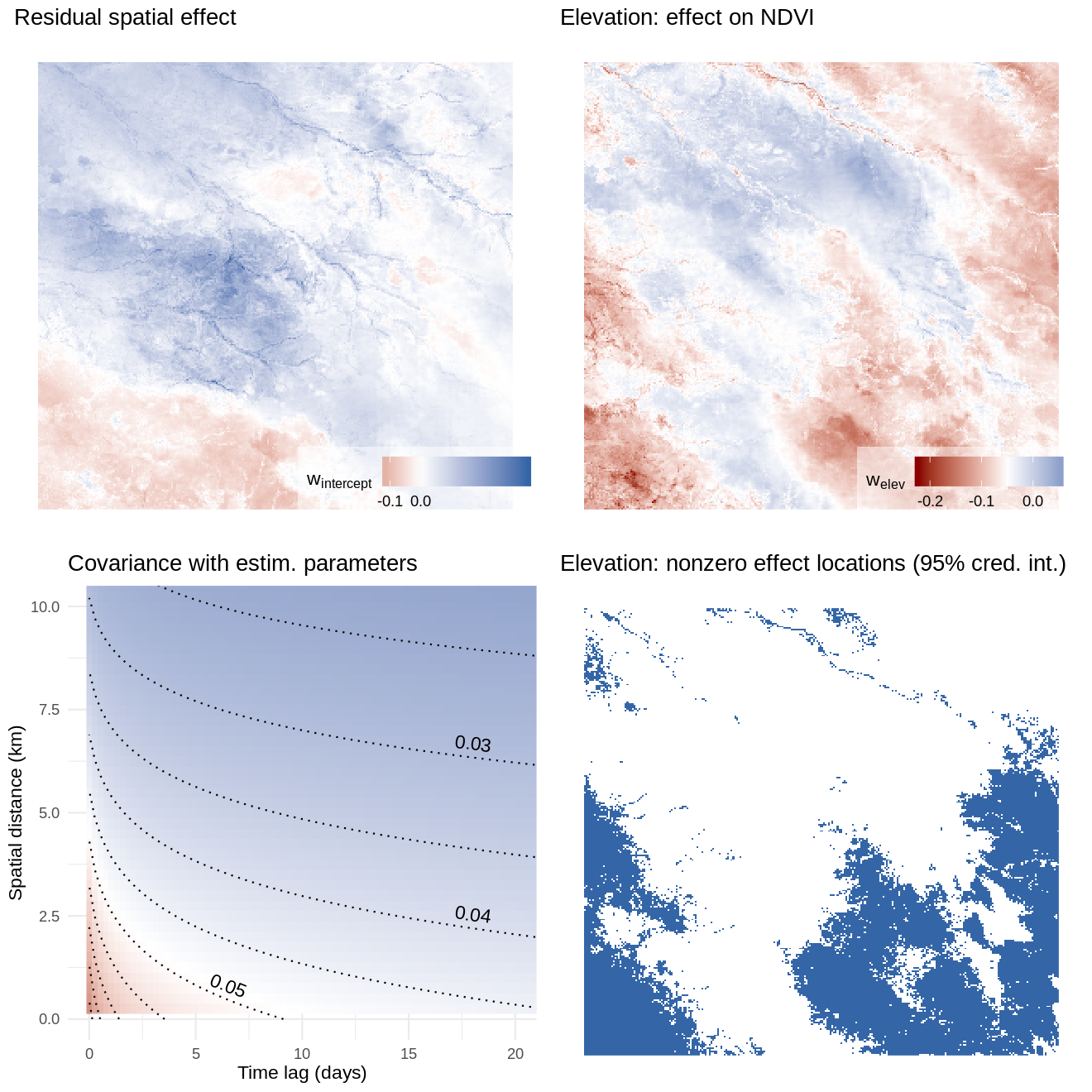}
		\caption{Top: the effect of elevation on NDVI output, evolving over five time periods. Middle left: effect on NDVI not explained by elevation; right: effect on NDVI attributable to elevation. Bottom left: estimated covariance at different space-time lags; right, in blue: locations with credibly nonzero effect of elevation on NDVI output.}
		\label{fig:ndvi_other} 
	\end{figure}
	
	In both cases, approximate posterior samples of the latent random effects and the other unknown parameters were obtained by running the proposed Gibbs sampler for a total of 25,000 iterations. A posterior sample of size 500 was obtained by using the first 22,000 iterations as burn-in, and thinning the remaining 3,000 by a factor of 6. Additional computational details are at Appendix F, available online. 
	Posterior summaries of the unknown parameters for these models are reported in Table \ref{tab:serengeti:postsummary}, along with RMSE in predicting NDVI at 10,000 left-out locations, 95\% posterior coverage at those locations, and run times. Both models achieved similar out-of-sample predictive performance and coverage. Figure \ref{fig:ndvi_predict} shows the NDVI predictions of model (2) at one of the 64 time points. This reveals that the varying effect of elevation on NDVI output (see e.g. Figure \ref{fig:ndvi_other}) is credibly different from zero at 42.54\% of the space-time locations (95\% c. i.). In particular, it highlights the extent to which higher elevation reduces vegetation. The spatial range is approximately 4km; the time range is about 8 days. The large estimated $\psi_2$ indicates that the correlation between the two covariates of the latent random process is very small at all spatial and temporal lags. The predicted NDVI and latent spatiotemporal effects are supplied as animated GIF images in the supplement.
	
	\begin{table}
    \centering
        \resizebox{.58\columnwidth}{!}{
    \begin{tabular}{|l|l|l|}
        \hline
     & \textsc{Q-MGP Model (1)} & 
        \textsc{Q-MGP Model (2)} \\ 
     \hhline{|=|=|=|}
     $n_{\text{all}}$ & $16,000,000$ 
     & $4,946,176 $ \\
     $n_{\text{obs}}$ & $12,755,856$ & $3,961,715$ \\
    $M$ & 328125 
    & 156800 \\
     $q$ & 1 & 2 \\
     $\beta_{\text{elevation}}$ & $0.0017_{(0.0014,0.0021)}$ 
        & $0.0415_{(0.0398, 0.0432)}$ \\
     $\beta_{\text{topoindex}}$ & $5.54\textsc{e-4}_{(4.72\textsc{e-4}, 6.30\textsc{e-4})} $ 
     & $-0.0011_{(-0.0012, -0.0008)}$ \\ 
     $\beta_{\text{accum}}$ & $-4.84\textsc{e-4}_{(-5.66\textsc{e-4}, -4.02\textsc{e-4})}$ 
     & $7.88\textsc{e-4}_{(6.94\textsc{e-4}, 9.06\textsc{e-4})}$ \\ 
     $\sigmasq$ & $0.0585_{(0.0583, 0.0587)}$ 
     & $0.0728_{(0.0711, 0.0749)}$ \\ 
     $\tau^2$ & $1.05\textsc{e-4}_{(1.05\textsc{e-4}, 1.05\textsc{e-4} )}$ 
     & $1.27\textsc{e-4}_{(1.21\textsc{e-4}, 1.32\textsc{e-4})}$ \\ 
     $c \sim U(0, 1\textsc{e+6})$ & $7.0331_{(7.0146, 7.0519)}$ 
     & $3.0710_{(3.0562, 3.0846)}$ \\ 
     $a_1 \sim U(0, 1\textsc{e+6})$ & $433.98_{(429.67, 439.50)}$ 
     & $3857.6_{(3492.6, 4154.7)}$ \\ 
     $\beta_1 \sim U(0, 1)$ & $0.0694_{(0.0690, 0.0697)}$ 
     & $0.1058_{(0.1043, 0.1080)}$ \\ 
     $\psi_2\sim U(0, 1\textsc{e+6})$
     & -- & $221.36_{(198.09, 240.57)}$ \\ 
     95\% coverage & $ 94.96 $ & $ 95.66 $ \\ 
     RMSE & $0.0175$ & $0.0253$ \\ 
     Time/it. (s) & $6.18$ 
     & $4.53$ \\ 
     Time (hours) & $42.9$ 
     & $31.5$ \\ 
        \hline
    \end{tabular}}
    \caption{Posterior summaries of Q-MGP models implemented on the Serengeti data.}
        \label{tab:serengeti:postsummary}
    \end{table}
    
	
	\section{Discussion}
	We have developed a class of Bayesian hierarchical models for large spatial and spatiotemporal datasets based on linking domain partitions to directed acyclic graphs. These models can be tailored for specific algorithmic needs, and we have demonstrated the advantages of using a cubic tessellation scheme (Q-MGP) when targeting the efficient recovery of spatial random effects in Bayesian hierarchical models using Gibbs samplers.
	
	When considering alternative computational strategies, the proposed Q-MGP may not be optimal. For example, Gaussian first stage models enable marginalization of the latent spatial effects; posterior sampling of unknown covariance parameters via MCMC is typically associated by better mixing. Future research may thus focus on identifying ``optimal'' DAGs for collapsed samplers. Furthermore, the blocked conditional independence structure of Q-MGPs may be suboptimal as it corresponds to possibly restrictive conditional independence assumptions in neighboring locations. While we have not focused on the effect of different tessellations or partitioning choices in this article, alternative tessellation schemes (e.g. hexagonal) may be associated to less stringent assumptions and possibly better performance, while retaining all the desirable features of Q-MGP.
	
	Other natural extensions to high-dimensional spatiotemporal statistics include settings where there are a very large number of spatiotemporal outcomes in addition to a large number of spatial and temporal locations. Here there are a few different avenues. One approach is in the same spirit of joint modeling pursued here, but instead of modeling the cross-covariance functions explicitly, as has been done here, we pursue dimension reduction using factor models \citep[see, e.g.,][]{Christensen2003,Lopes2008,Ren2013,taylor2019spatial}. The aforementioned references have attempted to model the factor models using spatial processes some of which have used scalable low-rank predictive processes or the NNGP. We believe that modeling latent factors using spatiotemporal MGPs will impart some of the computational and inferential benefits seen here. However, this will need further development especially with regard to identifiability of loading matrices \citep[][]{Lopes2008,Ren2013} and process parameters. 
	
	A different approach to multivariate spatial modeling has relied upon conditional or hierarchical specifications. This has been well articulated in the text by \cite{CressieWikle2011}; see also \cite{royle1999hierarchical} and the recent developments in \cite{cressie2016multivariate}. An advantage of the hierarchical approach is that the multivariate processes are valid stochastic processes, essentially by construction and without requiring spectral representations, and can also impart considerable computational benefits. It will be interesting to extend the ideas in \cite{cressie2016multivariate} to augmented spaces of DAGs to further introduce conditional independence, and therefore sparsity, in MGP models with high-dimensional outcomes.
	
	Finally, it is worth pointing out that alternate computational algorithms, particularly tuned for high-dimensional Bayesian models, should also be explored. Recent developments on algorithms based upon classes of piecewise deterministic Markov processes \citep[see e.g.,][and references therein]{fearnhead2018,bierkens2019} that avoid Gibbs samplers and even reversible MCMC algorithms are being shown to be increasingly effective for high-dimensional Bayesian inference. Adapting such algorithms to MGP and Q-MGP for scalable Bayesian spatial process models will constitute natural extensions of our current offering.  
	
	\if0\blind
	{
	\subsection*{Acknowledgements}
	Finley and Peruzzi were supported by National Science Foundation (NSF) EF-1253225 and DMS-1916395, and National Aeronautics and Space Administration's Carbon Monitoring System project. Peruzzi was supported in part by 1R01ES028804 of the National Institute of Environmental Health Sciences of the National Institutes of Health and European Union project 856506. Banerjee was supported by NSF DMS-1513654, IIS-1562303, and DMS-1916349.
	} \fi
	
%

\newpage
	\begin{center}
	    {\Huge Appendix}
	\end{center}
	
	\appendix

\section{Spatial Meshed Process} \label{section:general:process:kolmogorov}
	Let $\bw(\bs), \bs \in \mathcal{D}$ be the base process, $\calS$ the fixed reference set, $\calL = \{ \bl_1, \dots, \bl_n \} \subset \calD$ and $\calU = \calL \setminus \calS$. Then the joint density $\tp(\bw_{\calL})$ is proper. In fact, using the definitions of $\tp(\bw_{\calS})$ and $\tp(\bw_{\calU} \mid \bw_{\calS})$ 
	we get
	\begin{align*}
	\int \tp(\bw_{\calL}) \prod_{\bl_i \in \calL} d( \bw(\bl_i) ) &=  \int \int \tilde{p}(\bw_{\calU} \mid \bw_{\calS}) \tilde{p}(\bw_{\calS}) \prod_{ \bs_i \in \calS \setminus \mathcal{L}} d(\bw(\bs_i)) \prod_{\bl_i \in \calL} d( \bw(\bl_i) )\\
	& \int \tp(\bw_{\calU} \mid \bw_{\calS}) \tp(\bw_{\calS}) \prod_{\bl_i \in \calU} d( \bw(\bl_i) ) \prod_{\bl_i\in \calS} d( \bw(\bl_i) ) \\
	& \int \tp(\bw_{\calS}) \left( \int \tp(\bw_{\calU} \mid \bw_{\calS}) \prod_{\bl_i \in \calU} d( \bw(\bl_i) ) \right) \prod_{\bl_i\in \calS} d( \bw(\bl_i) ) = 1.
	\end{align*}
	We now verify the Kolmogorov consistency conditions. Take $\calL_{\pi} = \{ \bl_{\pi(1)}, \dots,  \bl_{\pi(n)} \}$ as any permutation of $\calL$. Then call $\calU_{\pi} = \calL_{\pi} \setminus \calS$ and we get
	\begin{align*}
	\tilde{p}(\bw_{\calL_\pi}) &= \int \tilde{p}(\bw_{\calU_\pi} \mid \bw_{\calS}) \tilde{p}(\bw_{\calS}) \prod_{ \bs_i \in \calS \setminus \calL_\pi} d(\bw(\bs_i)).
	\end{align*}
	Set membership is invariant to permutations, so $\calU_{\pi} = \calL_{\pi} \setminus \calS = \calL \setminus \calS = \calU$. and therefore in no way the order of the $\calL$ locations changes $\tp(\bw_{\calL})$. Therefore $\tilde{p}(\bw_{\calL_\pi}) = \tilde{p}(\bw_{\calL})$, i.e.
	\begin{align*}
	\tilde{p}(\bw(\bl_1), \dots, \bw(\bl_n)) &= \tilde{p}(\bw(\bl_{\pi(1)}), \dots, \bw(\bl_{\pi(n)})).
	\end{align*}
	Next, take another location $\bl_0 \in \calD$.  Call $\calL_1 = \calL \cup \{ \bl_0 \} $. We want to show that $\int \tp(\bw_{\calL_1}) d( \bw(\bl_0) ) = \tp(\bw_{\calL}) $. If $\bl_0 \in \calS$ then $\calL_1 \setminus \calS = \calL\setminus \calS = \calU$ hence 
	\begin{align*}
	\int \tp(\bw_{\calL_1}) d( \bw(\bl_0) ) & = \int  \tp( \bw_{\calL_1 \setminus \calS} \mid \bw_{\calS} ) \tp(\bw_{\calS})\prod_{\bs_i \in \calS\setminus \calL_1} d(\bw(\bs_i)) d( \bw(\bl_0) ) \\
	&= \int  \tp( \bw_{\calU} \mid \bw_{\calS} ) \tp(\bw_{\calS}) \prod_{\bs_i \in \calS\setminus \calL} d(\bw(\bs_i)) = \tp(\bw_{\calL})
	\end{align*}
	If $\bl_0 \notin \calS$
	\begin{align*}
	\int \tilde{p}(\bw_{\calL_1}) d( \bw(\bl_0) ) &= \int \left( \int \tilde{p}(\bw_{\calL_1 \setminus \calS} \mid \bw_{\calS}) \tilde{p}(\bw_{\calS}) \prod_{ \bs_i \in \calS \setminus \calL_1} d(\bw(\bs_i)) \right) d( \bw(\bl_0) )\\
	&= \int \left( \int \tilde{p}(\bw_{\calL \setminus \calS \cup \{ \bl_0 \}} \mid \bw_{\calS}) \tilde{p}(\bw_{\calS}) \prod_{ \bs_i \in \calS \setminus \calL} d(\bw(\bs_i)) \right) d( \bw(\bl_0) )\\
	&= \int \left( \int \tilde{p}(\bw_{\{ \bl_0 \}} \mid \bw_{\calL \setminus \calS}, \bw_{\calS}) \tilde{p}(\bw_{\calL \setminus \calS} \mid \bw_{\calS}) \tilde{p}(\bw_{\calS}) \prod_{ \bs_i \in \calS \setminus \calL} d(\bw(\bs_i)) \right) d( \bw(\bl_0) )\\
	&= \int \tilde{p}(\bw_{\calL \setminus \calS} \mid \bw_{\calS}) \tilde{p}(\bw_{\calS}) \prod_{ \bs_i \in \calS \setminus \calL} d(\bw(\bs_i))\int \tilde{p}(\bw_{\{ \bl_0 \}} \mid \bw_{\calS}) d( \bw(\bl_0) )\\
	&=\int \tilde{p}(\bw_{\calL \setminus \calS} \mid \bw_{\calS}) \tilde{p}(\bw_{\calS}) \prod_{ \bs_i \in \calS \setminus \calL} d(\bw(\bs_i))\\
	&=\tilde{p}(\bw_{\calL}).
	\end{align*}
	
\section{Meshed Gaussian Process} \label{section:general:gaussian:appendix}
	In graph $\mathcal{G}$, the number of parents of node $\bv_j$ is $b_j$, i.e. $b_j = | \pa{\bv_j} | $. Therefore  $\bH_j = \Cov_{\calS_j, \calS_{\pa{j}}} \Cov^{-1}_{\calS_{\pa{j}}}$ is a $q n_j \times q \sum_{r=1}^{b_j} n_r $ matrix which can be partitioned by column in $b_j$ blocks:
	\[ 
	\bH_j = \begin{bmatrix}[c:c:c] \bH_{j, 1} & \cdots & \bH_{j, b_j} \end{bmatrix},
	\]
	whose $r$th block is of size $q\times q$ and corresponds to the $r$th block of $\bw_{\pa{j}} $. 
	$\tilde{p}(\bw_{\calS}) = \prod_{j=1}^{M} N( \bw_j \mid \bH_{j} \bwpa{j}, \bR_{j})$ is then proportional to
	\begin{align*}
	\frac{1}{\prod_{j=1}^M \left| \bR_{j} \right|^{\frac{1}{2}}} \exp \left\{ -\frac{1}{2} \sum_{j=1}^M \left(\bw_j - \bH_j \bw_{\pa{j}} \right)^{\top} \bR_j^{-1} \left(\bw_j - \bH_j \bw_{\pa{j}} \right) \right\},
	\end{align*}
	and we can define $\bH^{*}_j$ as the $q n_j \times q n_{\calS} $ matrix such that $\bw_j - \bH_j \bw_{\pa{j}} =\bH^*_j \bw_{\calS}$. Analogously to $\bH_j$, we can partition $\bH^{*}_j$ by column in $M$ blocks,
	\begin{align*}
	\bH_j^* = \begin{bmatrix}[c:c:c:c:c] \bH^*_{j, 1} & \cdots & \bH^*_{j, h} & \cdots & \bH^*_{j, M} \end{bmatrix},
	\end{align*}
	where each block $h = 1, \dots, M$
	\begin{align*}
	\bH^*_{j, h} = \begin{cases} 
	\bO_{q n_h \times q n_h} & \text{if\ } \bv_h \notin \pa{\bv_j}  \\
	\bI_{q n_j \times q n_j} & \text{if\ } \bv_h = \bv_j \\
	-\bH_{j, r} & \text{if\ } \bv_h \text{ is $\bv_j$'s $r${th} parent, } 
	\end{cases}
	\end{align*}
	and notice that $h > j$ implies $\bv_h \notin \pa{\bv_j}$.
	Then,
	\begin{align*}
	\sum_{j=1}^M \left(\bw_j - \bH_j \bw_{\pa{j}} \right)^{\top} \bR_j^{-1} \left(\bw_j - \bH_j \bw_{\pa{j}} \right) = \sum_{j=1}^M \bw_{\calS}^\top\bH_j^{*\top} \bR_j^{-1} \bH^*_j \bw_{\calS} = \bw_{\calS}^\top \bH_{\calS}^{*\top} \bR_{\calS}^{-1} \bH^*_{\calS} \bw_{\calS},
	\end{align*}
	where $\bH^*_{\calS}$ is a $qn_{\calS} \times qn_{\calS}$ block-matrix whose $j$th block-row is $\bH_{j}^*$ for $j=1, \dots, M$, and $\bR_{\calS} = \bdiag ( \bR_1, \dots, \bR_M )$. The resulting precision matrix $\tilde{\Cov}^{-1}_{\calS} = \bH_{\calS}^{*\top} \bR_{\calS}^{-1} \bH^*_{\calS}$ is thus a $q n_{\calS} \times q n_{\calS}$ matrix made of $M \times M$ blocks of sizes $q n_j \times q n_j$ for $j=1, \dots, M$ such that $\sum n_j = n_{\calS}$. We denote its $(i,j)$ block as $\tilde{\Cov}^{-1}_{\calS}(i,j)$ and note that by symmetry its transpose is $(\tilde{\Cov}^{-1}_{\calS}(i,j))^{\top} = \tilde{\Cov}^{-1}_{\calS}(j,i)$. We find the nonzero blocks of $\tilde{\Cov}^{-1}_{\calS}$ in its block-diagonal, and for $i \neq j$ in the $(i,j)$ block such that $\bv_i$ and $\bv_j$ are connected in the moralized subgraph $\mathcal{G}^m(\bA)$ of $\mathcal{G}(\bA)$ obtained by subsetting $\calG$ to the $\bA$ nodes and linking those that share a child; in other words for all $\bv_i, \bv_j \in \bA$, either (1) $\bv_i \in \pa{\bv_j}$ or vice-versa, or (2) there exists $\bv_h$ such that $\{ \bv_i, \bv_j \} \subset \pa{\bv_h}$. The sparsity of $\tilde{\Cov}^{-1}_{\calS}$ thus depends on the specific choice for $\calG(\bA)$ and the corresponding moralized graph $\calG^m(\bA)$ \citep{graphs}. Suppose that $\calG(\bA)$ has $M$ vertices and induces $\calG^m(\bA)$ whose number of undirected edges is $k \ll M(M-1)/2$. This means that the number of ``missing'' connections is $\ell = M(M-1)/2 - k$. Assume that each subset $\calS_j$ of the reference set is of size $n_j = n/M$, so that each block of $\tilde{\Cov}^{-1}_{\calS}$ is of size $qn/M \times qn/M$. Then the precision matrix will have $\ell \left( \frac{qn}{M} \right)^2$ zeros. Given a partition for $\calD$ which induces a partition for $\calS$ into $M$ disjoint subsets, one can thus increase sparsity in the precision matrix at $\calS$ by choosing a graph which induces a lower $\ell$. But note that all nodes linked to locations outside the reference set, i.e. any $\bv \in \bB$, do not influence the precision matrix $\tilde{\Cov}^{-1}_{\calS}$.
	
	The covariance function of the new process will then be defined using $\tilde{\Cov}$: consider two locations $\bl_1$ and $\bl_2$ in $\calD$. If both are in $\calS$ then $\bl_1  = \bs_i$, $\bl_2 = \bs_j$ for some $i,j$, and $Cov_{\tp}(\bw(\bs_i), \bw(\bs_j)) = \widetilde{\Cov}_{\bs_i, \bs_j}$. If $\bl_1 = \bu_i \in \calU$ with parents $\pa{\eta(\bu_i)}$ and $\bl_2 = \bs_j \in \calS$ then
	\begin{align*}
	Cov_{\tp}(\bw(\bu_i), \bw(\bs_j)) &= 
	E_{\tp}( Cov_{\tp}(\bw(\bu_i), \bw(\bs_j) \mid \bw_{\calS}) ) +
	Cov_{\tp}( E_{\tp}(\bw(\bu_i) \mid \bw_{\calS}), E_{\tp}(\bw(\bs_j) \mid \bw_{\calS} ) ) \\
	&= 0 + Cov_{\tp}( \bH_{\bu_i} \bwpa{\bu_i}, \bw(\bs_j ) ) \\
	&= \bH_{\bu_i} \tilde{\Cov}_{\Spa{\bu_i}, \bs_j }
	\end{align*}
	Finally if $\bl_1 = \bu_i \in \calU$ and $\bl_2 = \bu_j \in \calU$ then 
	\begin{align*}
	Cov_{\tp}(\bw(\bu_i), \bw(\bu_j)) &= \delta_{(\bu_i = \bu_j)}\bR_{\bu_i} + \bH_{\bu_i} \tilde{\Cov}_{\Spa{ \bu_i}, \Spa{ \bu_j}} \bH_{\bu_j}^\top.
	\end{align*}
	Domain tessellations in meshed GPs will induce discontinuities in the covariance function when evaluated at locations that span different regions; however, $\tilde{\Cov}_{\bl_1, \bl_2}$ is continuous if $\bl_1$ and $\bl_2$ are in the same domain region. To show this, we adapt a result from \cite{nngp}. Let $\| A, B \|$ denote the Euclidean distance matrix at pairs of locations from sets $A, B \subset \calD \subset \Re^d$. Take $\calP = \{ \calD_1, \dots, \calD_M \}$ as the chosen domain partition, and let $\delta \calP = \cup_{i=1}^M \delta \calD_i$ be the finite union of their boundaries, each of which has measure zero. Then consider $(\bl_1, \bl_2) \in \calD \times \calD \setminus (\delta \calP \cup \{ (\bl, \bl) : \bl \in \calD \} )$ and let $\pa{\bl_i} = \{ \bl \in \calS : \eta(\bl) \in \pa{\eta(\bl_i)} \}$ be the parent set for $\bl_i$, i.e. the set of locations that are mapped to parents of the node to which $\bl_i$ is mapped. Now take $\bl_1 \in \calD_j \setminus \calS$ and $\bl_2 = \bs \in \calS_j$ for some $j=1, \dots, M$ (other cases follow). We assume that $\calS_j \subset \calD_j \setminus \delta \calD_j$ and $\Cov(\cdot, \cdot)$ is isotropic and continuous. Then $\lim_{\bh_i \to 0} \| \bl_i + \bh_i, \pa{\bl_i + \bh_i} \| \to \| \bl_i, \pa{\bl_i} \|$ for $i =1, 2$ and $\lim_{\bh_1 \to 0, \bh_2 \to 0} \| \pa{\bl_1 + \bh_1}, \pa{\bl_2 + \bh_2} \| \to \| \pa{\bl_1}, \pa{\bl_2}\|$, implying that $\bH_{\bl_1 + \bh_1} = \Cov_{\bl_1+\bh_1, \pa{\bl_1 + \bh_1}} \Cov^{-1}_{\pa{\bl_1 + \bh_1}} \to \Cov_{\bu, \pa{\bl_1}} \Cov^{-1}_{\pa{\bl_1}} = \bH_{\bl_1}$. Taking $\bh_2$ small enough implies $\bl_2+\bh_2 \in \calD_j \setminus \calS$ thus $\bl_2 \in \pa{\bl_2 + \bh_2}$. Then suppose $\bs = \pa{\bl_2 + \bh_2}_1$ i.e. $\bs$ is the first element of the parent set of $\bl_2 + \bh_2$. Consequently $\Cov_{\bl_2 + \bh_2, \pa{\bl_2 + \bh_2}} \Cov^{-1}_{\pa{\bl_2 + \bh_2}} \to \bolds{e}_1 = (1, 0, \cdots, 0)$ and therefore
	\begin{align*}
	    \lim_{\bh_1\to 0, \bh_2\to 0} \tilde{\Cov}(\bl_1 + \bh_1, \bl_2 + \bh_2) &= \bH_{\bl_1} \lim_{\bh_1\to 0, \bh_2\to 0} \widetilde{Cov}(\bw_{\pa{\bl_1 + \bh_1}}, \bw_{\pa{\bl_2 + \bh_2}}) \bolds{e}_1 \\
	    &= \bH_{\bl_1} \lim_{\bh_1\to 0} \widetilde{Cov}(\bw_{\pa{\bl_1 + \bh_1}}, \bw(\bs)) \\
	    &= \bH_{\bl_1} \tilde{\Cov}_{\pa{\bl_1}, \bs} = \tilde{\Cov}(\bl_1, \bl_2).
	\end{align*}
	
\section{Choosing the tessellation, $M$, $\calS$ and $\calU$} \label{section:qmeshgp:choiceSU}
	\begin{figure}[H]
	\centering
	\includegraphics[width=.9\textwidth]{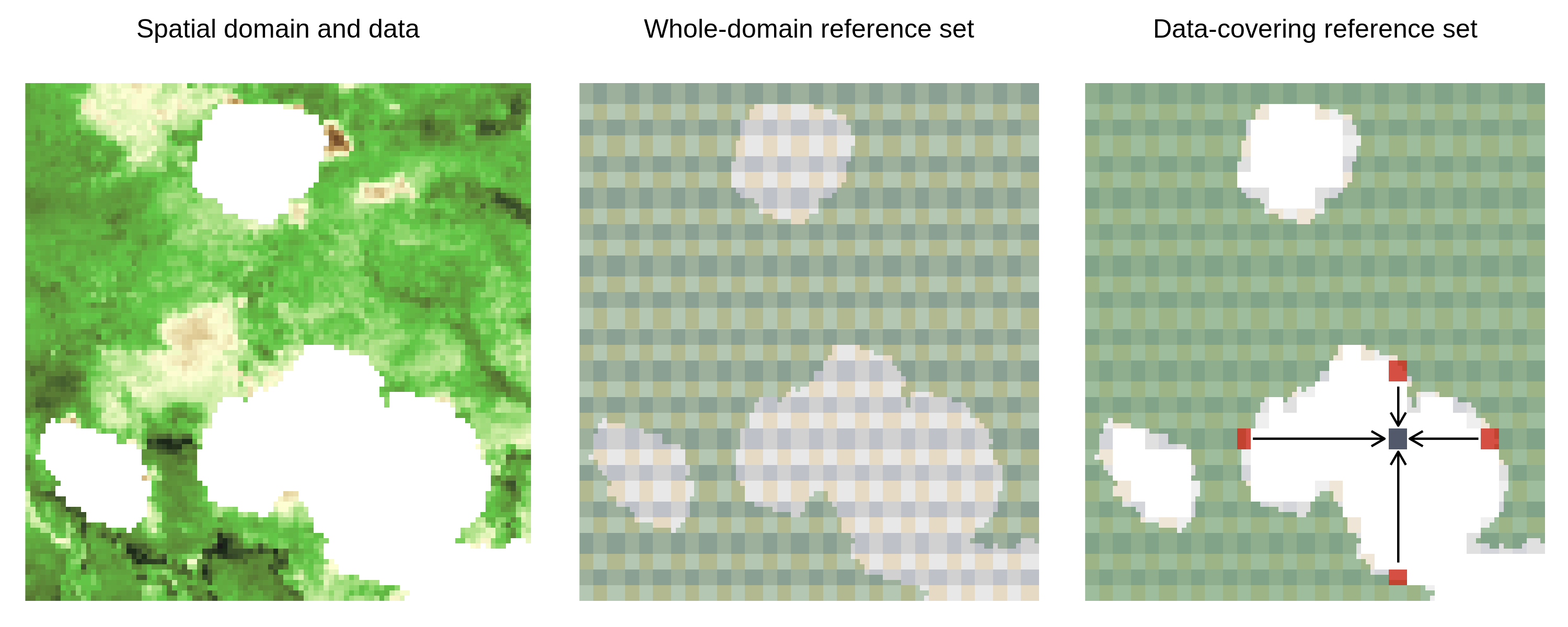}
	\caption{Reference sets}
	\label{fig:Figure_refset} 
	\end{figure}

\noindent\textbf{Tessellation}

\noindent Meshed GPs based on tessellations correspond to fixed DAGs constructed from the immediate neighbors bordering each region. For this reason they are unlike ``nearest-neighbor'' models in which the number of neighbors is a parameter to be fixed. Increasing the parent set in tessellated meshed GPs amounts to changing the tessellation or increasing the size of each region. As a reviewer points out, estimating the unknown covariance parameter may be facilitated by choosing parent sets that include more distant points \citep[see e.g.][]{stein2014}. Doing so in tessellated meshed GPs amounts to choosing a tessellation with elongated regions. With axis-parallel partitioning, one can partition one of the axis in a much smaller number of intervals than the others, resulting in parent sets that are elongated in that directions. Other shapes, such as a ``herringbone'' tessellation pattern, might be better at capturing variation in more spatial directions.

Multivariate data pose additional challenges, as one can potentially elect to use different tessellations on different margins of the multivariate process. However, considering different tessellations may induce complicated connections in the underlying DAG; this goes against our modeling approach that is to keep a simple, manageable graph structure in order to speed up computations. In cases in which one seeks to maintain good spatial coverage and is willing to make conditional independence assumptions on the variables, we suggest using a single tessellations, but perhaps dropping inter-variable edges across nodes. To clarify, we may assume that $w_j(\bl)$, the $j$th margin of the multivariate spatial process at location $\bl$, is conditionally independent of $w_{j'}(\bl')$ if $j'\neq j$ and $\bl$ and $\bl'$ are not in the same region. This design choice maintains long range spatial dependence along the same margin of $\bw$, but reduces it across different margins.
\\

\noindent\textbf{M -- the number of regions}

\noindent In a meshed GP based on axis-parallel domain partitioning, the number of partitioning regions $M$ is defined as $\prod_{j=1}^{d+1} M_j$ where $M_j$ is the number of intervals into which axis $j$ is partitioned. As seen above, $M$ determines the number of Cholesky factors to compute, as well as the complexity of each factor computation (given by the size of the individual matrices). A small $M$ corresponds to larger regions, which in turn will enlarge the parent sets of each block. For this reason, $M$ should be chosen as small as possible, given the allotted computational budget.
\\
	
\noindent $\calS$ \textbf{and} $\calU$

\noindent The two sets of locations $\calS$ and $\calU$ differ in that the former are mapped to nodes $\bA$ of graph $\calG$, whereas the latter to $\bB$ with no children. Since $\bB$ nodes are independent of each other given realizations of $\bA$ they provide little information on spatial dependence; in fact, the set $\calS$ is akin to the set of knots in a predictive process model. Unlike those models, $\calS$ can be chosen to be as large as the set of observations, and $\calU$ left empty. Generally, it is possible to choose $\calS\supset \calT$ to smoothly model spatial dependence across larger regions of unobserved locations, but convergence of the Gibbs sampler may be slower at locations that are very far from the nearest observed ones. Therefore, a convenient choice which is possibly safer for MCMC convergence is to let them coincide, i.e. $\calS = \calT$. 
	
	Matters are less straightforward when redundancies occur.
	Consider locations $\bt = ( t_1 ,\dots, t_d ) \in \calD$ such that $t_j = t^*_j \delta_j $ where $t^*_j = 1,\dots, N_j$ and $\delta_j \in [0, 1/N_j]$ for $j=1, \dots, d$. Denote the set of such locations as $\calT^*$; this is a $N_1 \times \cdots \times N_d$ grid of equally-spaced locations. If observed locations are on a quantized grid of coordinates, i.e. there are $\{ t^*_j, \delta_j \}_{j=1}^d $ such that $\calT \subset \calT^*$, choosing $\calS = \calT$ may be inefficient as it may reduce the number of redundant $\bR_j$'s and thus increase computation time.
	Instead, choosing $\calS$ such that $\calT\subset \calS \subset \calT^*$ (possibly $\calS = \calT^*$) may be much more efficient. 
	
	In Figure \ref{fig:Figure_refset} we show how a small subsample of a time slice of the Serengeti data. 
	In that case, we extend $\calS$ to the whole domain (Fig. \ref{fig:Figure_refset}, center) or to extend it just enough to cover all observed locations (Fig. \ref{fig:Figure_refset}, right). For predictions, 
	in the former case we use samples from the full conditional distribution $p(\bw_j \mid \bw_{-j}, \mathscr{D})$. In the latter case, we map empty blocks to nodes $\bB$: predictions thus easily proceed in parallel since nodes $\bB$ are independent of each other given nodes in $\bA$, with $\pa{\bolds{b}}$ being composed of the nearest $\bA$ nodes along each axis-parallel direction.

\section{Application: Q-MGP on multivariate response}
	We consider the multivariate regression model we previously defined:
	\begin{equation}
	\by( \bl ) = \bX(\bl)^\top \bbeta + \bZ(\bl)^\top \bw(\bl) + \beps(\bl),
	\end{equation}
	with $\bl \in \Re^{d}$ and $\beps(\bl) \iidsim N(0, \bD)$ and $\bD = \text{diag}(\tau^2_1, \dots, \tau^2_l)$, and where $\by(\bl) \in \Re^l$ is the multivariate point-referenced outcome. In this section, we specifically consider  the case $l=q$, $\bZ(\bl) = I_q$, and drop $\bX$ and $\bbeta$ from the model, resulting in 
	\begin{equation}
	\by( \bl ) = \bw(\bl) + \beps(\bl),
	\end{equation}
	where some of the $q$ components of $\by(\bl)$ may be unobserved. We construct a cross-covariance function based on latent distances among the $q$ variables. To do so, we set
	\begin{align*}
	    C(\bh, \bv) = \frac{\exp\left\{ - \phi \| \bh \|/\exp\left\{\frac{1}{2} \beta \log(1+\alpha \| \bv \| \right\} \right\} ) }{ \exp\left\{ \beta \log(1+\alpha \| \bv \| ) \right\}},
	\end{align*}
	and for $j=1, \dots, q$, $C_j(\bh) = \exp\left\{ - \phi_j \| \bh \|\right\}$. We then model $\bw(\bl)$ as a Q-MGP with base cross-covariance 
	\begin{equation}\label{eq:apanasovich_genton_covariance2}
    \begin{aligned}
    \Cov_{ij}(\bh, \bv) &= \begin{cases}
    \sigmasq_{i1} C(\bh, \bv) + \sigmasq_{i2} C_i(\bh) & \text{if } i=j \\
    \sigma_{i1} \sigma_{j1} C(\bh, \bv) & \text{if } i\neq j,
    \end{cases}
    \end{aligned}
	\end{equation}
which is a valid cross-covariance derived from eq. (7) in \cite{apanasovich_genton2010}. 

\subsection{Synthetic data}
\begin{figure}[H]
	\centering
	\includegraphics[width=\textwidth]{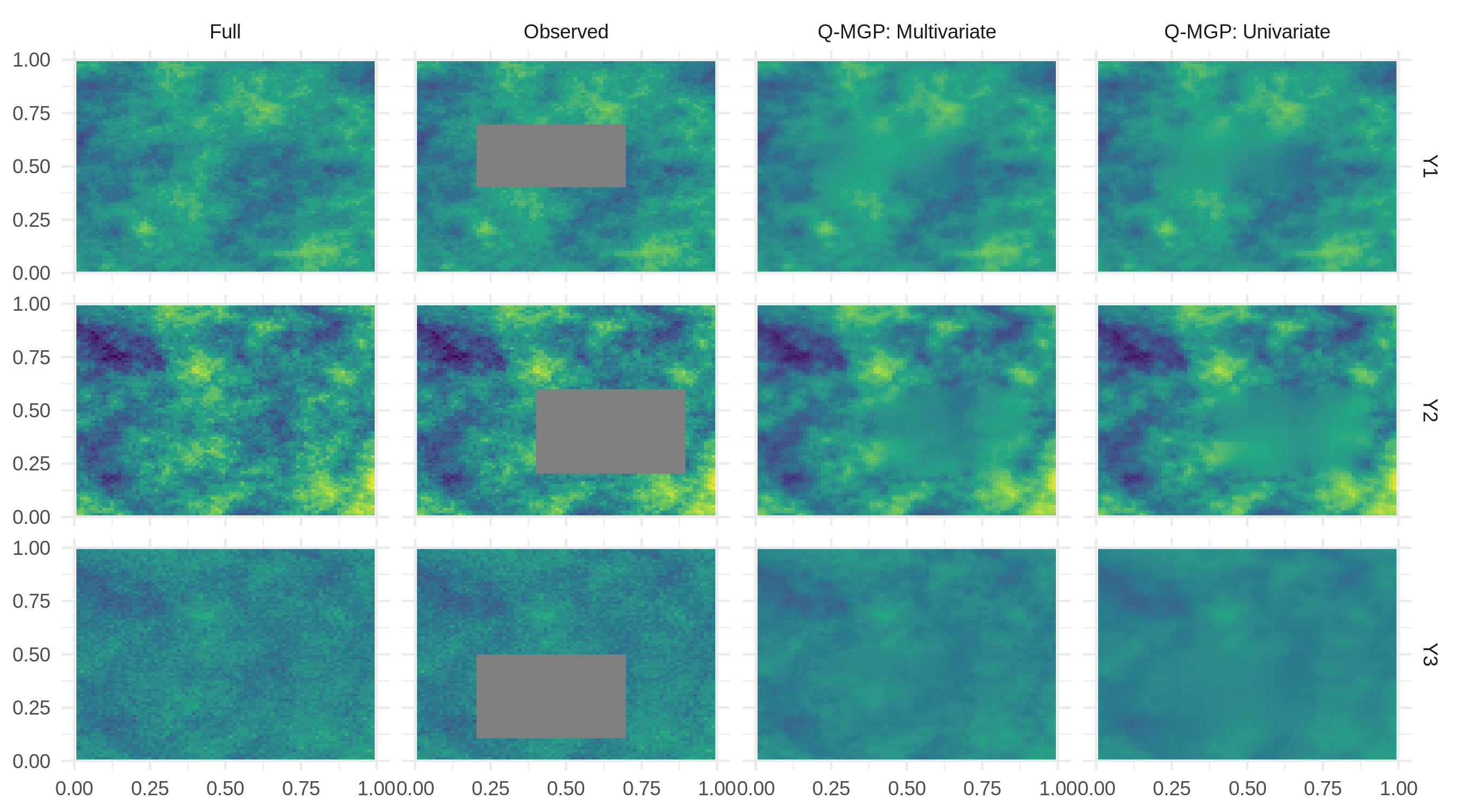}
	\caption{Simulated data and predictions from multivariate and univariate Q-MGPs.}
	\label{fig:mv:simulated} 
\end{figure}

We generate multivariate outcomes with $q=3$ at $n=6,400$ spatial locations on a regular grid on $\calD = [0, 1]^2$ for a total sample size of $19,200$, sampling the latent effects from a Gaussian Process with cross-covariance (\ref{eq:apanasovich_genton_covariance2}) and setting the parameters as
\begin{equation*}
    \begin{matrix} 
        \sigma_{11} = 1 & \sigma_{21} = 1.5 & \sigma_{31} = 0.5 \\
        \sigma_{12} = 0.1 & \sigma_{22} = 0.05 & \sigma_{32} = 0.1 \\
        \phi_1 = 0.5 & \phi_2 = 2 & \phi_3 = 1\\
        \alpha = 1 & \beta = 0.9 & \phi = 5 & \\
        \|\bv_{12}\| = 1 & \|\bv_{13}\| = 1 & \|\bv_{23}\| = 0.05,
    \end{matrix}
\end{equation*}
with measurement errors fixed at $\tau^2_1 = 0.01, \tau^2_2 = 0.05, \tau^2_3 = 0.1$. For predictions, we set $y_1(\bl) = \texttt{NA}$ if $\bl \in [0.2, 0.7] \times [0.4, 0.7]$,  $y_2(\bl) = \texttt{NA}$ if $\bl \in [0.4, 0.9] \times [0.2, 0.6]$, and $y_3(\bl) = \texttt{NA}$ if $\bl \in [0.2, 0.7] \times [0.1, 0.5]$ be observed at all $10,000$ spatial locations. This setup will highlight the advantages of a multivariate approach given the small latent distance between the second and third margin (given by $\|\bv_{23} \|$), and the non-separability of the covariance function on the variables, as given by $\beta$.
We fit a Q-MGP model on $M = 25^2$ regions. 
We compare the predictions of this multivariate specification with those obtained from univariate Q-MGP models applied on each margin independently. Domain partitioning for the univariate models along each margin was chosen in order to match the total computing time of the multivariate model. We run MCMC for 7000 iterations, of which 5000 discarded as burn-in and thinning the chain with a 1:1 ratio resulting in an approximated posterior sample of size 1000. 
The results are reported in Table \ref{table:mv:simulated}, whereas Figure \ref{fig:mv:simulated} shows the data along with the results from the two approaches.

\begin{table}[H]
	\centering
	\resizebox{.8\columnwidth}{!}{
		\begin{tabular}{ccccc}    
			 & \textbf{MAE}    & \textbf{RMSE}   & \textbf{Coverage} & \textbf{Run Time} \\
			\hhline{=====}
			Q-MGP (multivariate) & \textbf{0.4439} & \textbf{0.3200} & 96.39 & 1.04 min  \\
			Q-MGP (univariate) & 0.5056 & 0.4297 & \textbf{95.14} & 1.11 min   
		\end{tabular}
	}
	\caption{Multivariate and univariate Q-MGPs compared on multivariate synthetic data.} \label{table:mv:simulated}
\end{table}

\subsection{Spatial multivariate analysis of NDVI in the Serengeti region}
We consider the Serengeti dataset previously described; in this section, we build a spatial dataset by considering the NDVI measurement on December 17, 2016, along with elevation and soil moisture at the same spatial locations. The resulting dataset corresponds to $q=3$ variables and $n=1,000,000$ spatial locations, for a total size of $3,000,000$; cloud cover obstructed the sensor view, resulting in NDVI being missing at $322,555$ spatial locations. The goal of is thus to fill those gaps. Unlike in the previous analysis, we are not interested in mapping the varying effect of covariates on the outcome. Instead, we are modeling all variables jointly, aiming to recover the conditional distribution of NDVI given the others. We ran a total of 20000 MCMC iterations, thinning the last 5000 at a 4:1 ratio to eventually obtain a posterior sample of size 1000. Run-time averaged about 1 second per iteration using 50 threads.
	
\begin{figure}[H]
	\centering
	\includegraphics[width=.75\textwidth]{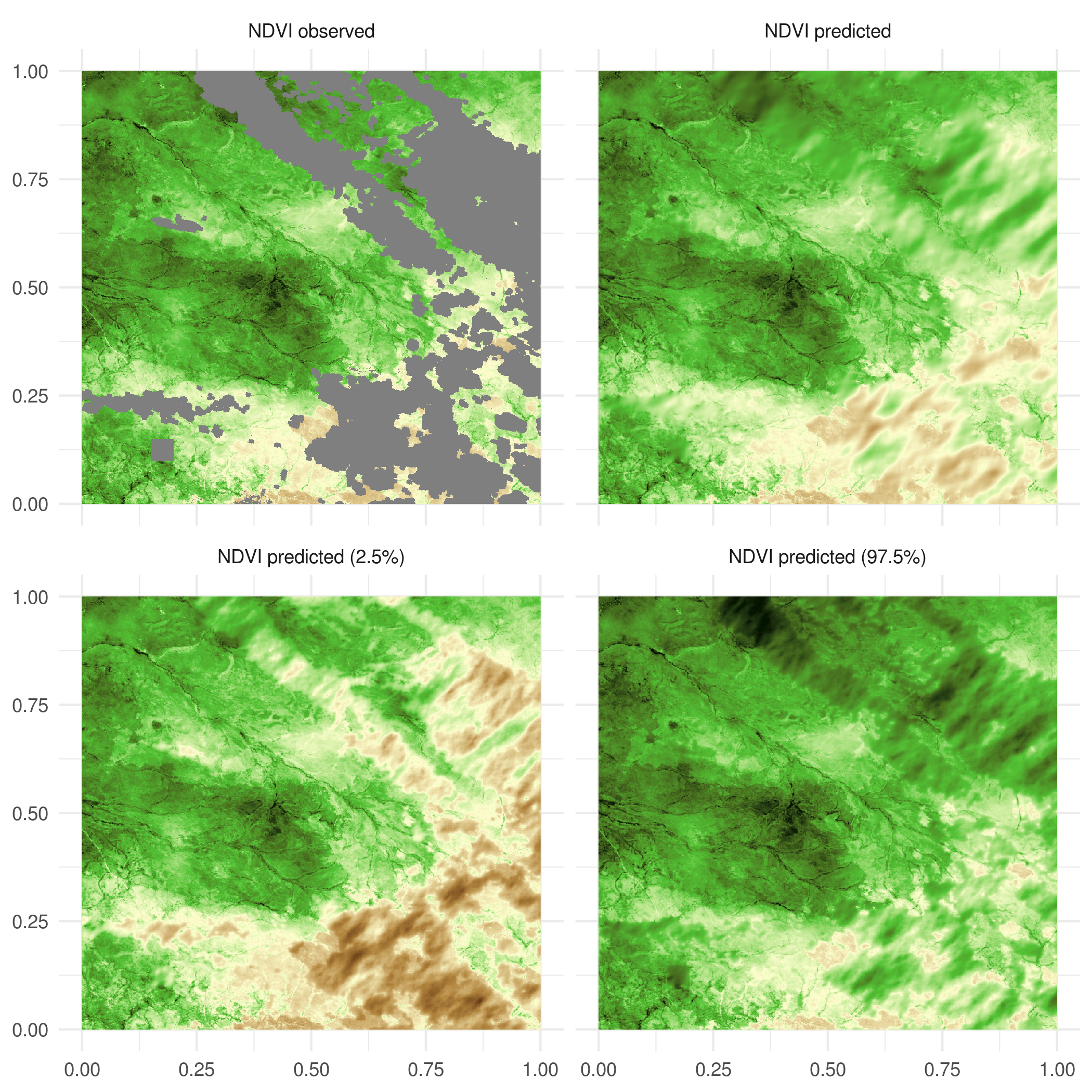}
	\caption{NDVI data and multivariate Q-MGP predictions for the Serengeti region on 2016-12-17. 2.5\% and 97.5\% quantiles are in the bottom row.}
	\label{fig:mv:serengeti} 
\end{figure}

\begin{table}[ht]
\centering
\begin{tabular}{rrrr}
& $\btheta$ \\
  \hline
$\sigma_{11}$ & $3.1889_{(3.1965, 3.2049)}$ & 
  $\sigma_{12}$ & $0.0081_{(0.0081, 0.0082)}$ \\ 
  $\sigma_{13}$ & $0.3015_{(0.3019, 0.3023)}$ & 
  $\sigma_{21}$ & $7.1617_{(7.1846, 7.2075)}$\\ 
  $\sigma_{22}$ & $0.2697_{(0.2701, 0.2705)}$ &
  $\sigma_{23}$ & $0.2731_{(0.2738, 0.2747)}$ \\ 
  $\phi_1$ & $26.5273_{(26.6811, 26.8183)}$ &
  $\phi_2$ & $3.4674_{(3.4781, 3.4867)}$ \\ 
  $\phi_3$ & $0.7512_{(0.7520, 0.7527)}$ & 
  $\alpha$ & $3.1094_{(3.1153, 3.1215)}$\\ 
  $\beta$ & $0.0024_{(0.0024, 0.0024)}$ & 
  $\phi$ & $1.0615_{(1.0632, 1.0651)}$ \\ 
  $v_{12}$ & $0.6281_{(0.6292, 0.6306)}$ & & \\
  $v_{13}$ & $0.2997_{(0.3008, 0.3019)}$ & 
  $v_{23}$ & $1.0079_{(1.0092, 1.0107)}$ \\ 
   \hline
\end{tabular}
\caption{Posterior means with 95\% credible intervals for the covariance parameters.}
	\label{tab:mv:serengeti} 
\end{table}

\section{Comparisons with methods in \cite{Heaton2019}} \label{heaton}
In recent work, \cite{Heaton2019} have reviewed and compared 13 state-of-the-art models for large spatial datasets in a predictive challenge involving (1) simulated and (2) real-world spatial data (daytime land surface temperatures as measured by the Terra instrument onboard the MODIS satellite on August 4, 2016, Level-3 data). The two datasets are available at \url{github.com/finnlindgren/heatoncomparison}. The total number of locations is the same in both cases ($n_{\text{all}} = 150,000$), with the goal of predicting outcomes when missing. Both datasets include $n_{\text{train}} = 105,569$ available locations; the test set is of size $n_{\text{test}} = 44,431$ for the simulated data, $n_{\text{test}} = 42,740$ for the MODIS data due to cloud cover. 
\begin{figure}
	\centering
	\includegraphics[width=.7\textwidth]{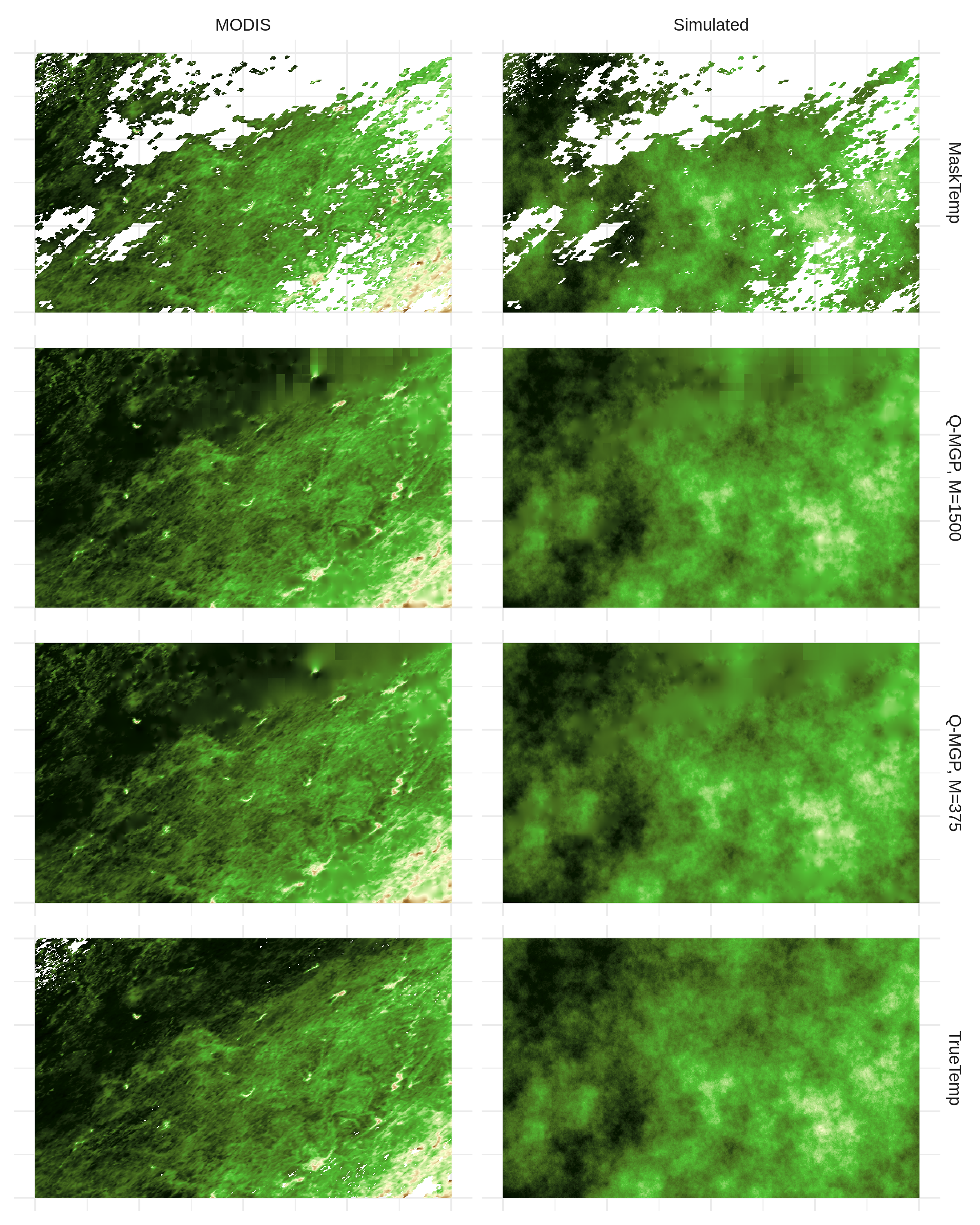}
	\caption{Spatial data and predicted temperatures for Q-MGP models. Top: grey areas correspond to missing observations. Middle rows: spatial predictions. Bottom: the true outcomes. Left: MODIS data; Right: simulated data.}
	\label{heaton:figure} 
\end{figure}

We estimate two \gpname \ models for each dataset by partitioning the spatial domain into $M = 1500$ (resp. $375$) rectangular regions for an average block size of $100$ (resp. $400$) spatial locations, and fix $\calG$ as a cubic mesh. We assign blocks to $\calS$ to cover the observed locations; the remaining ones will be used for prediction as in the right plot of Figure \ref{fig:Figure_refset}. Finally, we use an exponential base covariance function. We ran our Gibbs sampler 
using $\sigmasq \sim Inv.Gamma(2.01, 1)$, $\tau^2 \sim Inv.Gamma(2.01, 1)$, $\phi \sim U[1/10, 30]$ as priors, and respectively $10, 1, 10$ as starting values. A log-Normal proposal function was used for the Metropolis update of $\phi$. We ran a total of 6,000 Monte Carlo iterations, with 4,000 dropped as burn-in, and thinning 2:1 the remaining ones to obtain a sample of size 1,000 that approximates the joint posterior distribution $p(\bw, \phi, \sigmasq, \tau^2 \mid \by)$ which we then use to obtain predictions. In both cases, the algorithm ran on 40 threads. We report the results of our analysis at the top of Table \ref{heaton:table} along with results from select models tested in \cite{Heaton2019} to facilitate comparisons, whereas prediction plots for both datasets are in Figure \ref{heaton:figure}. 

Refer to the cited work for a more in-depth overview. We selected: nearest-neighbor Gaussian process (NNGP), conjugate \citep{nngp_algos} and response \citep{nngp} algorithms; the multiresolution approximation (MRA) model of \cite{katzfuss_jasa17}; a stochastic partial differential equations (SPDE) approach estimated via integrated nested Laplace approximations \citep{inla, spde}; metakriging \citep{metakriging}.

\begin{table}[]
	\centering
	\resizebox{.95\columnwidth}{!}{
		\begin{tabular}{ccccccc}
			\textbf{Simulated data} & \textbf{MAE}    & \textbf{RMSE}   & \textbf{Coverage} & \textbf{Run Time} & \textbf{N. Cores} & \textbf{Code Lang.} \\
			\hhline{=======}
			
			\hline
			Q-MGP, $M=1500$ & 0.6270 & 0.8721 & \textbf{95.46} & 16.4 min  & 40 & C++ (Armadillo)       \\
			Q-MGP, $M=375$ & \textbf{0.6136} & 0.8407 & \textbf{95.62} & 133.2 min  & 40 & C++ (Armadillo)       \\
			\hline
			NNGP Conjugate & 0.65 & 0.88 & 96 & 1.99 min & 10 & C++ \\
			NNGP Response & 0.65 & 0.88 & 96 & 45.56 min & 10 & C++ \\
			MRA & \textbf{0.61} & \textbf{0.83} & 93 & 13.57 min  & 1  & Matlab      \\
			SPDE & 0.62 & 0.86 & 100 & 138.34 min & 2 & R (\texttt{inla})\\
			Metakriging & 0.74 & 97 & 99 & 2888.89 min & 30 & R (\texttt{spBayes}) 
			\vspace{.5cm}
	\end{tabular}}
	\resizebox{\columnwidth}{!}{
		\begin{tabular}{ccccccc}
			\textbf{MODIS data} & \textbf{MAE}    & \textbf{RMSE}   & \textbf{Coverage} & \textbf{Run Time} & \textbf{N. Cores} & \textbf{Code Lang.} \\
			\hhline{=======}
			Q-MGP, $M=1500$ & 1.1151 & 1.5598 & \textbf{95.14} & 16.5 min  & 40 & C++ (Armadillo)   \\
			Q-MGP, $M=375$ &\textbf{1.0729} & \textbf{1.5034} & \textbf{95.20} & 133.7 min  & 40  & C++ (Armadillo)   \\  
			\hline
			NNGP Conjugate & 1.21 & 1.64 & \textbf{95} & 2.06 min & 10 & C++ \\
			NNGP Response & 1.24 & 1.68 & 94 & 42.85 min & 10 & C++ \\
			MRA & 1.33 & 1.85 & 92 & 15.61 min  & 1  & Matlab      \\
			SPDE & 1.10 & 1.53 & 97 & 120.33 min & 2 & R (\texttt{inla}) \\
			Metakriging & 2.08 & 2.50 & 89 & 2888.52 min & 30 & R (\texttt{spBayes}) 
		\end{tabular}
	}
	\caption{Comparative performance of Q-MGP and select methods in \cite{Heaton2019}. All values related to methods other than Q-MGPs have been taken from Tables 2 and 3 of \cite{Heaton2019}. Refer to the cited article for additional comparisons on the same datasets.} \label{heaton:table}
\end{table}

In terms of predictive performance, coverage, and computation time, both implemented models are competitive with the top-ranking ones in \cite{Heaton2019}. In particular, our approach (with $M=1500$) is much faster than Bayesian models estimated via MCMC -- note that \textit{NNGP Conjugate} uses cross-validation to select suitable covariance parameters. The slower alternative with $M=375$ -- corresponding to a much coarser partitioning, i.e. larger regions -- was possible by caching the redundant matrix operations given that the data are on a regular lattice. We chose $M=375$ by targeting a total computing time similar to the SPDE model, which achieved the lowest predictive error in the MODIS dataset.
Computing times should not differ dramatically on a dual-CPU Intel Xeon server as in \cite{Heaton2019}.

\section{C++ implementation of Q-MGP in \texttt{meshgp}} \label{codeinfo}
    All applications were run on a 512GB-memory, dual-CPU server with two Intel Xeon E5-2699v3 processors, each with 18 cores at 2.3Ghz each, running Debian linux and R version 3.6 linked to the OpenBLAS libraries. Source code for implementing Q-MGP models is available as R package \texttt{meshgp} (\meshgpurl). The \texttt{meshgp} package is written in Armadillo (\citealt{armadillo}, a C++ library for linear algebra) which can easily be linked to high performance libraries. Parallelization of all high-dimensional operations is implemented via OpenMP \citep{dagum1998openmp}. R packages \texttt{Rcpp} \citep{rcpp} and \texttt{RcppArmadillo} \citep{rcpparmadillo} are used for interfacing the C++ source with R. 
	
\section{Caching algorithm} \label{appx:caching}
\vspace{0.8cm}
\footnotesize
\begin{algorithm}[H] \label{algorithm:caching}
\SetKwInput{KwStart}{Start}
\SetKwInput{KwInitialize}{Initialize}
\SetKwInput{KwInput}{Input}
\SetKwInput{KwOutput}{Output}

\KwInput{$\bGamma = \{ \Gamma_i  \}_{i=1}^g$: a collection of matrices of size $n_{\Gamma_i} \times d$ for $i=1, \dots, g$. Each row is denoted  $\matrow{\Gamma_i}{r}$ for $r = 1, \dots, n_{\Gamma_i}$.}
\For{$i \in \{1, \dots, g \}$}{
\begin{itemize}
\setlength\itemsep{-.5em}
\item Sort $\Gamma_i$ using column 1; resolve ties using columns 2 to $d$.
\item Calculate $\Gamma_{0,i}$ as the matrix of size $n_{\Gamma_i} \times d$ such that\\ {\hfil $\matrow{\Gamma_{0,i}}{r} = \matrow{\Gamma_i}{1}$ for $r = 1, \dots, n_{\Gamma_i}$.}
\item Calculate $\tilde{\Gamma}_i = \Gamma_i - \Gamma_{0, i}$. 
\end{itemize}
}
\KwInitialize{$\Delta_\Gamma = \{ \delta_1, \dots, \delta_g \}$ where $\delta_i = i$ for all $i=1, \dots, g$.}
\For{$i \in \{1, \dots, g \}$}{
\For{$j \in \{1, \dots, i\}$}{
\uIf{$\tilde{\Gamma}_i = \tilde{\Gamma}_j$ (element by element)}{
Set $\delta_i = j$.
}
}
}
\KwOutput{ $\Delta_\Gamma$: a dictionary with $g$ keys and $g^*$ unique values. If $f$ is a function such that $f(\Gamma + t) = f(\Gamma)$ then $f(\Gamma_{\delta_i}) = f(\Gamma_j)$ for all $i$ such that $\delta_i = j$.}
\vspace{0.5cm}
\caption{Caching algorithm.}
\end{algorithm}
\vspace{0.4cm}

\normalsize
Algorithm \ref{algorithm:caching} can be used by taking $\bGamma = \{ (\calS_j, \Spa{j}) \}_{j=1:M}$ for $\Cov_{\calS_j, \Spa{j}}$,  $\bGamma = \{ \calS_j \}_{j=1:M}$ for $\Cov_{\calS_j}$, and $\bGamma = \{ \Spa{j} \}_{j=1:M}$ for $\Cov_{\Spa{j}}^{-1}$. The resulting dictionaries can be used to build a dictionary $\Delta_{\bR}$ for $\{ \bR_j \}_{j=1:M}$ with $M^*$ unique values. Each iteration of the Gibbs sampler becomes cheaper if the number of unique values in $\Delta_{\bR}$ is $M^* \ll M$.

\section{Compute time comparisons with NNGPs} \label{appx:cache_parallel}
\begin{figure}[H]
	\centering
	\includegraphics[width=.6\textwidth]{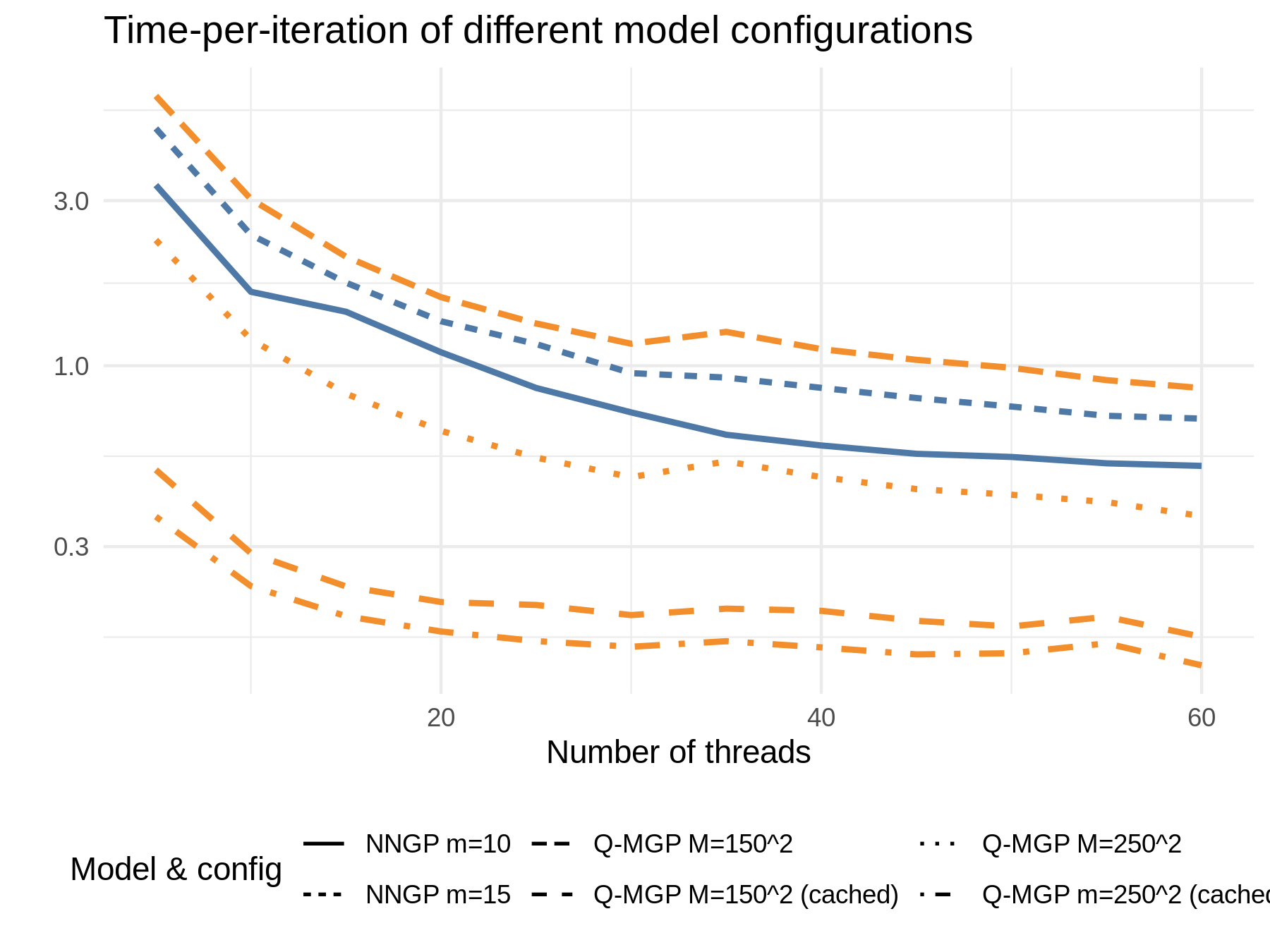}
	\caption{Computation time per iteration, in logarithmic scale, of Q-\gpname\ and NNGP at different configurations.}
	\label{parallel:figure} 
\end{figure}

Figure \ref{parallel:figure} shows the time-per-iteration of Q-\gpname, on a fully-observed slice of the Serengeti data, $n=1,000,000$, choosing $M \in \{ 150^2 , 250^2 \}$, and with or without caching. $M=250^2$ corresponds to blocks with $16$ locations (all blocks are of dimension $4\times4$), whereas $M=150^2$ to blocks of dimension $36, 42, 49$ corresponding to squares of size $6\times 6, 6\times7 \text{ or } 7\times 6, 7\times 7$, respectively. Caching is optional; in this case we showcase the speed differential between a favorable case (regular lattice) and the worst case (completely irregularly-spaced observations) scenarios by disabling caching.
We compare our results with the NNGP sequential sampler of \cite{nngp} as implemented in the \texttt{spNNGP} package for \texttt{R}, choosing $m\in \{10, 15\}$. This fully Bayesian NNGP recovers the spatial random effects and is also estimated via a Gibbs sampler, but unlike a Q-\gpname, it does not use caching or sample $\bw$ in parallel. Computing times of Q-MGP models without caching are comparable to NNGP models; the Q-MGP model with block size $16$ is about 30\% faster than the NNGP model with 10 neighbors. 

The optional caching feature results in a much smaller computational penalty when considering large regions. The large-region model is 25\% slower than the small-region model when caching is activated, but 250\% slower if caching is disabled because many more large matrix inverses are computed. The caching feature of Q-MGP models allowed us to design the best-performing model for the real-world application of \cite{Heaton2019}, see Table \ref{heaton:table}.

\section{Effective sample size of MCMC posterior samples} \label{appx:mixing}
Markov-chain Monte Carlo algorithms output correlated samples from the posterior distribution, and their efficiency is decreased when successive samples are highly autocorrelated. The effective sample size (ESS) is the size of an independent sample equivalent to the correlated sample at hand. Highly autocorrelated Markov chains result in lower ESS. This scenario arises in geostatistical settings with large-scale spatial dependency, as the high-dimensional latent spatial random effects are highly correlated at nearby locations. For this reason, Gibbs samplers such as the one proposed in this article and the sequential NNGP sampler of \cite{nngp} are expected to be negatively affected by slowly-decaying spatial correlation. 

\begin{figure}[]
	\centering
	\includegraphics[width=.45\textwidth]{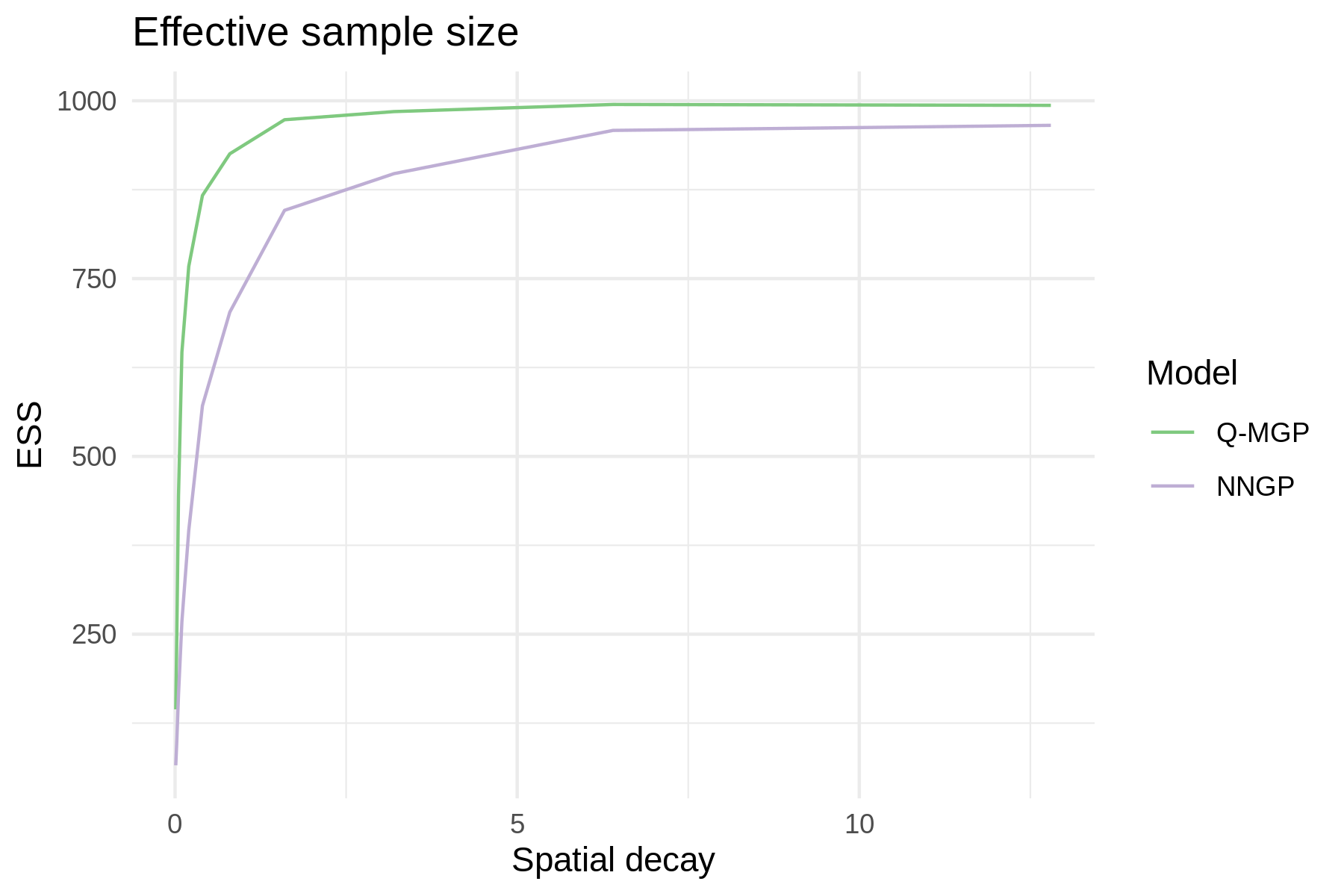}
	\includegraphics[width=.45\textwidth]{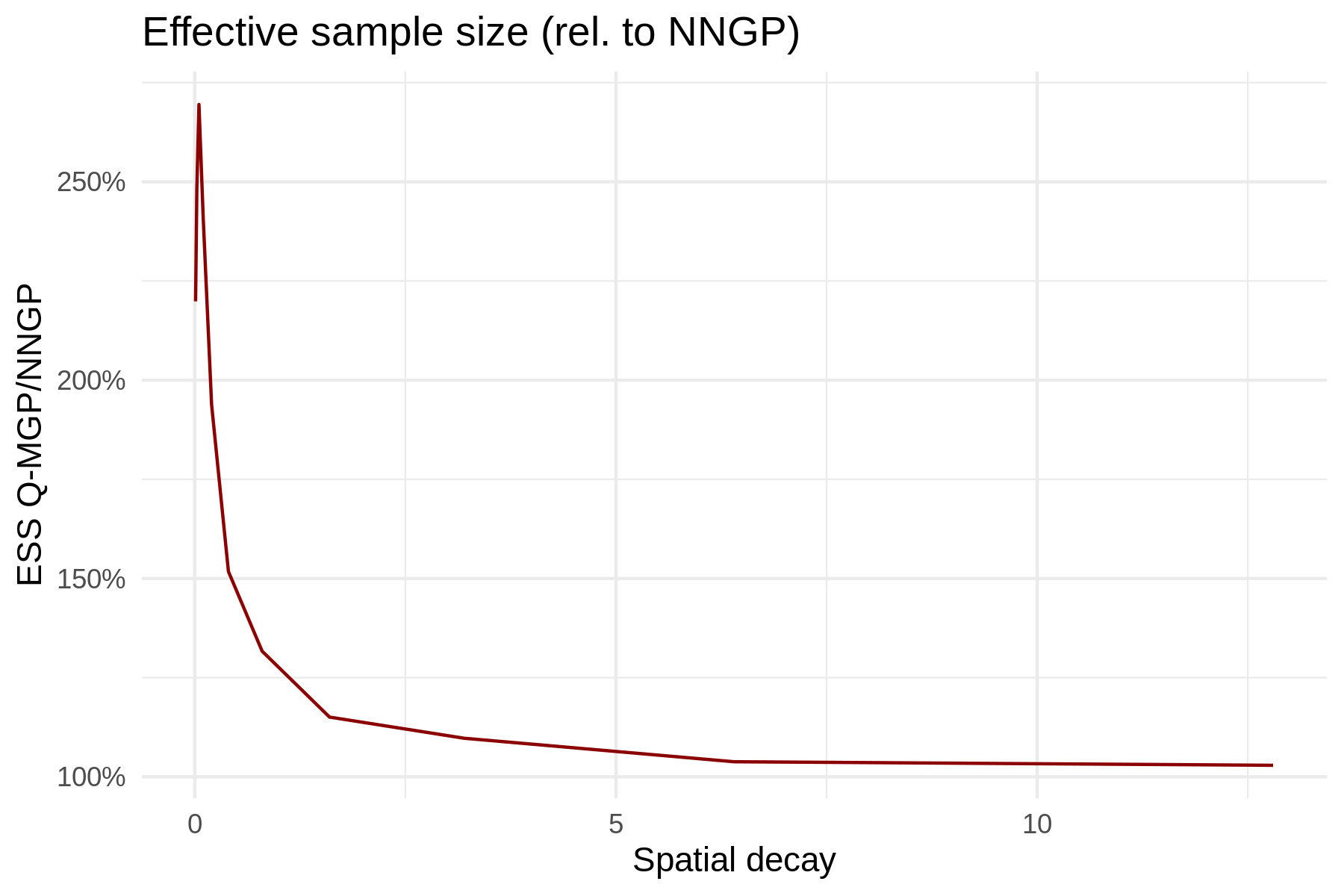}
	\caption{Effective sample size (ESS) of Q-MGP and NNGP for different values of the spatial exponential decay parameter $\phi$.}
	\label{fig:ESS} 
\end{figure}

With this in mind, we now compare the parallel sampler of Q-MGPs to a sequential NNGP sampler (from R package \texttt{spNNGP}, \citealt{spnngp_rpack}) in terms of ESS of the latent random effects. We generate $\by = \bZ \bw + \beps$ on a regular $80\times80$ grid where $\bZ = I_n$, $n=6400$, the spatial process is $\bw \sim GP(0, \Cov)$, $\Cov(\| \bh \|) = \sigmasq \exp \{-\phi \|\bh\| \}$ for $\bh = \bl'-\bl$, $\beps(\bl) \sim N(0, \tau^2)$ for all $\bl$ and using parameters $\sigmasq = 1$ and $\phi \in \{0.01, 0.025, 0.05, .1, .2, .4, .8, 1.6, 3.2, 6.4, 12.8\}$, and nugget $\tau^2 = 0.01$. The goal is to sample $\bw$ a posteriori. We assign priors $\phi \sim U[.01, 30]$, $\sigmasq \sim Inv.G.(2.01, 1)$, $\tau^2 \sim Inv.G(2.01, 1)$ and use the true values of $\{\sigmasq, \phi, \tau^2\}$ as starting values for the Gibbs sampler of both NNGP and Q-MGP. The starting value of $\bw$ was set to a vector of zeros of size $n$. We let the Markov chain run for a total of 2000 iterations, discarding the first 1000. The NNGP model used $m=10$ neighbors, whereas the Q-MGP model used $M=270$ regions obtained by axis-parallel partitioning the spatial domain along the two dimensions into $15$ and $18$ intervals, respectively. We chose $M=270$ to achieve about the same computation time when caching was disabled (actual Q-MGP run times were about 2.6 to 3 times faster than NNGP-sequential once caching was turned on).
Figure \ref{fig:ESS} shows the effective sample size of the resulting approximated posterior sample of $\bw$, averaged across all spatial locations. While both models evidence a degradation of performance with long-range spatial dependence (small $\phi$), the NNGP model shows significantly worse performance. With low spatial decay, the Q-MGP model exhibits effective sample sizes up to 2.5 times larger than the equivalent NNGP model. The smaller effective sample size of NNGPs implies a longer effective runtime to achieve similar performance. Coupled with speed advantages due to caching, we estimate Q-MGP models to result in an effective improvement in compute times up to one order of magnitude over sequential NNGPs. 

\section{Using MGPs with $d=1$}
\begin{figure}[H]
	\centering
	\includegraphics[width=.8\textwidth]{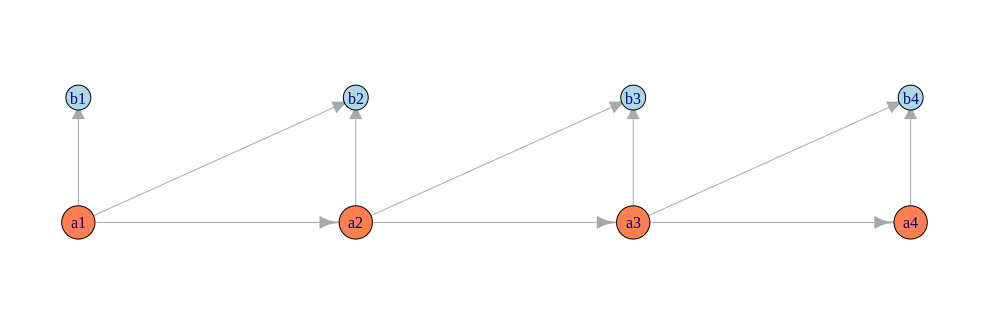}
	\caption{Graph $\calG$ for interval-MGP in unidimensional domains.}
	\label{appx:figure1d} 
\end{figure}
MGPs can be implemented on unidimensional domain by splitting the real line into disjoint intervals, then choosing a finite reference set of locations $\calS \subset \Re$, linking each of them to a reference node based on the intervals they falls into, and using the DAG (e.g. see Figure \ref{appx:figure1d}) to establish a process at all other locations. Grouping the reference locations is not an appealing strategy if the chosen covariance exhibits the screening effect (see e.g. \cite{stein_screening} for a more general discussion), in which case there may be numerical errors in calculating $\bR_j$. In fact, suppose $\bl < \bl'$ for all $\bl \in \calS_j$ and $\bl \in \calS_{j'}$ and $\ba \rightarrow \ba'$ in $\calG$. If $\Cov$ exhibits the screening effect, $p(\bw_{\calS'} \given \bw_{\calS}) = p(\bw_{\calS'} \given \bw(\bl^*))$ where $\bl^* = \max_{\bl}(\calS)$. Therefore, one must either choose a valid covariance function, or reduce the size of the reference sets to 1. 

\section{Recovering missing pixels of an animated GIF image} \label{lion}
\begin{table}[H]
	\centering
	\begin{tabular}{cccc}
		\textbf{Lion GIF} & \textbf{MAE}    & \textbf{RMSE}   & \textbf{Coverage} (90\%) \\
		\hhline{====}
		Q-MGP, $M=1050$ & \textbf{0.0341} & \textbf{0.0715} & \textbf{93.28}  \\
		Gapfill & 0.0480 & 0.0923 & 62.76  \\ 
	\end{tabular}
	\caption{Comparative performance of Q-MGP and Gapfill on recovering missing pixels of an animated GIF image.} \label{lion:table}
\end{table}

\begin{figure}[H]
	\centering
	\includegraphics[width=.85\textwidth]{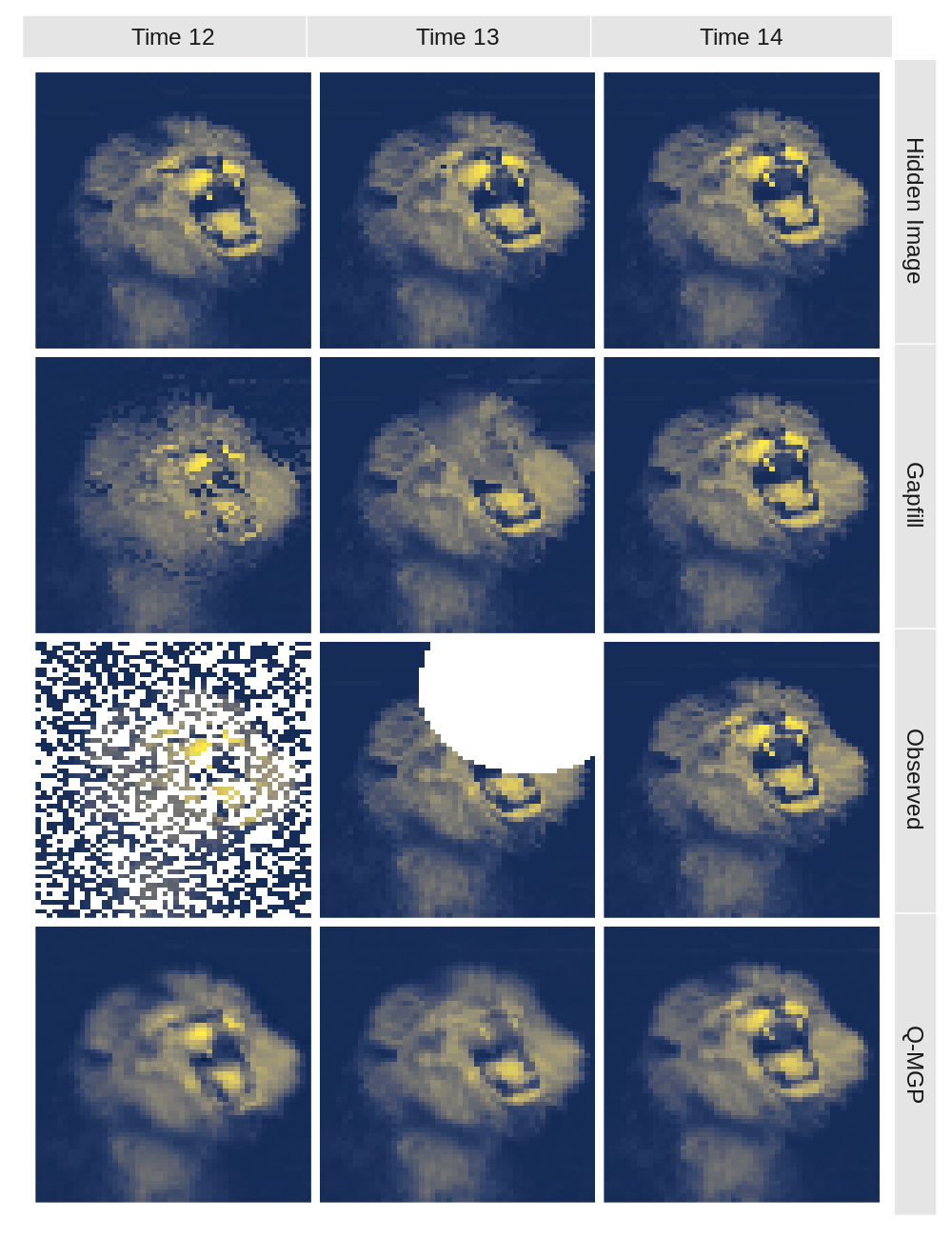}
	\caption{From the top: full original image frames to be recovered, predictions by Q-MGP and Gapfill, and observed image frames, at three select times (of 30 total) of the GIF image.}
	\label{lion:figure} 
\end{figure}

We compare Q-MGPs with Gapfill \citep{gapfill} on non-Gaussian data in the form of an animated GIF image. The original GIF image collects 30 frames, each of size $200\times250$, for a total data size of 1.5 million locations. We subsample the data size to $n_{\text{all}}=94,500$ by downsampling each of the 30 frames at a 16:1 ratio. We simulate cloud cover by placing random clouds in 5 of the 30 frames, covering all but 10 locations in 5 of 30 frames, and covering 50\% random locations in 5 of the 30 frames; 15 frames were thus untouched. See Figure \ref{lion:figure}. The resulting dataset includes $n_{\text{obs}} = 67,363$ non-missing observations and $n_{\text{test}} = 27,137$ missing ones to be predicted. We implement a Q-MGP model with $M=1050$ found by partitioning the spatial coordinates into $5$ and $21$ intervals, and time in $10$ intervals. We run the Gibbs sampler for 15000 iterations; we discard the first 10000 and keep one every five of the remaining 5000 to obtain a sample of size 1000 from the approximated posterior distribution, which we use for predictions. We used the same base covariance, prior distributions, and starting values as in previous sections, 
for a total run time of 35 minutes (0.14s/iteration) on 11 threads.  
As seen in Table \ref{lion:table} and from a visual inspection of Figure \ref{lion:figure}, Q-MGP outperforms Gapfill. 

	\spacingset{1.35}
	\bibliographystyle{hapalike}
	\bibliography{biblio}

\begin{thebibliography}{}

\bibitem[Apanasovich and Genton, 2010]{apanasovich_genton2010}
Apanasovich, T.~V. and Genton, M.~G. (2010).
\newblock {Cross-covariance functions for multivariate random fields based on
  latent dimensions}.
\newblock {\em Biometrika}, 97:15--30.
\newblock \doib{10.1093/biomet/asp078}.

\bibitem[Banerjee, 2017]{sudipto_ba17}
Banerjee, S. (2017).
\newblock High-dimensional {B}ayesian geostatistics.
\newblock {\em Bayesian Analysis}, 12(2):583--614.
\newblock \doib{10.1214/17-BA1056R}.

\bibitem[Banerjee, 2020]{sudipto_ss20}
Banerjee, S. (2020).
\newblock Modeling massive spatial datasets using a conjugate {Bayesian} linear
  modeling framework.
\newblock {\em Spatial Statistics}, in press.
\newblock \doib{10.1016/j.spasta.2020.100417}.

\bibitem[Banerjee et~al., 2010]{gp_pp_biasadj}
Banerjee, S., Finley, A.~O., Waldmann, P., and Ericsson, T. (2010).
\newblock Hierarchical spatial process models for multiple traits in large
  genetic trials.
\newblock {\em Journal of American Statistical Association}, 105(490):506--521.
\newblock \doib{10.1198/jasa.2009.ap09068}.

\bibitem[Banerjee et~al., 2008]{gp_predictive_process}
Banerjee, S., Gelfand, A.~E., Finley, A.~O., and Sang, H. (2008).
\newblock {Gaussian predictive process models for large spatial data sets}.
\newblock {\em Journal of the Royal Statistical Society, Series B},
  70:825--848.
\newblock \doib{10.1111/j.1467-9868.2008.00663.x}.

\bibitem[Bierkens et~al., 2019]{bierkens2019}
Bierkens, J., Fearnhead, P., and Roberts, G. (2019).
\newblock The zig-zag process and super-efficient sampling for bayesian
  analysis of big data.
\newblock {\em The Annals of Statistics}, 47(3):1288--1320.
\newblock \doib{10.1214/18-AOS1715}.

\bibitem[Chen et~al., 2008]{cholmod}
Chen, Y., Davis, T.~A., Hager, W.~W., and Rajamanickam, S. (2008).
\newblock {Algorithm 887: CHOLMOD, Supernodal Sparse Cholesky Factorization and
  Update/Downdate}.
\newblock {\em ACM Trans. Math. Softw.}, 35(3).
\newblock \doib{10.1145/1391989.1391995}.

\bibitem[Christensen and Amemiya, 2003]{Christensen2003}
Christensen, W.~F. and Amemiya, Y. (2003).
\newblock Modeling and prediction for multivariate spatial factor analysis.
\newblock {\em Journal of Statistical Planning and Inference}, 115(2):543 --
  564.
\newblock \doib{10.1016/S0378-3758(02)00173-8}.

\bibitem[Cover and Thomas, 1991]{coverthomas91}
Cover, T.~M. and Thomas, J.~A. (1991).
\newblock {\em Elements of information theory}.
\newblock Wiley Series in Telecommunications and Signal Processing. Wiley
  Interscience.

\bibitem[Cowell et~al., 1999]{graphs}
Cowell, R.~G., Dawid, A.~P., Lauritzen, S.~L., and Spiegelhalter, D.~J. (1999).
\newblock {\em Probabilistic Networks and Expert Systems}.
\newblock Springer-Verlag, New York.

\bibitem[Cressie and Johannesson, 2008]{frk}
Cressie, N. and Johannesson, G. (2008).
\newblock Fixed rank kriging for very large spatial data sets.
\newblock {\em Journal of the Royal Statistical Society, Series B},
  70:209--226.
\newblock \doib{10.1111/j.1467-9868.2007.00633.x}.

\bibitem[Cressie and Zammit-Mangion, 2016]{cressie2016multivariate}
Cressie, N. and Zammit-Mangion, A. (2016).
\newblock {Multivariate spatial covariance models: a conditional approach}.
\newblock {\em Biometrika}, 103(4):915--935.
\newblock \doib{10.1093/biomet/asw045}.

\bibitem[Cressie and Wikle, 2011]{CressieWikle2011}
Cressie, N. A.~C. and Wikle, C.~K. (2011).
\newblock {\em Statistics for spatio-temporal data}.
\newblock Wiley series in probability and statistics. Hoboken, N.J. Wiley.

\bibitem[Dagum and Menon, 1998]{dagum1998openmp}
Dagum, L. and Menon, R. (1998).
\newblock {OpenMP}: an industry standard api for shared-memory programming.
\newblock {\em Computational Science \& Engineering, IEEE}, 5(1):46--55.

\bibitem[Datta et~al., 2016a]{nngp}
Datta, A., Banerjee, S., Finley, A.~O., and Gelfand, A.~E. (2016a).
\newblock Hierarchical nearest-neighbor gaussian process models for large
  geostatistical datasets.
\newblock {\em Journal of the American Statistical Association}, 111:800--812.
\newblock \doib{10.1080/01621459.2015.1044091}.

\bibitem[Datta et~al., 2016b]{nngp_aoas}
Datta, A., Banerjee, S., Finley, A.~O., Hamm, N. A.~S., and Schaap, M. (2016b).
\newblock Nonseparable dynamic nearest neighbor gaussian process models for
  large spatio-temporal data with an application to particulate matter
  analysis.
\newblock {\em The Annals of Applied Statistics}, 10:1286--1316.
\newblock \doib{10.1214/16-AOAS931}.

\bibitem[Davis, 2006]{davis}
Davis, T.~A. (2006).
\newblock {\em Direct Methods for Sparse Linear Systems}.
\newblock SIAM, Philadelphia, PA.
\newblock \doib{10.1137/1.9780898718881}.

\bibitem[Desanker et~al., 2019]{Desanker2019}
Desanker, G., Dahlin, K.~M., and Finley, A.~O. (2019).
\newblock Environmental controls on landsat-derived phenoregions across an east
  african megatransect.
\newblock {\em Ecosphere}, In press.

\bibitem[Eddelbuettel and Fran\c{c}ois, 2011]{rcpp}
Eddelbuettel, D. and Fran\c{c}ois, R. (2011).
\newblock {Rcpp}: Seamless {R} and {C++} integration.
\newblock {\em Journal of Statistical Software}, 40(8):1--18.
\newblock \doib{10.18637/jss.v040.i08}.

\bibitem[Eddelbuettel and Sanderson, 2014]{rcpparmadillo}
Eddelbuettel, D. and Sanderson, C. (2014).
\newblock {RcppArmadillo}: Accelerating {R} with high-performance {C++} linear
  algebra.
\newblock {\em Computational Statistics and Data Analysis}, 71:1054--1063.
\newblock \doib{10.1016/j.csda.2013.02.005}.

\bibitem[Eidsvik et~al., 2014]{block_composite_likelihood}
Eidsvik, J., Shaby, B.~A., Reich, B.~J., Wheeler, M., and Niemi, J. (2014).
\newblock Estimation and prediction in spatial models with block composite
  likelihoods.
\newblock {\em Journal of Computational and Graphical Statistics}, 23:295--315.
\newblock \doib{10.1080/10618600.2012.760460}.

\bibitem[Fearnhead et~al., 2018]{fearnhead2018}
Fearnhead, P., Bierkens, J., Pollock, M., and Roberts, G.~O. (2018).
\newblock Piecewise deterministic {Markov} processes for continuous-time {Monte
  Carlo}.
\newblock {\em Statistical Science}, 33(3):386--412.
\newblock \doib{10.1214/18-STS648}.

\bibitem[Finley et~al., 2012]{pp_spacetime}
Finley, A.~O., Banerjee, S., and Gelfand, A.~E. (2012).
\newblock Bayesian dynamic modeling for large space-time datasets using
  {Gaussian} predictive processes.
\newblock {\em Journal of Geographical Systems}, 14:29--47.
\newblock \doib{10.1007/s10109-011-0154-8}.

\bibitem[Finley et~al., 2020]{spnngp_rpack}
Finley, A.~O., Datta, A., and Banerjee, S. (2020).
\newblock {R package for Nearest Neighbor Gaussian Process models}.
\newblock \arXiv{2001.09111}.

\bibitem[Finley et~al., 2019]{nngp_algos}
Finley, A.~O., Datta, A., Cook, B.~D., Morton, D.~C., Andersen, H.~E., and
  Banerjee, S. (2019).
\newblock Efficient algorithms for {Bayesian} nearest neighbor {Gaussian}
  processes.
\newblock {\em Journal of Computational and Graphical Statistics}, 28:401--414.
\newblock \doib{10.1080/10618600.2018.1537924}.

\bibitem[Furrer et~al., 2006]{taper1}
Furrer, R., Genton, M.~G., and Nychka, D. (2006).
\newblock Covariance tapering for interpolation of large spatial datasets.
\newblock {\em Journal of Computational and Graphical Statistics}, 15:502--523.
\newblock \doib{10.1198/106186006X132178}.

\bibitem[Genton and Kleiber, 2015]{genton_ccov}
Genton, M.~G. and Kleiber, W. (2015).
\newblock Cross-covariance functions for multivariate geostatistics.
\newblock {\em Statistical Science}, 30:147--163.
\newblock \doib{10.1214/14-STS487}.

\bibitem[Gerber et~al., 2018]{gapfill}
Gerber, F., Furrer, R., Schaepman-Strub, G., de~Jong, R., and Schaepman, M.~E.
  (2018).
\newblock Predicting missing values in spatio-temporal remote sensing data.
\newblock {\em IEEE Transactions on Geoscience and Remote Sensing},
  56(5):2841--2853.
\newblock \doib{10.1109/TGRS.2017.2785240}.

\bibitem[Gneiting, 2002]{gneiting2002}
Gneiting, T. (2002).
\newblock Nonseparable, stationary covariance functions for space-time data.
\newblock {\em Journal of the American Statistical Association}, 97:590--600.
\newblock \doib{10.1198/016214502760047113}.

\bibitem[Gonzalez et~al., 2011]{parallel_gibbs}
Gonzalez, J., Low, Y., Gretton, A., and Guestrin, C. (2011).
\newblock Parallel {Gibbs} sampling: From colored fields to thin junction
  trees.
\newblock In Gordon, G., Dunson, D., and Dudík, M., editors, {\em Proceedings
  of the Fourteenth International Conference on Artificial Intelligence and
  Statistics}, volume~15 of {\em Proceedings of Machine Learning Research},
  pages 324--332, Fort Lauderdale, FL, USA. PMLR.

\bibitem[Gramacy and Apley, 2015]{gramacy_apley14}
Gramacy, R.~B. and Apley, D.~W. (2015).
\newblock Local {Gaussian} process approximation for large computer
  experiments.
\newblock {\em Journal of Computational and Graphical Statistics}, 24:561--578.
\newblock \doib{10.1080/10618600.2014.914442}.

\bibitem[Gramacy and Lee, 2008]{treed}
Gramacy, R.~B. and Lee, H. K.~H. (2008).
\newblock {Bayesian} treed {Gaussian} process models with an application to
  computer modeling.
\newblock {\em Journal of the American Statistical Association},
  103:1119--1130.
\newblock \doib{10.1198/016214508000000689}.

\bibitem[Guhaniyogi and Banerjee, 2018]{metakriging}
Guhaniyogi, R. and Banerjee, S. (2018).
\newblock Meta-kriging: Scalable bayesian modeling and inference for massive
  spatial datasets.
\newblock {\em Technometrics}, 60(4):430--444.
\newblock \doib{10.1080/00401706.2018.1437474}.

\bibitem[Guhaniyogi et~al., 2011]{pp_adaptive_knots}
Guhaniyogi, R., Finley, A.~O., Banerjee, S., and Gelfand, A.~E. (2011).
\newblock {Adaptive Gaussian predictive process models for large spatial
  datasets}.
\newblock {\em Environmetrics}, 22:997--1007.
\newblock \doib{10.1002/env.1131}.

\bibitem[Guinness, 2018]{guinness_techno}
Guinness, J. (2018).
\newblock Permutation and grouping methods for sharpening {Gaussian} process
  approximations.
\newblock {\em Technometrics}, 60(4):415--429.
\newblock \doib{10.1080/00401706.2018.1437476}.

\bibitem[Heaton et~al., 2019]{Heaton2019}
Heaton, M.~J., Datta, A., Finley, A.~O., Furrer, R., Guinness, J., Guhaniyogi,
  R., Gerber, F., Gramacy, R.~B., Hammerling, D., Katzfuss, M., Lindgren, F.,
  Nychka, D.~W., Sun, F., and Zammit-Mangion, A. (2019).
\newblock A case study competition among methods for analyzing large spatial
  data.
\newblock {\em Journal of Agricultural, Biological and Environmental
  Statistics}, 24(3):398--425.
\newblock \doib{10.1007/s13253-018-00348-w}.

\bibitem[Katzfuss, 2017]{katzfuss_jasa17}
Katzfuss, M. (2017).
\newblock A multi-resolution approximation for massive spatial datasets.
\newblock {\em Journal of the American Statistical Association}, 112:201--214.
\newblock \doib{10.1080/01621459.2015.1123632}.

\bibitem[Katzfuss and Guinness, 2017]{katzfuss_vecchia}
Katzfuss, M. and Guinness, J. (2017).
\newblock {A general framework for Vecchia approximations of Gaussian
  processes}.
\newblock \arXiv{1708.06302}.

\bibitem[Kaufman et~al., 2008]{taper2}
Kaufman, C.~G., Schervish, M.~J., and Nychka, D.~W. (2008).
\newblock Covariance tapering for likelihood-based estimation in large spatial
  data sets.
\newblock {\em Journal of the American Statistical Association},
  103:1545--1555.
\newblock \doib{10.1198/016214508000000959}.

\bibitem[Lauritzen, 1996]{lauritzen}
Lauritzen, S., L. (1996).
\newblock {\em {Graphical Models}}.
\newblock Clarendon Press, Oxford, UK.

\bibitem[Lewis, 2016]{lewis2016}
Lewis, R. (2016).
\newblock {\em A guide to graph colouring}.
\newblock Springer International Publishing.
\newblock \doib{10.1007/978-3-319-25730-3}.

\bibitem[Lindgren et~al., 2011]{spde}
Lindgren, F., Rue, H., and Lindström, J. (2011).
\newblock {An explicit link between Gaussian fields and Gaussian Markov random
  fields: the stochastic partial differential equation approach}.
\newblock {\em Journal of the Royal Statistical Society: Series B},
  73:423--498.
\newblock \doib{10.1111/j.1467-9868.2011.00777.x}.

\bibitem[Lopes et~al., 2008]{Lopes2008}
Lopes, H.~F., Salazar, E., and Gamerman, D. (2008).
\newblock Spatial dynamic factor analysis.
\newblock {\em Bayesian Analysis}, 3(4):759 -- 792.
\newblock \doib{10.1214/08-BA329}.

\bibitem[Molloy and Reed, 2002]{molloyreed2002}
Molloy, M. and Reed, B. (2002).
\newblock {\em Graph colouring and the probabilistic method}.
\newblock Springer-Verlag Berlin Heidelberg.
\newblock \doib{10.1007/978-3-642-04016-0}.

\bibitem[Quiroz et~al., 2019]{prates}
Quiroz, Z.~C., Prates, M.~O., and Dey, D.~K. (2019).
\newblock Block nearest neighboor {Gaussian} processes for large datasets.
\newblock \arXiv{1908.06437}.

\bibitem[Ren and Banerjee, 2013]{Ren2013}
Ren, Q. and Banerjee, S. (2013).
\newblock Hierarchical factor models for large spatially misaligned data: A
  low-rank predictive process approach.
\newblock {\em Biometrics}, 69(1):19--30.
\newblock \doib{10.1111/j.1541-0420.2012.01832.x}.

\bibitem[Royle and Berliner, 1999]{royle1999hierarchical}
Royle, J.~A. and Berliner, L.~M. (1999).
\newblock A hierarchical approach to multivariate spatial modeling and
  prediction.
\newblock {\em Journal of Agricultural, Biological, and Environmental
  Statistics}, 1(4):29--56.
\newblock \doib{10.2307/1400420}.

\bibitem[Rue and Held, 2005]{grmfields}
Rue, H. and Held, L. (2005).
\newblock {\em Gaussian Markov Random Fields: Theory and Applications}.
\newblock Chapman \& Hall/CRC.
\newblock \doib{10.1201/9780203492024}.

\bibitem[Rue et~al., 2009]{inla}
Rue, H., Martino, S., and Chopin, N. (2009).
\newblock {Approximate Bayesian inference for latent Gaussian models by using
  integrated nested Laplace approximations}.
\newblock {\em Journal of the Royal Statistical Society: Series B},
  71:319--392.
\newblock \doib{10.1111/j.1467-9868.2008.00700.x}.

\bibitem[Sanderson and Curtin, 2016]{armadillo}
Sanderson, C. and Curtin, R. (2016).
\newblock Armadillo: a template-based {C++} library for linear algebra.
\newblock {\em Journal of Open Source Software}, 1:26.

\bibitem[Sang and Huang, 2012]{fsa}
Sang, H. and Huang, J.~Z. (2012).
\newblock A full scale approximation of covariance functions for large spatial
  data sets.
\newblock {\em Journal of the Royal Statistical Society, Series B},
  74:111--132.
\newblock \doib{10.1111/j.1467-9868.2011.01007.x}.

\bibitem[Stein, 2002]{stein_screening}
Stein, M.~L. (2002).
\newblock The screening effect in kriging.
\newblock {\em The Annals of Statistics}, 30(1):298--323.
\newblock \doib{10.1214/aos/1015362194}.

\bibitem[Stein, 2014]{stein2014}
Stein, M.~L. (2014).
\newblock Limitations on low rank approximations for covariance matrices of
  spatial data.
\newblock {\em Spatial Statistics}, 8:1--19.
\newblock \doib{doi:10.1016/j.spasta.2013.06.003}.

\bibitem[Stein et~al., 2004]{steinetal2004}
Stein, M.~L., Chi, Z., and Welty, L.~J. (2004).
\newblock {Approximating likelihoods for large spatial data sets}.
\newblock {\em Journal of the Royal Statistical Society, Series B},
  66:275--296.
\newblock \doib{10.1046/j.1369-7412.2003.05512.x}.

\bibitem[Sun et~al., 2011]{sunligenton}
Sun, Y., Li, B., and Genton, M. (2011).
\newblock {Geostatistics for large datasets}.
\newblock In Montero, J., Porcu, E., and Schlather, M., editors, {\em Advances
  and Challenges in Space-time Modelling of Natural Events}, pages 55--77.
  Springer-Verlag, Berlin Heidelberg.
\newblock \doib{10.1007/978-3-642-17086-7}.

\bibitem[Taylor-Rodriguez et~al., 2019]{taylor2019spatial}
Taylor-Rodriguez, D., Finley, A.~O., Datta, A., Babcock, C., Andersen, H.~E.,
  Cook, B.~D., Morton, D.~C., and Banerjee, S. (2019).
\newblock Spatial factor models for high-dimensional and large spatial data: An
  application in forest variable mapping.
\newblock {\em Statistica Sinica}, 29(3):1155--1180.
\newblock \doib{10.5705/ss.202018.0005}.

\bibitem[Vecchia, 1988]{vecchia88}
Vecchia, A.~V. (1988).
\newblock Estimation and model identification for continuous spatial processes.
\newblock {\em Journal of the Royal Statistical Society, Series B},
  50:297--312.
\newblock \doib{10.1111/j.2517-6161.1988.tb01729.x}.

\end{thebibliography}
	
\end{document}